\documentclass{amsart}
\usepackage[cp1250]{inputenc}
\usepackage[T1]{fontenc}
\usepackage{amsfonts}
\usepackage{indentfirst}
\usepackage{graphicx}
\graphicspath{{figures/}}
\usepackage[dvips]{epsfig}
\usepackage{amsmath, amsthm}
\usepackage{mathtools}  
\usepackage{multirow}
\usepackage{latexsym}
\usepackage{cite} 

\usepackage[font=small,labelfont=bf, width =0.9\linewidth]{caption} 

\newcounter{expcounter}			
\newenvironment{experiment}
{\refstepcounter{expcounter}
	{~\\\textbf{Experiment \theexpcounter~---~}}
}

\usepackage{parskip}

\def\D{\mathbb{D}}
\def\R1{\mathbb{R}}
\def\Rn{\mathbb{R}^n}
\makeatletter
\newcommand{\doublewidetilde}[1]{{%
  \mathpalette\double@widetilde{#1}%
}}
\newcommand{\double@widetilde}[2]{%
  \sbox\z@{$\m@th#1\widetilde{#2}$}%
  \ht\z@=.9\ht\z@
  \widetilde{\box\z@}%
}
\makeatother

\usepackage[colorlinks=true, citecolor=blue, linkcolor=blue]{hyperref}
\author{Zuzanna~Szyma\'nska}
\address{{\it Zuzanna Szyma\'nska:} ICM, University of Warsaw, ul. Tyniecka 15/17, 02-630 Warsaw, Poland}
\email{z.szymanska@icm.edu.pl}
\author{Miros\l{}aw Lachowicz}
\address{{\it Miros\l{}aw Lachowicz:} Institute of Applied Mathematics and Mechanics\\ Faculty of Mathematics, Informatics and Mechanics, University of Warsaw\\
ul. Banacha 2, 02-097 Warsaw, Poland}
\email{lachowic@mimuw.edu.pl}
\author{Nikolaos Sfakianakis}
\address{{\it Nikolaos Sfakianakis:} School of Mathematics and Statistics \\ University St Andrews \\ North Haugh, St Andrews, United Kingdom}
\email{n.sfakianakis@st-andrews.ac.uk}
\author{Mark Chaplain}
\address{{\it Mark Chaplain:} School of Mathematics and Statistics, University of St Andrews \\ North Haugh, St Andrews, United Kingdom}
\email{majc@st-andrews.ac.uk}

\begin{document} 
\title[Modelling invasive foci formation]
{{Mathematical modelling of cancer invasion: Phenotypic transitioning provides insight into multifocal foci formation}}

\begin{abstract}
The transition from the epithelial to mesenchymal phenotype and its reverse (from mesenchymal to epithelial) are crucial processes necessary for the progression and spread of cancer. In this paper, we investigate how phenotypic switching at the cancer cell level impacts on behaviour at the tissue level, specifically on the emergence of isolated foci of the invading solid tumour mass leading to a multifocal tumour. To this end, we propose a new mathematical model of cancer invasion that includes the influence of cancer cell phenotype on the rate of invasion and metastasis. The implications of model are explored through numerical simulations revealing that the plasticity of tumour cell phenotypes appears to be crucial for disease progression and local invasive spread. The computational simulations show the progression of the invasive spread of a primary cancer reminiscent of in vivo multifocal breast carcinomas, where multiple, synchronous, ipsilateral neoplastic foci are frequently observed and are associated with a poorer patient prognosis. 
\end{abstract}


\maketitle


\section{Introduction}\label{introduction}


The epithelial to mesenchymal transition (EMT) is a reversible process of cell phenotype transition from, as the name indicates, the epithelial phenotype to the mesenchymal phenotype. This transition is perceived as a crucial developmental program characterised by loss of cell-cell adhesion, and increased cell mobility~\cite{Kavya2021,Pinzani2011}. The reverse process, called mesenchymal-to-epithelial transition (MET), involves transforming mesenchymal-type cells into epithelial-like cells~\cite{Kavya2021,Pinzani2011}. Both processes are present in at least three different biological contexts~\cite{Kalluri2009}. The first one, often referred to as type 1, is associated with embryonic development and aims to generate diverse cell types that share common mesenchymal phenotypes. This type neither causes fibrosis nor induces an invasive phenotype (which can result in systemic spread via the circulation system). Essentially type 1 EMT generates mesenchymal cells that have the potential to subsequently undergo a reverse transition to regain the epithelial phenotype~\cite{Kalluri2009}. Type 2 EMT is associated with inflammation and occurs during wound healing, tissue regeneration, and organ fibrosis. During these processes, otherwise non-motile epithelial cells pass through EMT, acquire motility properties, and migrate to a wound site. Once inflammation is attenuated, as is seen during most cases of wound healing and tissue regeneration, EMT ceases and cells revert to the epithelial state to restore the integrity of the epithelial barrier~\cite{Kalluri2009}. Finally, type 3 EMT occurs in neoplastic cells that have previously undergone genetic and/or epigenetic changes promoting clonal outgrowth and the development of localised tumours~\cite{Kalluri2009}. Excessive epithelial cell proliferation and tumour-induced angiogenesis are recognised as hallmarks of the early stage of growth of primary epithelial cancers~\cite{Hanahan2011,Kalluri2009}. At some point in the development of a solid tumour cells from the primary cancerous mass start to invade the local tissue surrounding the tumour. This initial invasion of the local tissue is the first step in the complex process of secondary spread where the cancer cells migrate to other locations in the body and set up new life-threatening tumour foci called metastases. These secondary tumours are responsible for around 90\% of all (human) deaths from cancer~\cite{Hanahan2011}.

Within the process of invasion, one can distinguish four basic components: the adhesion of tumour cells to the extracellular matrix (the extracellular matrix being the substance that fills the space between cells that binds cells to tissues and organs), the secretion of matrix-degrading enzymes by tumour cells, the migration of tumour cells and finally their proliferation. The migration of tumour cells is the result of several different mechanisms, the most important of which is haptotaxis - the movement of cells towards a higher density of non-diffusive substances such as collagen or fibronectin which are contained in the extracellular matrix. Migratory tumour cells produce enzymes that digest the extracellular matrix, which in turn leads to the formation of gradients in its density. It is haptotaxis that is most responsible for tissue infiltration by tumour cells. Thus, the phenomenon of invasion is largely understood, but despite many efforts we lack a deeper understanding of the relevant details~\cite{Lowengrub2010}. Nowadays, many efforts are focused on better understanding the mechanisms governing the phenotypic changes leading to increased invasion and the emergence of new tumour foci. That means chiefly creating more sophisticated in vitro assays suitable to investigate scenarios like releasing growth factors during the invasion process~\cite{Nii2029,Poon2022}. 


In this paper, we propose a new mathematical model of cancer invasion explicitly taking into account the processes of EMT and its reverse MET. To investigate the importance of these phenomena on the formation of isolated neoplastic foci, we modify a generic cancer invasion model by incorporating the phenotype as a structured variable describing the density of cancer cells. We chose to model the tumour cell phenotype as a continuous structure because observations indicate that cancer cells may undergo EMT differently, with some cells retaining many epithelial traits while acquiring only some mesenchymal ones, and other cells dropping most vestiges of their epithelial origin and becoming fully mesenchymal~\cite{Kiesslich2013}. In line with the biological evidence, we model transitions of phenotypes using switch functions that, depending on the local concentration of TGF-$\beta$, force an appropriate change of phenotype. To the best of our knowledge, epigenetic changes in phenotypes associated with EMT and MET have not been modelled in this way. The model behaviour suggests a possible mechanism responsible for the emergence of isolated focal colonies of cancer cells, which is clearly illustrated by the numerical simulations we perform.


\section{Related work}\label{related_works}


The first mathematical models describing the phenomenon of cancer invasion appeared in the literature more than 25 years ago immediately gaining much attention as they accurately depict the invasion process qualitatively~\cite{Orme1996,perumpanani1996}. Most invasion models have the form of systems of nonlinear partial differential equations describing the proliferation and migration of cancer cells into the surrounding extracellular matrix. Typically, these models also take into account the activity of matrix-degrading enzymes which are produced by cancer cells, break down the components of extracellular matrix thus facilitating the cancer cells migration. Other variants of the model were proposed depending on the particular area of interest of researchers. Therefore in the literature, there are models, which, in addition to the main variables take into account the effects of other factors~\cite{Andasari2011,Bitsouni2017,Chaplain2005,Meral2013,szymanska2009a}. For instance, an interesting extension of the invasion model was proposed in 2005 by Lolas and Chaplain who described the effect of the  dynamics of the urokinase enzyme on the invasion speed~\cite{Chaplain2005}. Another type of modification was presented in the paper devoted to the study of the effect of heat shock proteins on cancer invasion~\cite{szymanska2009a}. So far, this effect has been linked to the activation of metalloproteinase, one of the major extracellular matrix-degrading enzymes~\cite{Eustace2004}. Based on the results of the simulations and experiments, the authors proposed an alternative hypothesis whereby the effect of heat shock proteins on cancer invasion is achieved by influencing cellular flexibility.

Apart from taking into account various biochemical factors, the structure of the generic model has been also modified to better reflect chosen features of the biological system under consideration. A good example here may be modifications of models to hybrid or non-local ones. Within the hybrid approach, cancer cells are usually modelled as discrete individuals which interact with each other via some type of potential function, while spatiotemporal dynamics of the other variables in the model, such as extracellular matrix, matrix-degrading enzymes or stroma are governed by partial differential equations~\cite{RamisConde2008}. Treating biological cells as discrete individuals has several advantages. First of all, the dynamics of the system are being investigated from the level of a cell, linking its behaviour with the macroscopic condition. Cells are influenced by signals from other cells, external factors, such as nutrient or chemical factors,  stress, and of course their own internal signalling pathways. It provides an easy framework to include different cell types, intracellular dynamics and spatial location of individual cells. The literature involving hybrid models is vast and led to obtaining many interesting results, including both mathematical modelling and methodology development~\cite{Lowengrub2010}. For instance, very interesting results concern the influence of the environment on the emergence of the malignant glycolytic phenotype~\cite{Gerlee2008}, while others examine how individual-based cell interactions can affect the tumour shape~\cite{Anderson2005}. Due to the extreme complexity of the global system, such models often require so-called hybrid approaches to computational implementation, with systems enabling coarse-grain and fine-grain parallel implementation. This has led to the development of several high-performance simulation platforms with cutting-edge, tailored-to-the-problem algorithmic solutions~\cite{Cytowski2014,Cytowski2015,Cytowski2017,szymanska2018,Macklin2018}.

The non-local approach is usually used to model processes where the interacting entities are some distance apart, which leads to partial integro-differential equations. Usually, this means simplifying the description of the investigated processes because even if there are known information-transmitting factors, taking them into account directly would in most cases complicate the model leading to analysis becoming unfeasible. In the past, non-local approaches were used to model the invasion process, for instance, to capture adhesive properties of {\it in vivo} cells~\cite{Wrzosek,gerisch2008}, cell motion~\cite{Surulescu}, cell-matrix interactions \cite{CMR}, variable cell-cell and cell-matrix adhesion \cite{trucu2014}, and cell proliferation \cite{szymanska2021,gwiazda2023}. These models address the problem using non-local integral terms for both cell-cell and cell-matrix adhesion and the concepts of {\it adhesive flux} and {\it cell sensing radius}. Innovative multiscale modelling along with a moving boundary approach has also been pioneered to examine interactions between the cancer cells and the components of the extracellular matrix \cite{trucu2013, trucu2019, trucu2020}. 

More recently raised the idea of describing different or even changing, properties of cancer cells with a variable describing phenotype. The cellular phenotypes were modelled using both discrete and continuous approaches~\cite{Chaplain2019,Chisholm2016,Macfarlane2022}. Focusing on the continuous description which appears to be more suitable to capture epigenetic changes, we note that a continuous structural variable was used to model, for instance, the selection of chemoresistant phenotypes within cancer treatment~\cite{Almeida2019,Lorz2013}. Later that concept was also adopted to model the emergence of phenotypic heterogeneity within a solid tumour~\cite{Lorenzi2018,Villa2021}.

The activation of the EMT program is perceived as the critical mechanism for acquiring malignant phenotypes by epithelial cancer cells. Therefore, the genetic and biochemical mechanisms underlying the acquisition of the invasive phenotype and the subsequent systemic dissemination of the cancer cell is a focus of intensive studies~\cite{Kalluri2009}. The importance of the topic, and the fact that many aspects of the EMT transition associated with cancer are still unclear, the issue also gained interest from the mathematical modelling community. Therefore, in the recent past, several mathematical models aiming to shed more light on the phenomenon have also been proposed \cite{Mark_Linnea_1,Mark_Linnea_2,SfakKolbe2017,Chauviere2010,hybrid_1st, hybrid_2nd, deutsch2012, deutsch2012b}.


\section{The mathematical model}\label{Model}


We consider three functions: $c(t,x,\alpha)$, $v(t,x)$, and $f(t,x)$ describing the cancer cells, tumour micro-environment densities, and growth factors concentrations, respectively. More precisely, the function $c=c(t,x,\alpha )$ describes the density of cancer cells at the instant of time $t\geq 0$, at position $x\in \D$, where $\D$ is a bounded domain in $\Rn$ and $\alpha\in [0,1]$ is a parameter that describes the cell's phenotype along the epithelial-mesenchymal axis. We assume that $\alpha=0$ corresponds to purely epithelial-like cancer cells, $\alpha=1$ corresponds to purely mesenchymal-like cancer cells, and $0<\alpha<1$ stands for intermediate phenotypes. Function $v(t,x)$ describes the density of the tissue surrounding the primary tumour at the instant of time $t\geq 0$, at position $x\in \D$. For simplicity, we represent the surrounding tissue using one function that comprises the description of the extracellular matrix (ECM) and tissue-resident fibroblasts. Finally, to complete the essential description of the system regulating the change in cancer cells' phenotype we consider the main signalling factor responsible involved for the epithelial and mesenchymal transitions, namely transforming growth factor beta (TGF-$\beta$). The function $f(t,x)$ stands for the concentration of TGF-$\beta$ at the instant of time $t\geq 0$, at position $x\in \D$, whose concentration conditions the phenotype of cancer cells, i.e. it shifts towards a more epithelial phenotype or mesenchymal one.

\textbf{Cancer cells:} 
Epithelial cells are characterised by the stable cell-cell junction, apical-basal polarity and interactions with the basement membrane. During EMT cells acquire mesenchymal characteristics, that is fibroblast-like morphology and cytoskeleton architecture, as well as invasive properties including increased migratory capacity~\cite{Yang2020}. There is growing evidence that EMT is triggered in response to signals that cells receive from their micro-environment and the activation and execution of EMT do not require changes in DNA sequence and is as a result of complex epigenetic regulatory programmes. Recent studies support this hypothesis as it was demonstrated that some cell populations might undergo multiple rounds of EMT and MET, indicating substantial phenotypic plasticity~\cite{Yang2020}. Moreover, EMT programmes rarely function as binary switches that toggle between the two extremes of the epithelial-mesenchymal spectrum~\cite{Dongre}. On the contrary, an important feature of the ETM type 3 that was observed {\it in vivo} is that EMT is often incomplete, which results in the emergence of cells that are in intermediate states having both epithelial and mesenchymal characteristics~\cite{Sha_2019}.

Induction of EMT is regulated by extracellular signals and frequently involves cooperation between various signalling pathways among which TGF-$\beta$ plays a dominant role~\cite{Fazilaty2019,Lee2022b}. It was found that TGF-$\beta$ signalling is required not only for inducing EMT but also for maintenance of in vitro invasiveness and metastasis~\cite{Lee2022b,Friedl2011,Ort1998,Xu2009}. Precisely, inhibition of TGF-$\beta$ signalling prevents EMT in epithelial cells and causes MET in mesenchymal tumour cells~\cite{Ort1998,Xu2009,Santos2022}. Phenotypic reverted epithelial cells obtained after MET induction presented epithelial morphologies and proliferation rates resembling E cells~\cite{Santos2022}. TGF-$\beta$ binds to complexes of receptors and activates the crucial factors that trigger EMT, namely proteins from Snail and Prrx1 families~\cite{Fazilaty2019,Du2021,Giampieri2009,Hardin2014,Ocana2012}. In particular, Prrx1 was identified as an EMT inducer that can trigger the whole process independently on Snail1 and in a dose-dependent manner~\cite{Ocana2012}. Upon the expression of Prrx1 cells acquire the mesenchymal-like phenotype and invasive properties, while the loss of Prrx1 allows mesenchymal cells a complete reversion to the epithelial phenotype even in the presence of the classical EMT inducers~\cite{Ocana2012}. Knowing that a drop in concentration of Prrx1 causes MET and that the level of Prrx1 depends on TGF-$\beta$ activation, we assume that MET is also TGF-$\beta$ dependent, however, this dependence is inverse~\cite{Ocana2012}. As both processes have limited intensity we assume that TGF-$\beta$-dependent stimulation of the EMT process is described by  
\begin{equation}\label{EMT-switch}
S_{EMT}(f) = \dfrac{L^{\text{emt}}}{ 1+e^{-\lambda^\text{emt} \left(f-f_0^\text{emt}\right)}},
\end{equation}
where $\lambda^\text{emt}$ and $f_0^\text{emt}$ define the intensity (slope) of the process, whereas $L^{\text{emt}}$ is its maximal strength. Similarly, for the TGF-$\beta$ dependent MET switch we have 
\begin{equation}\label{MET-switch}
S_{MET}(f) = \dfrac{L^{\text{met}}}{ 1+e^{\lambda^\text{met} \left(f-f_0^\text{met}\right)}},
\end{equation}
with analogical interpretation of parameters $\lambda^\text{met}$, $f_0^\text{met}$ and $L^{\text{met}}$. The integral terms describe the change in the phenotype, with the kernel $k(\alpha,\alpha')$ describing the nature of these changes. We require that the kernel $k(\alpha,\alpha')$ is symmetrical, that is $k(\alpha,\alpha')=k(\alpha',\alpha)$. To sum up, we assume that the influx and outflux of the cancer cells of phenotype $\alpha$ due to the EMT are given by 
\begin{equation}\label{EMT}
\dfrac{L^\text{emt}}{ 1+e^{\lambda^\text{emt} \left(f-f_0^\text{emt}\right)}} \left( \int_0^\alpha k(\alpha, \alpha') c(t,x,\alpha') \, d\alpha' - c(t,x,\alpha)\int_\alpha^1 k(\alpha,\alpha') \,d\alpha'\right),
\end{equation}
whereas the influx and outflux of the cancer cells of phenotype $\alpha$ due to the MET is given by
\begin{equation}\label{MET}
\dfrac{L^\text{met}}{ 1+e^{\lambda^\text{met} \left(f-f_0^\text{met}\right)}} \left( \int_\alpha^1 k(\alpha, \alpha') c(t,x,\alpha') \, d\alpha' - c(t,x,\alpha)\int_0^\alpha k(\alpha,\alpha') \,d\alpha'\right).
\end{equation}
respectively. We may note that both the terms given by Eq.\eqref{EMT} and Eq. \eqref{MET} are conservative in the sense that the integrals with respect to $\alpha$ over [0,1] give 0. This means that these terms do not cause any growth - they just describe the changes of phenotype $\alpha$. To approximate the character of EMT and MET processes we choose the kernel 
\begin{equation}
k(\alpha,\alpha') = e ^{-\gamma(\alpha - \alpha' )^2} 
\end{equation}
as it gives more probability to the phenotypic change to a similar rather than a distant one.

\begin{figure}
    \centering
    \footnotesize
    \begin{tabular}{p{20em}p{20em}}
        \centering\includegraphics[height=12em]{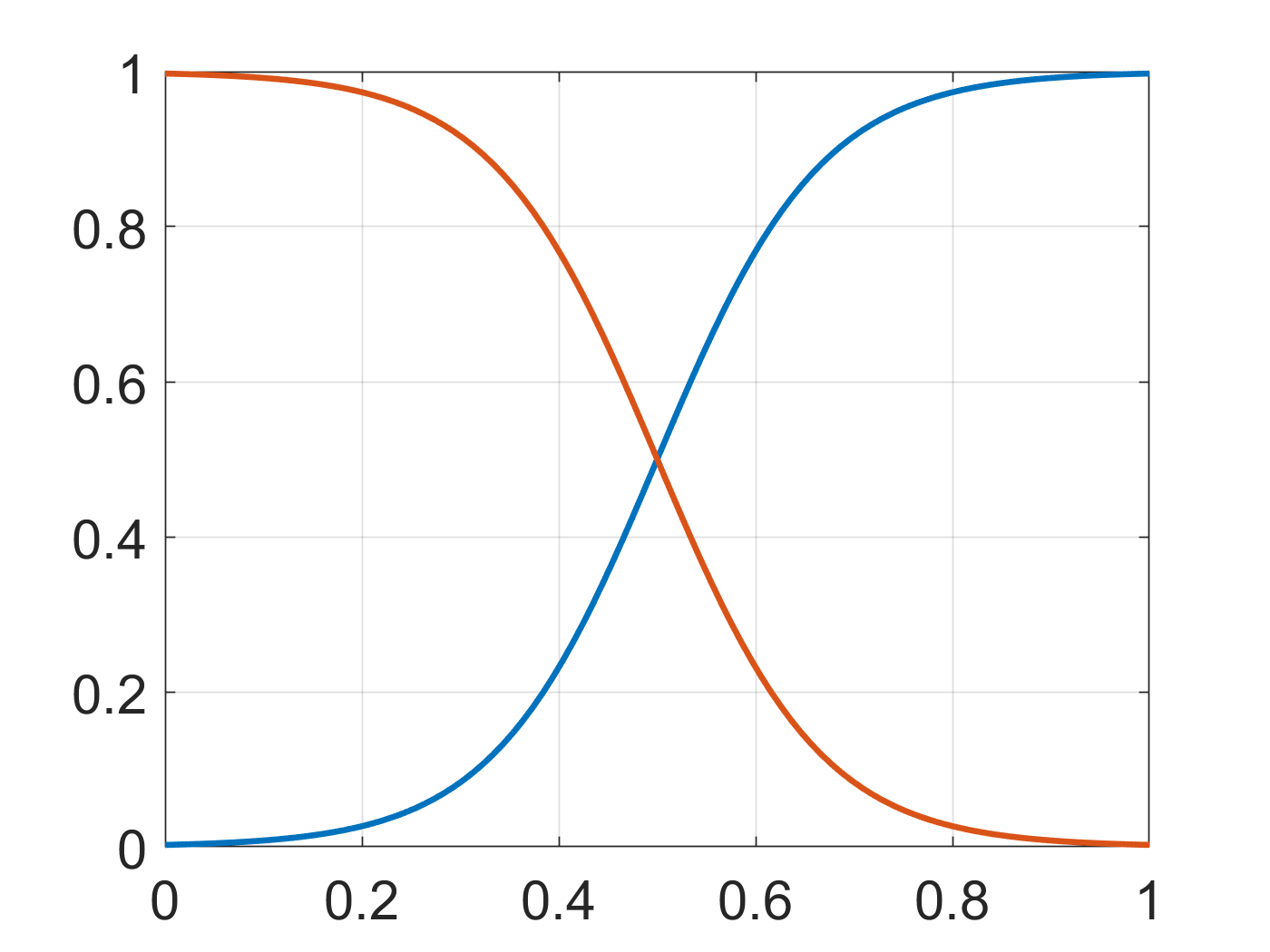}&
        \includegraphics[height=12em]{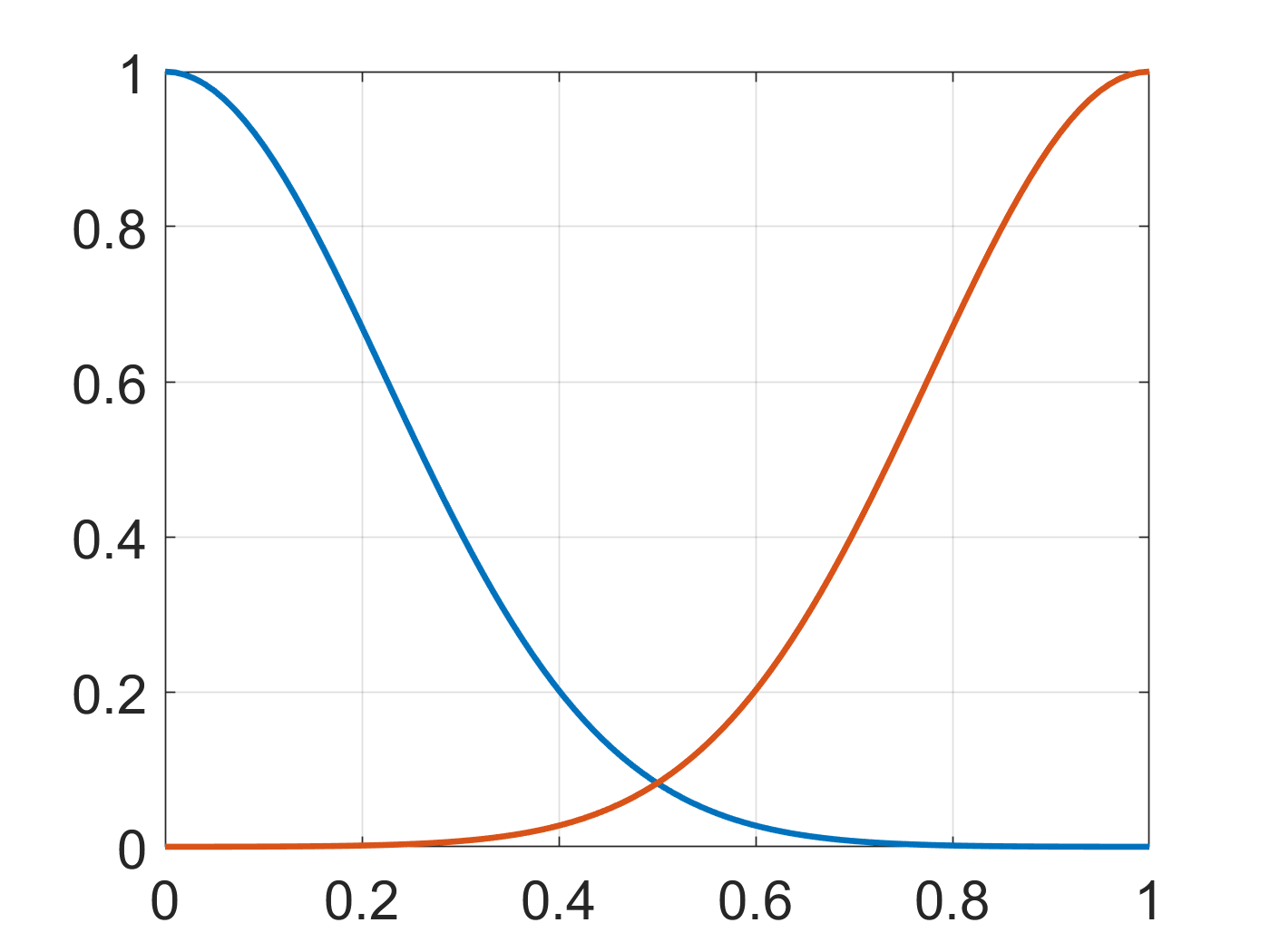}\\
        (a) $S(f)=L/\left(1+\exp(-\lambda(f-f_0))\right)$ for $L=1$, $f_0=0.5$, and $\lambda=\pm12$ (increasing and decreasing respectively) over $f\in[0,1]$ &(b) $k(a)=\exp(-\gamma(a-a_0)^2)$ for $\gamma=10$ and $a_0=0,1$ (decreasing and increasing respectively) over $\alpha\in[0,1]$.
    \end{tabular}
    \caption{(a) Examples of the sigmoid function used as ($f$ dependent) EMT and MET switches, cf. \eqref{EMT-switch} and \eqref{MET-switch}. They are also used for the ($a$ dependent) cancer cell diffusivity \eqref{diffusivity} and chemotaxis sensitivity \eqref{chemotaxis}. (b) Examples of the exponential functions used as the ($a$ dependent) proliferation switch, cf. \eqref{prol_switch}.}
    \label{fig:two_functions}
\end{figure}

We assume that the cancer cells migrate into the surrounding tissue through a combination of diffusion and haptotaxis, with haptotaxis being the dominant mechanism. Therefore we consider some (relatively small) diffusion of cancer cells and a more pronounced cancer cell haptotaxis towards higher densities of ECM components that co-constitute the surrounding tissue. We assume that both types of movement depend on the cell phenotype, that is, epithelial cells due to their tight junctions with neighbouring cells tend to be more sessile, whereas mesenchymal cells are highly motile. These features are captured through the  diffusion coefficient $D_c$ and haptotaxis sensitivity $\chi_c$ being functions of the phenotype $\alpha$. The diffusivity of the cells is assumed to be correlated with the migratory properties of the cancer cells (i.e. with $\alpha$), with epithelial cells having low diffusion and mesenchymal cells having higher diffusion. Furthermore, in order to take into account the interactions of the cancer cells and the ECM, we model this using a porous medium-like diffusion coefficient. This has been used and discussed previously in \cite{Vazquez2006, Byrne1999, Mascheroni2016, Mascheroni2017} and has the added benefit of providing a biologically relevant finite propagation speed \cite{Winkler2011,Byrne2001}. Altogether we take the diffusivity of the cancer cells to be 
\begin{equation}\label{diffusivity}
  D_c(c,\alpha) = {c^2(t,x,\alpha)} \frac{D_c^M}{1+e^{-D_c^{sl}(\alpha-D_c^{a_0})}},
\end{equation}
where $D_c^M$ is the maximum diffusivity, and where the dependence on $\alpha$ is a typical sigmoid curve with $D_c^{sl}$ it's maximum slope and $D_c^{a_0}$ the corresponding middle point of the sigmoid. In a similar fashion, the haptotactic response $\chi_c$ of the cancer cells has been modelled to have a dependence on the phenotype $\alpha$. Namely, we assume a sigmoid-like dependence
\begin{equation}\label{chemotaxis}
    \chi_c (\alpha) = \frac{\chi_c^M}{1+e^{-\chi_c^{sl}(\alpha-\chi_c^{\alpha_0})}}
\end{equation}
where $\chi_c^M$ represents the maximum haptotactic sensitivity achieved by larger values of $\alpha$, $\chi_c^{sl}$ the slope of sigmoid curve, and $\chi_c^{\alpha_0}$ the corresponding inflection point. We refer to Figure~\ref{fig:two_functions} for a graphical representation for both \eqref{diffusivity} and \eqref{chemotaxis}.

Finally, the last term to include in the equation describing the cancer cell dynamics is the proliferation of cancer cells. Focusing on phenotypic transition we employ the simple logistic type of proliferation with an inhibition term for overcrowded regions. However, as the proliferation rate strongly depends on cell phenotype, with epithelial cells, contrary to mesenchymal ones, being highly proliferative, we use an alpha-dependent proliferation coefficient. The switching function 
\begin{equation}\label{prol_switch}
    \gamma_c(\alpha)=\bar{\gamma_c}e^{-\gamma^\ast_c \alpha^2}
\end{equation}
assigns the maximum proliferation value for endothelial cells, and for cells with a pure mesenchymal phenotype, the value is close to none. Altogether we get the following equation for cancer cells dynamics
\begin{align}
\dfrac{\partial}{\partial\, t} c(t,x,\alpha ) = &  \underbrace{\dfrac{L^\text{emt}}{ 1+e^{\lambda^\text{emt} \left(f-f_0^\text{emt}\right)}} \left( \int_0^\alpha k(\alpha, \alpha') c(t,x,\alpha') \, d\alpha' - c(t,x,\alpha)\int_\alpha^1 k(\alpha,\alpha') \,d\alpha'\right)}_{\textrm{EMT transition}} \\\nonumber
& +  \underbrace{\dfrac{L^\text{met}}{ 1+e^{\lambda^\text{met} \left(f-f_0^\text{met}\right)}} \left( \int_\alpha^1 k(\alpha, \alpha') c(t,x,\alpha') \, d\alpha' - c(t,x,\alpha)\int_0^\alpha k(\alpha,\alpha') \,d\alpha'\right)}_{\textrm{MET transition}}\\\nonumber
& + \underbrace{\nabla\cdot\left( D_c(\alpha) \, c^2(t,x,\alpha ) \nabla c(t,x,\alpha )\right)}_{\textrm{diffusion}} -  \underbrace{\nabla\cdot 
\Big( \chi_c (\alpha) \, c(t,x,\alpha ) \nabla v(t,x) \Big)}_{\textrm{haptotaxis}} \\\nonumber
& + \underbrace{\gamma_c (\alpha ) c(t,x,\alpha )\left(1-\int\limits_0^1 c(t,x,\alpha)\, d\alpha -v(t,x) \right)}_{\textrm{proliferation}}.
\end{align}

\textbf{Tumour micro-environment:} 
Although cancer arises from mutations accruing within cancer cells the disease progression and responses to therapy are strongly modulated by non-mutant cells residing in the tumour microenvironment~\cite{Sahai2020}. In a simplified framework, the tumour microenvironment is composed of an extracellular matrix (ECM), which is a large network of macro-molecules that surround cells giving them support and structure, and stroma cells. Among all the stromal cells, a group of activated fibroblasts with significant heterogeneity and plasticity, which are referred to as cancer-associated fibroblasts (CAFs), are one of the most abundant and critical components~\cite{Liu2019,Ping2021}. CAFs secrete a variety of active factors to participate in the generation and maintenance of cancer cell stemness, immune regulation, angiogenesis, metabolic response, therapeutic resistance, and other biological processes~\cite{Ping2021,Yu2014}. They are considered the most effective cell within the tumour micro-environment at depositing and remodelling the ECM, and, according to recent studies, might promote disease malignant progression as they are a substantial source of growth factors including TGF-$\beta$~\cite{Sahai2020,Ping2021,Bornes2021,Yu2014,Winkler2020}.\\

While there is consensus that type III EMT is associated with increased cell migration, invasion, metastasis, and hence cancer aggressiveness, it has only recently been shown that the epigenetic changes that occur also involve the regulation of matrix-degrading enzymes~\cite{Peixoto2019}. These new data confirm that the higher expression of genes involved in the degradation of ECM, such as MMP9, is not specific to a particular experimental model, tissue, or pathology but constitutes a reliable biomarker of EMT in different cancers in vivo. Therefore, we account for the phenotype-dependent ability to degrade the ECM using the integral term with the kernel $\eta (\alpha)$ that we assume to be an increasing function of $\alpha$, more precisely
\begin{equation}\label{eta}
\eta (\alpha ) = \eta_v \dfrac{\alpha^2}{K_\eta + \alpha^2}.
\end{equation} 
Altogether we have the following equation for the density of the tumour microenvironment
\begin{equation}
\frac{\partial}{\partial\, t} v(t,x) = - \underbrace{v(t,x) \, \int_0^1 \eta (\alpha) c(t,x,\alpha )\,\mathrm{d}\alpha}_{\textrm{degradation}} \underbrace{ + \;\; \lambda_v v(t,x) \, \Big( 1 - \int_0^1 c(t,x,\alpha)\, d\alpha - v(t,x) \Big)}_{\textrm{ECM remodelling and tissue regrowth}}\,.
\end{equation}
\textbf{TGF-$\beta$:} The phenotypic change from epithelial to mesenchymal one is governed by extracellular signals that activate a plethora of EMT transcription factors~\cite{Fazilaty2019}. However, there is a  consensus that TGF-$\beta$ is a major inducer and modulator of EMT~\cite{Friedl2011,Yu2014}. Because TGF-$\beta$ is mainly produced by CAFs, and CAFs are stromal fibroblast cells activated by tumour cells through stimulation with paracrine growth factors, we assume that it is produced within the tumour microenvironment upon the activation by cancer cells~\cite{Yu2014,Gallego-Rentero2021}. \\

Because signalling factors are small molecules, particularly compared to the cell size we assume its linear diffusion with the coefficient $D_f$. We assume that TGF-$\beta$ is produced by CAFs, which we do not explicitly model, therefore, we assume that it is produced by the microenvironment $v$ in the presence of cancer cells, independently of their phenotype. The TGF-$\beta$ inflow is therefore described by 
\begin{equation*}
    \gamma_f {v(t,x) \int_0^1 c(t,x,\alpha)\, d\alpha},    
\end{equation*}
where $\bar\gamma_f\in\mathbb R$ is a constant. We also assume some linear degradation of molecules proportional to some constant $d_f$. Altogether we have the following equation for the concentration of the growth factors
\begin{equation}
\dfrac{\partial}{\partial\, t} f(t,x) = \underbrace{D_f \Delta f(t,x)}_{\textrm{diffusion}} \, + \underbrace{\gamma_f \, v(t,x) \int_0^1 c(t,x,\alpha)\, d\alpha}_{\textrm{production}} - \underbrace{d_ff(t,x)}_{\textrm{decay}} \,.
\end{equation}

The whole system reads as
\begin{equation}\label{mdl3}
\begin{dcases}
    \frac{\partial}{\partial\, t} c(t,x,\alpha ) =          \dfrac{L^\text{emt}}{ 1+e^{\lambda^\text{emt} \left(f-f_0^\text{emt}\right)}} \left( \int_0^\alpha k(\alpha, \alpha') c(t,x,\alpha') \, d\alpha' - c(t,x,\alpha)\int_\alpha^1 k(\alpha,\alpha') \,d\alpha'\right) \\
     \hspace{6em}+ \dfrac{L^\text{met}}{ 1+e^{\lambda^\text{met} \left(f-f_0^\text{met}\right)}} \left( \int_\alpha^1 k(\alpha, \alpha') c(t,x,\alpha') \, d\alpha' - c(t,x,\alpha)\int_0^\alpha k(\alpha,\alpha') \,d\alpha'\right) \\
     \hspace{6em}+ \nabla\cdot\left( D_c(\alpha) \, c^2(t,x,\alpha ) \nabla c(t,x,\alpha )\right) - \nabla\cdot \Big( \chi_c (\alpha) \, c(t,x,\alpha ) \nabla v(t,x) \Big) \\
     \hspace{6em}+ \gamma_c (\alpha ) c(t,x,\alpha )\left(1-\int\limits_0^1 c(t,x,\alpha)\, d\alpha -v(t,x) \right)\\
    \frac{\partial}{\partial\, t} v(t,x) = - {v(t,x) \, \int_0^1 \eta (\alpha) c(t,x,\alpha )\,\mathrm{d}\alpha} { + \;\; \lambda_v v(t,x) \, \left( 1 - \int_0^1 c(t,x,\alpha)\, d\alpha - v(t,x) \right)} \\
    \frac{\partial}{\partial\, t} f(t,x)  =  D_f \Delta f(t,x) \, + \gamma_f \, v(t,x) \int_0^1 c(t,x,\alpha)\, d\alpha - d_ff(t,x)
\end{dcases}
\end{equation}


\section{Numerical method and model parameters}\label{sec:num-method}


The advection-reaction-diffusion system \eqref{mdl3} is solved numerically using a specific second-order {Implicit-Explicit Runge-Kutta Finite Differences, Finite Volumes} (IMEX-RK FD-FV) numerical method. This method constitutes an extension of a previous method developed and employed in \cite{SfakKolb2016,SfakKolbe2017,glioma_2022,SfakKolb2022,Kolbe_2021, hybrid_1st, hybrid_2nd} where we refer for most of the details. Here, we only discuss some of its components. 

We consider, at first, a generic advection-reaction diffusion system of the form 
\begin{equation}\label{ARD}
    {w}_{t} = A({w}) + R({w}) + D({w}),
\end{equation}
where ${w}$ represents the analytical solution vector of the system, and $A,R,$ and $D$ are the advection, reaction and diffusion operators respectively. After spatial discretisations have taken place, we denote the corresponding semi-discrete  approximation by ${w}_{h}$, where the index $h$ denotes the (maximal, if the space discretisation is non uniform) spatial grid diameter. The semi-discrete solution $w_h$ satisfies the following system of ODEs
\begin{equation}\label{numARD}
    \partial_{t}{w}_{h} = \mathcal{A}({w}_h) + \mathcal{R}({w}_h) + \mathcal{D}({w}_h),
\end{equation}
where the numerical operators $\mathcal{A}, \mathcal{R}$ and $\mathcal{D}$ are (spatially) discrete approximations of the advection-reaction, diffusion operators, $A,R$ and $D$ in equation \eqref{ARD}. Moreover, as the numerical scheme we employ is (partially) FV, raising its accuracy to the second order necessitates the use of {flux limiters} for the interface reconstruction of the numerical fluxes. Out of the large number of flux limiter options, we have found that the {Monotonized Central Difference} (MC) limiter, cf. \cite{van1977towards}, constitutes a robust and efficient choice. 

Before solving \eqref{numARD}, we re-organise its terms in implicit and explicit (IMEX splitting) and, accordingly, \eqref{numARD} takes the form
\begin{equation}\label{IMEX}
    \partial_{t}{w}_{h} = \mathcal{I}({w}_{h}) + \mathcal{E}({w}_{h}).
\end{equation} 
The actual IMEX splitting depends on the problem at hand, but in a typical case the advection terms $\mathcal{A}$ are treated explicitly in time, the diffusion terms $\mathcal{D}$ implicitly, and the reaction $\mathcal{R}$ terms either explicitly or implicitly, depending on their stiffness. In the problems that we encounter in this paper, all reaction terms have been resolved explicitly in time.

The semi-discrete problem \eqref{IMEX} is now solved using a diagonally implicit RK method for the implicit part $\mathcal{I}({w}_{h})$, and an explicit RK for the explicit part $\mathcal{E}({w}_{h})$. Altogether, we solve \eqref{IMEX} using the additive RK scheme
\begin{equation}\label{IMEX-RK}
    \begin{dcases}
	{w}^{*}_{i} = {w}^{n}_{h} + \tau_{n} \sum^{i-1}_{j=1}\Bar{a}_{i,j}{E}_{j}, &\quad i = 1 \ldots s, \\
	{w}_{i} = {w}^{*}_{i} + \tau_{n} \sum^{i-1}_{j=1}a_{i,j}{I}_{j} + \tau_{n}\Bar{a}_{i,i}{I}_{i}, &\quad i = 1 \ldots s, \\
	{w}^{n+1}_{h} = {w}^{n}_{h} + \tau_{n} \sum^{s}_{i=1}\Bar{b}_{i}{E}_{i} + \tau_{n}\sum^{s}_{i=1}b_{i}{I}_{i},
	\end{dcases}
\end{equation}
where $s = 4$ are the stages of the IMEX-RK method, ${E}_{i} = \mathcal{E}({w}_{i})$, ${I}_{i} = \mathcal{I}({w}_{i})$$, $\; $i = 1\ldots s$, and $\{\bar{b},\bar{A}\}, \; \{b,A\}$ are the coefficients for the explicit and implicit part of the scheme respectively. These coefficients can be found in the Butcher Tableau in Table \ref{tbl:IMEX},  cf. \cite{christopher2001additive}. As a final stage of this method, we solve the linear system in \eqref{IMEX-RK} using the {Iterative Biconjugate Gradient Stabilised Krylov subspace} method, see \cite{Krylov.1931, vdVorst.1992}.

For the discretisation of the domain we have employed a computational grid of $128\times 128$ grid cells for the spatial variable $x$, and $11$ grid cells for the phenotypic variable $a$. The time step of the numerical method was imposed by the CFL condition that was set at $0.25$ and was moreover bound from above by $0.05$ which corresponds to $1.2\, {\rm h}$.

\begin{table}[t]
    \caption{Butcher tableaux for the explicit (upper) and the implicit (lower) parts of the third order IMEX scheme \eqref{IMEX-RK}, see also \cite{christopher2001additive}.}\label{tbl:IMEX}
    \centering
    \footnotesize
    \begin{tabular}{c|cccc}
        $0$&&&&\\[0.5em]
	$\frac{1767732205903}{2027836641118}$&$\frac{1767732205903}{2027836641118}$&&&\\[0.5em]
	$\frac{3}{5}$&$\frac{5535828885825}{10492691773637}$&$\frac{788022342437}{10882634858940}$&&\\[0.5em]
	$1$&$\frac{6485989280629}{16251701735622}$&$-\frac{4246266847089}{9704473918619}$&$\frac{10755448449292}{10357097424841}$&\\[0.5em]
	\hline\\[-0.5em]
	&$\frac{1471266399579}{7840856788654}$ & $-\frac{4482444167858}{7529755066697}$ & $\frac{11266239266428}{11593286722821}$ & $\frac{1767732205903}{4055673282236}$ 
    \end{tabular}
    
    \begin{tabular}{c|cccc}
        $0$&0&&&\\[0.5em]
	$\frac{1767732205903}{2027836641118}$&$\frac{1767732205903}{4055673282236}$&$\frac{1767732205903}{4055673282236}$&&\\[0.5em]
	$\frac{3}{5}$&$\frac{2746238789719}{10658868560708}$&$-\frac{640167445237}{6845629431997}$&$\frac{1767732205903}{4055673282236}$&\\[0.5em]
	$1$&$\frac{1471266399579}{7840856788654}$&$-\frac{4482444167858}{7529755066697}$&$\frac{11266239266428}{11593286722821}$&$\frac{1767732205903}{4055673282236}$\\[0.5em]
	\hline\\[-0.5em]
	&$\frac{1471266399579}{7840856788654}$ & $-\frac{4482444167858}{7529755066697}$ & $\frac{11266239266428}{11593286722821}$ & $\frac{1767732205903}{4055673282236}$ 
    \end{tabular}
\end{table}


\subsection{Model parameters and their estimation} \label{sec:param-estim}


In general, the accurate approximation of parameters for models describing biological processes is a very challenging task. That is because establishing proper experimental setups for most cases is very difficult or for now impossible. The latter is the case of the cancer invasion models \textit{in vivo}. Despite this fact, we sought the best approximation of the relevant values using {\it in vitro} experiments data or results obtained for cognate cell lines.

We start by specifying the space and time units relevant to the process. Considering the maximum distance that cancer cells may achieve at the early stage of invasion, we settle the space unit to be in $\rm cm$. In light of the average duration of the healing process, we set the time unit to be $1$ day.

TGF-$\beta$1 (transforming growth factor) is secreted by fibroblasts as inactive precursors and deposited into the extracellular matrix~\cite{TGFb}. Active TGF-$\beta$1 half-life is about two minutes whereas latent TGF-$\beta$1 half-life is about 90 minutes~\cite{TGFb}.

We approximate the range of cellular diffusion rate with values measured for the cells originating from the mesodermal cell line, precisely for microvessel endothelial cells (MEC)~\cite{Stokes1991a}. The measured mean value of the diffusion parameter of MEC cells is $7.1 \pm 2.7 \times 10^{-9} {\rm cm^2 s^{-1}}$, which leads us to assume $D_c = 6.1 \times 10^{-2} {\rm mm^2 s^{-1}}$. The same authors in another paper investigated the chemotactic coefficient of migrating endothelial cells in gradients of acidic fibroblast growth factor (aFGF)~\cite{Stokes1990}. The maximum chemotactic response was measured for concentrations of aFGF around $10^{-10}{\rm M}$ (with a concentration gradient equal to $3.5 10^{-15} {\rm M {\mu m}^{-1}}$) giving a chemotactic coefficient of $2600\, {\rm cm^2 s^{-1}M^{-1}}$, which gives $2.25\, {\rm mm^2 d^{-1} M^{-1}}$. Collagens type I, III, IV, V, and FGF were found capable of inducing a similar chemotactic response among human endothelial cells however, their stimulation capacity was estimated to be one-third to one-half of the maximal capacity of factors such as ECGF~\cite{Terranova1985}. Summarising we set the value of haptotaxis sensitivity $\chi_c$ to be equal to $0.71\, {\rm mm^2 d^{-1}M^{-1}}$ which is consistent with previous studies \cite{Anderson1998,Chaplain1996,Orme1996}. 


\section{Numerical experiments and computational simulations} \label{numerical_results}


Using numerical simulations we verify whether the proposed simplified mechanism of phenotype transition under the influence of the TGF-$\beta$ enables qualitative reconstruction of metastases formation in the form of isolated neoplastic foci. To this end, we present three sets of numerical experiment results. At first, focusing on the investigated mechanism of phenotype changes and its consequences, we assume that the tissue surrounding the primary tumour is homogeneous. 

The initial conditions are set unless otherwise specified, as follows
\begin{equation}\label{eq:ICs}
    \begin{cases}
    c(t=0,x,\alpha) = 2e^{-\|x\|^2/0.3}\mathcal{X}_{\{\|x\|^2<0.1\}}(x), &\alpha=0\\
    c(t=0,x,\alpha) = 0,&0<\alpha\leq 1\\
    f(t=0,x)=0.8 \mathcal{X}_{\{\|x\|^2<0.1\}}(x)
    \end{cases}
\end{equation}
for $x\in \mathbb D$ and where $\mathcal X_S (x)$ is the characteristic function over the set $S\subset \mathbb D$ and $\|\cdot\|$ the Euclidean norm. The initial density of the ECM will be considered to be either constant or randomly structured as in Figure \ref{fig:ecm}, cf. \cite{hybrid_2nd}.


\begin{experiment}\textbf{Uniform ECM}\label{exp:uni-ecm}

\begin{table} 
\footnotesize 
\begin{tabular}{|c|c|c|c|c|c|}
\hline
    {}&{Param.}& {Name} & {Value} & {Units} & {Reference} \\ 
\hline
\multirow{14}{*}{\rotatebox{90}{$c$-eq.}}
    &$D_c^M$&CC diffusion rate&$10^{-6}$&$\rm {mm^2}d^{-1}$& \cite{Stokes1991a} \\
    &$D_c^{sl}$&CC diffusion scale &$20$&$\rm M^2{mm^{-4}}$&(our estimate)\\
    &$D_c^{\alpha_0}$& CC diffusion midpoint&$0.5$&---&(our estimate)\\
    &$\chi_c^M$&CC haptotactic response&$0.05$&$\rm {mm^2}M^{-1}d^{-1}$& cf. \cite{Anderson1998,Chaplain1996,Orme1996}\\
    &$\chi_c^{sl}$&CC haptotaxis scale&$20$&---&(our estimate)\\
    &$\chi_c^{\alpha_0}$&CC haptotaxis midpoint&$0.5$&---&(our estimate)\\
    &$L^\text{emt}$&EMT rate&$0.001$&$\rm {M}{mm^{-2}}d^{-1}$&(our estimate)\\
    &$\lambda^\text{emt}$&EMT scale&$30$&$\rm M^{-1}{mm^{-2}}$&(our estimate)\\
    &$f_0^\text{emt}$&midpoint EMT&$0.01$&$\rm M{mm^2}$&(our estimate)\\
    &$L^\text{met}$&MET rate&$10^{-4}$&$\rm {M}{mm^{-2}}d^{-1}$&(our estimate)\\
    &$\lambda^\text{met}$&MET scale&$20$&$\rm M^{-1}{mm^{-2}}$&(our estimate)\\
    &$f_0^\text{met}$&midpoint MET&$0.001$&$\rm M{mm^2}$&(our estimate)\\
    &$\bar{\gamma_c}$&CC proliferation rate&$1$&$\rm d^{-1}$&\cite{Stokes1991a}\\
    &$\gamma_c^\ast$&CC proliferation exponent&$7$&---&(our estimate)\\
\hline
    \multirow{3}{*}{\rotatebox{90}{$f$-eq.}}
    &$D_f$&TGF-$\beta$ diffusion rate&$0.01$ &$\rm {mm^2}d^{-1}$&\cite{Stokes1991a}\\
    &$\gamma_f$&TGF-$\beta$ production rate&$0.3$&$\rm {mm^2}M^{-1}d^{-1}$&(our estimate)\\
    &$d_f$&TGF-$\beta$ decay&$3$&$\rm d^{-1}$&cf. \cite{TGFb}\\
\hline
    \multirow{3}{*}{\rotatebox{90}{$v$-eq.}}
        &$\eta_v$&ECM degradation rate&$0.02$&$\rm {mm^2}M^{-1}d^{-1}$&cf. \cite{hybrid_2nd, Chaplain2005}\\
        &$K_\eta$&ECM degradation midpoint&$10$&---&(our estimate)\\
        &$\lambda_v$&ECM reconstruction rate& $0.005$&$\rm {mm^2}M^{-1}d^{-1}$&cf. \cite{hybrid_2nd}\\
\hline
\end{tabular}
\caption{Parameters used in \textbf{Experiment \ref{exp:uni-ecm} ---Uniform ECM}. They are organised by equation and by system component. The acronym ``CC'' represents the cancer cells.} \label{tbl:uni-ecm}
\end{table}

To illustrate the dynamics of the system without the impact of the non-uniformity of the extratumoural tissue, we first perform a numerical experiment with initially uniform ECM density. 

Our simulations show that originally, EMT takes place in the centre of the spatial domain where the concentration of the  TGF-$\beta$ is highest, which gives rise to cancer cells of higher phenotype $\alpha$, that is cells having more mesenchymal features. These cells degrade the ECM and create a local gradient in the tissue which, in turn, facilitates their migration away from their original formation location. Due to the uniformity of the ECM, and the radially symmetric initial conditions, the gradient in the tissue remains radially symmetric. In effect, the migration of the cancer cells takes the form of invasion rings. These results are presented in Figure \ref{fig:uniECM} where they are organised as follows: in two major panels, representing the time instances $t=0$ and $t=150$, are shown a) the densities of the cancer cells (first row, columns 1-3) along with the corresponding isolines (second row, columns 1-3) for the cancer-cell phenotypes $\alpha=0, 0.5, 1$ respectively, and the densities of TGF-$\beta$ (upper row, column 4) and the ECM (bottom row, column 4). 
The invasion rings, consisting of cells which are of higher phenotypic value $\alpha$, grow away from the origin, and move to regions of lower concentration of TGF-$\beta$. In these regions, the MET transition is, accordingly, triggered, resulting in the appearance of cells having more epithelial features including the proliferative potential. As a result, there emerge rings of higher-density of cancer cells, disconnected from the pre-existing cluster of cancer cells.

\begin{figure}\label{fig:uni-ecm}
\setlength{\tabcolsep}{1pt}
\centering
\begin{tabular}{l|ccc|c}
    \phantom{X}&\footnotesize{Cancer cells}&\footnotesize{Cancer cells}&\footnotesize{Cancer cells}&\footnotesize{TGF-$\beta$ \& ECM}\\[-0.3em]    &\footnotesize{$c(t,\cdot,\alpha=0)$}&\footnotesize{$c(t,\cdot,\alpha=0.5)$}&\footnotesize{$c(t,\cdot,\alpha=1)$}&\footnotesize{$f(t, \cdot),\ v(t,\cdot)$}\\
    \hline
    \multirow{2}{*}{\rotatebox{90}{\footnotesize{$t=0$}}}&
    \includegraphics[width=0.23\linewidth]{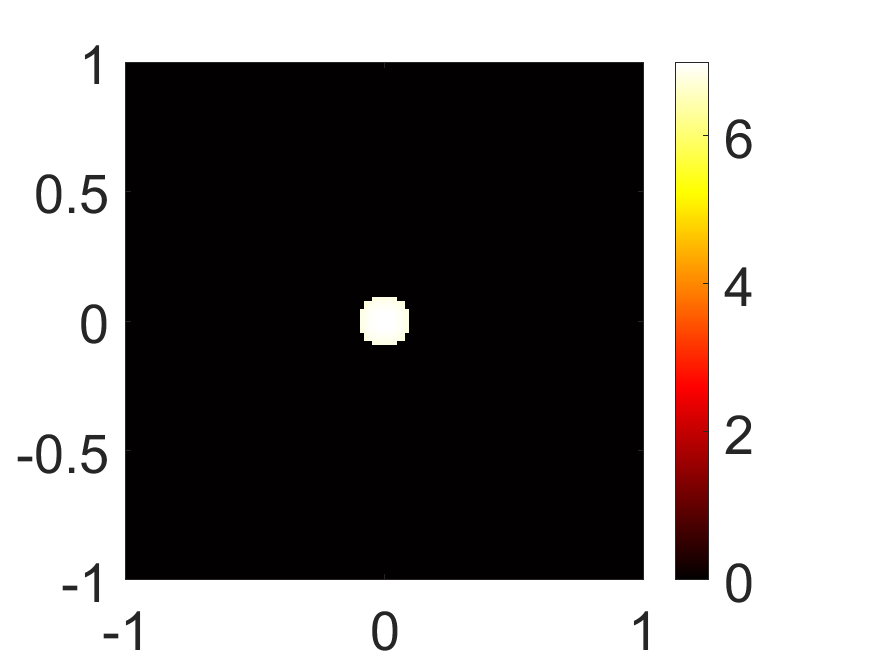} & \includegraphics[width=0.23\linewidth]{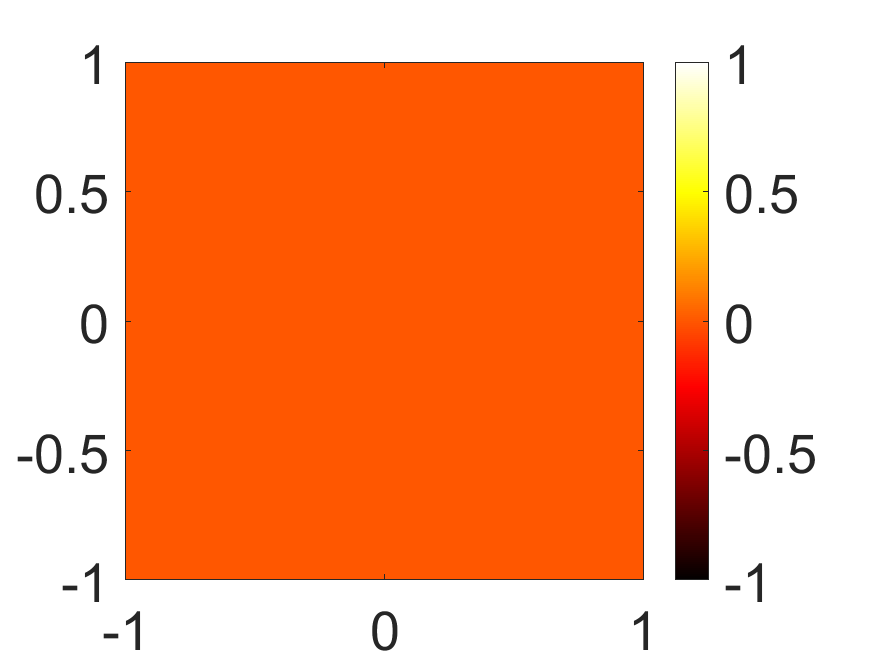} & \includegraphics[width=0.23\linewidth]{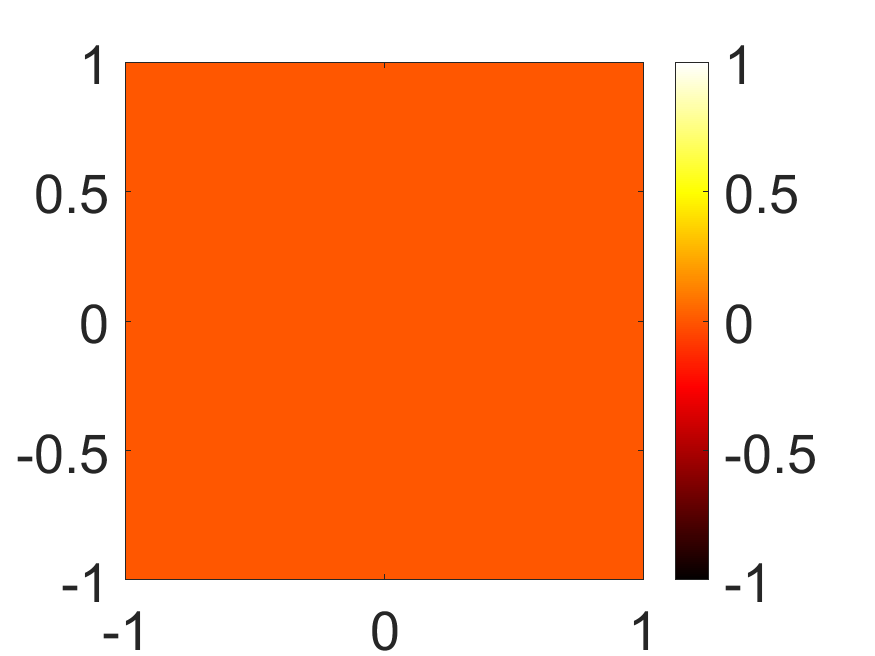} & 
    \includegraphics[width=0.23\linewidth]{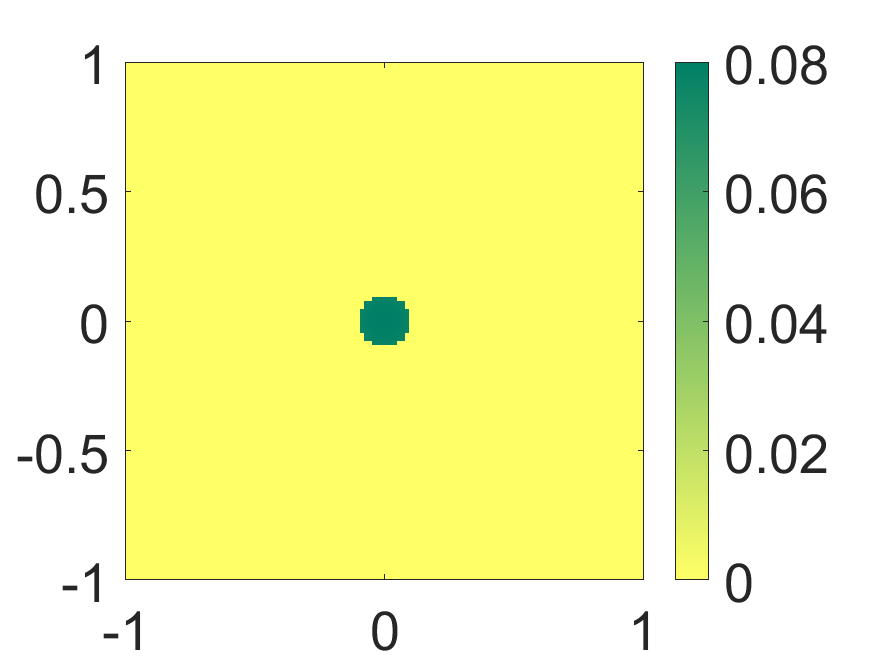}
    \\[-0.3em]
    &\includegraphics[width=0.23\linewidth]{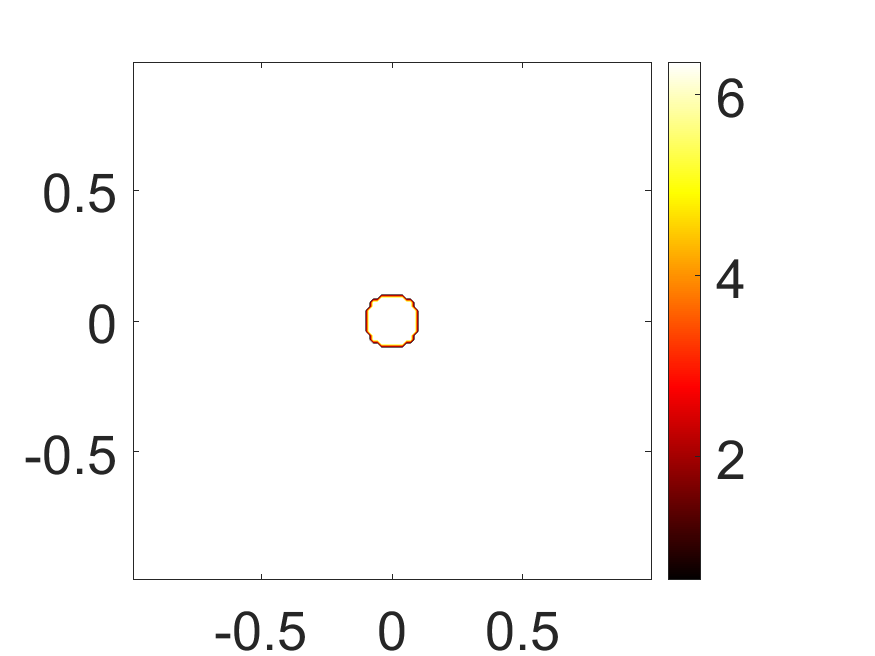} & \includegraphics[width=0.23\linewidth]{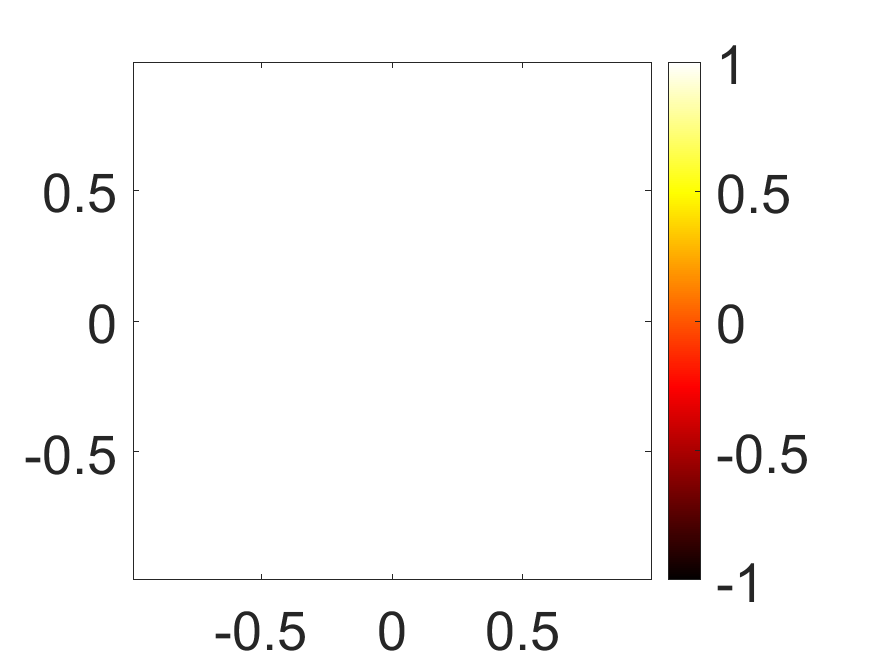} & \includegraphics[width=0.23\linewidth]{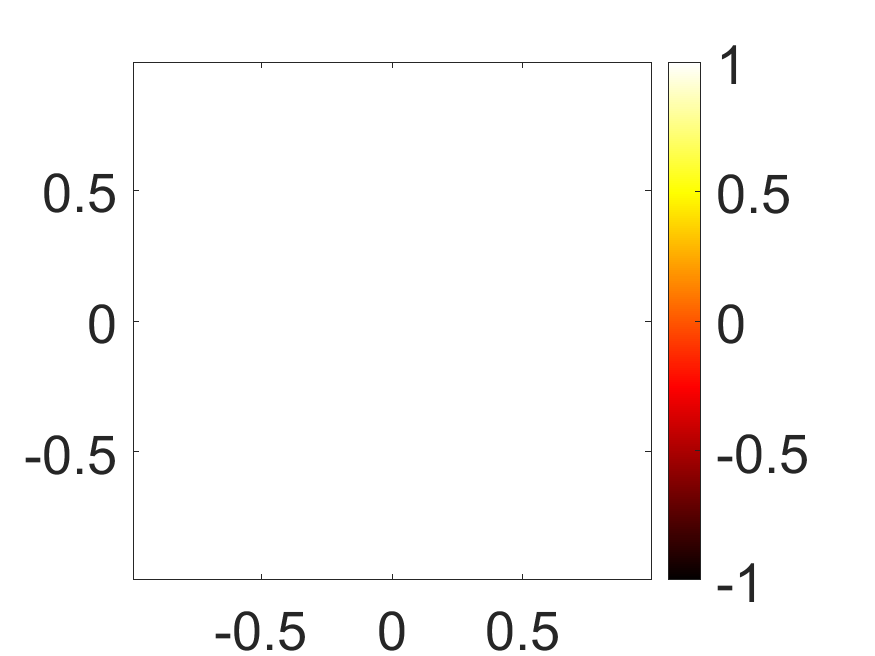} &
    \includegraphics[width=0.23\linewidth]{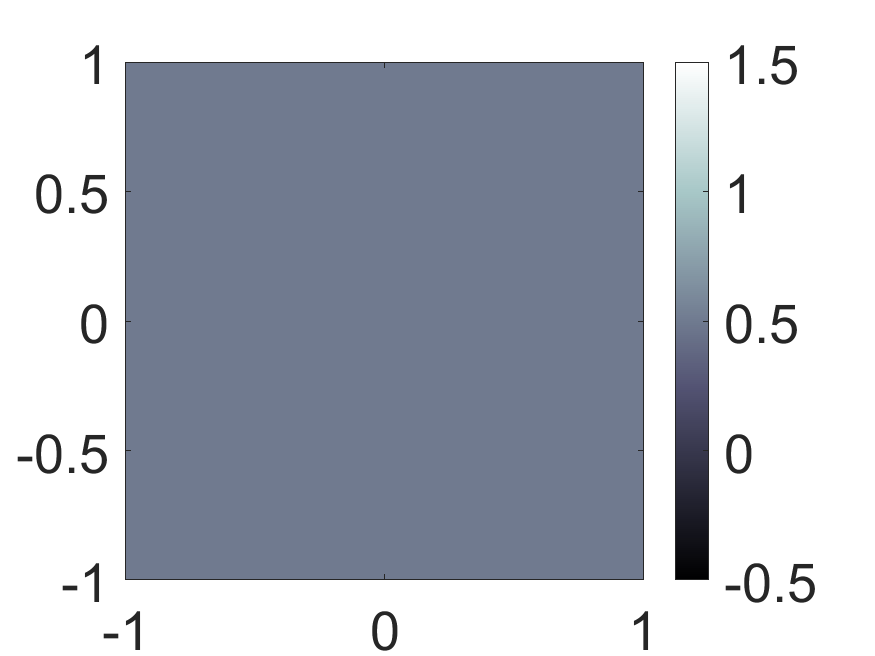} 
    \\
    \hline
    \multirow{2}{*}{\rotatebox{90}{\footnotesize{$t=150$}}}&
    \includegraphics[width=0.23\linewidth]{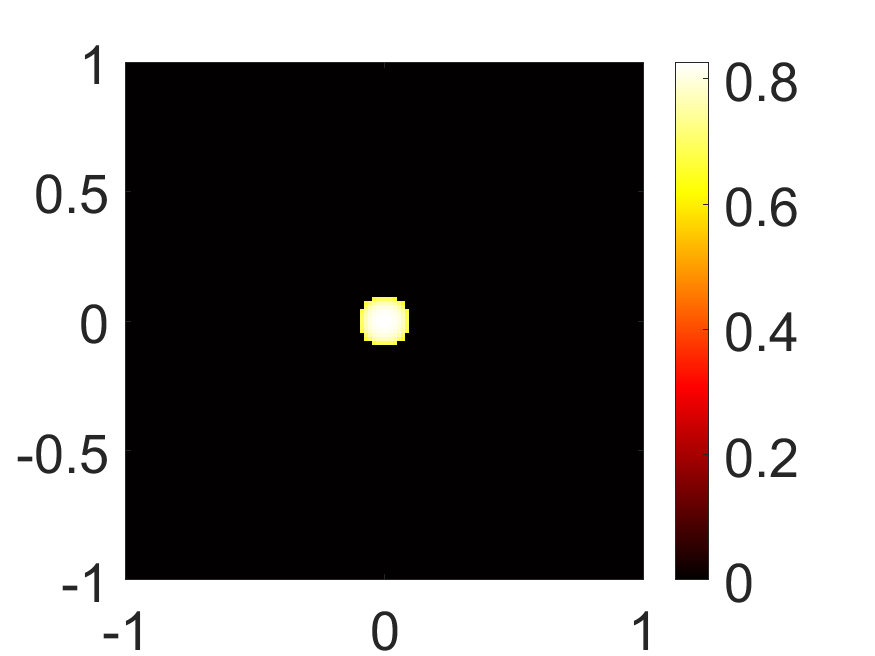} & \includegraphics[width=0.23\linewidth]{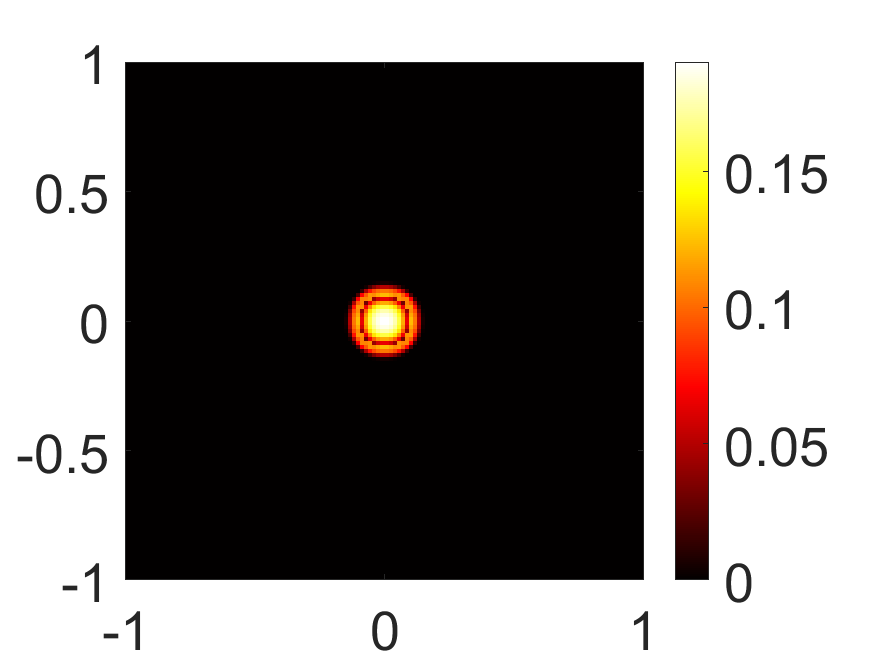} & \includegraphics[width=0.23\linewidth]{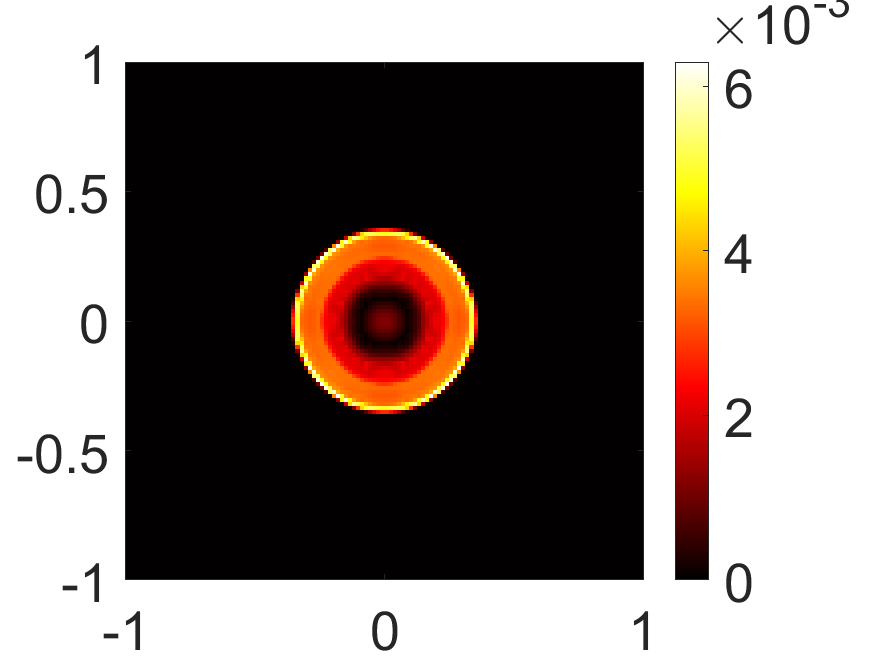} &
    \includegraphics[width=0.23\linewidth]{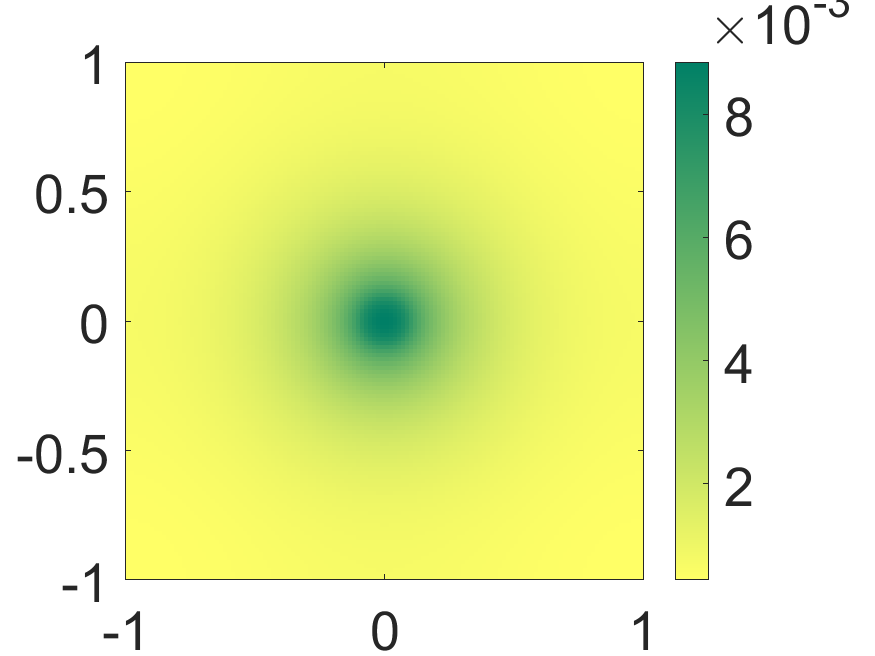}
    \\[-0.3em]
    &\includegraphics[width=0.23\linewidth]{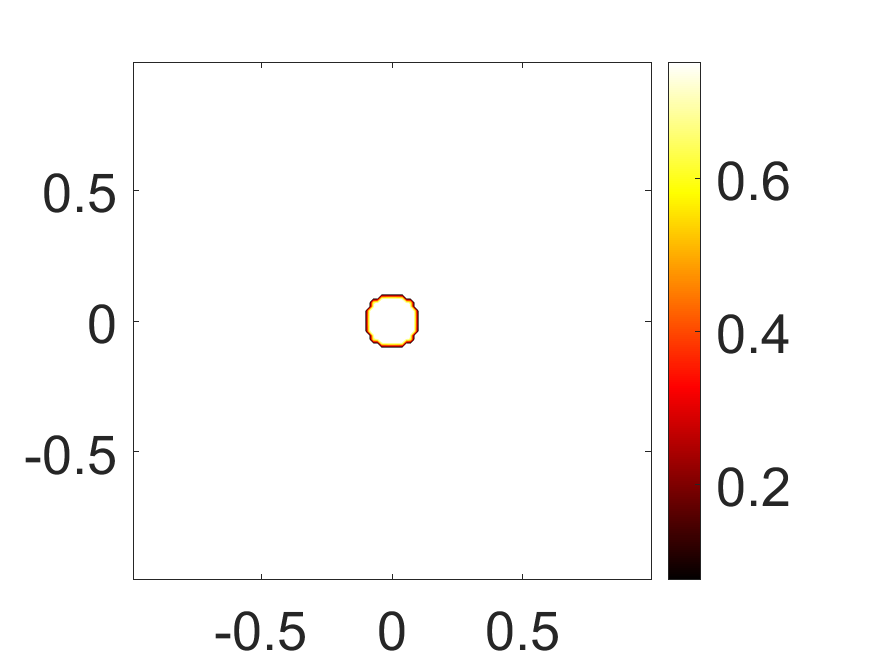} & \includegraphics[width=0.23\linewidth]{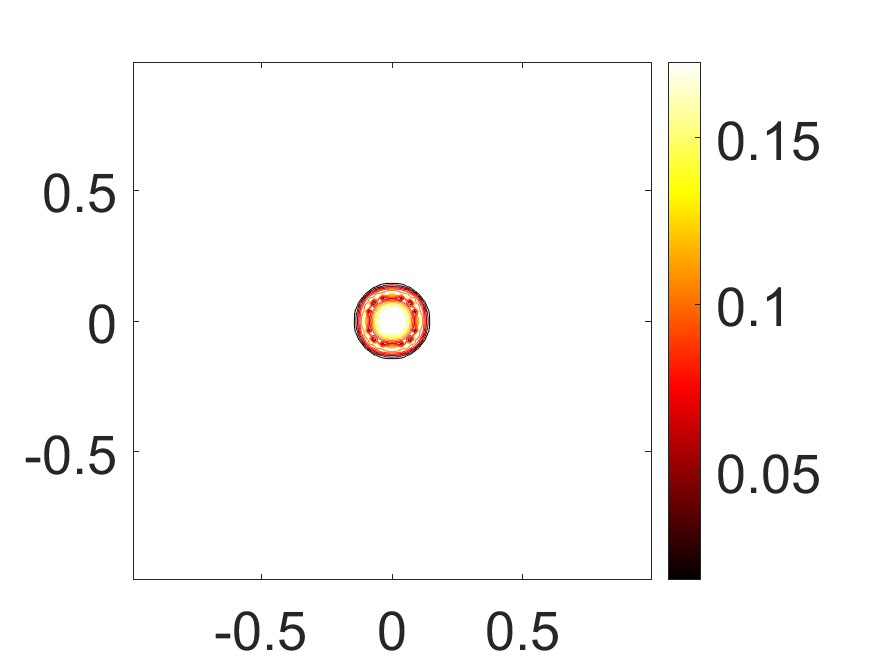} & \includegraphics[width=0.23\linewidth]{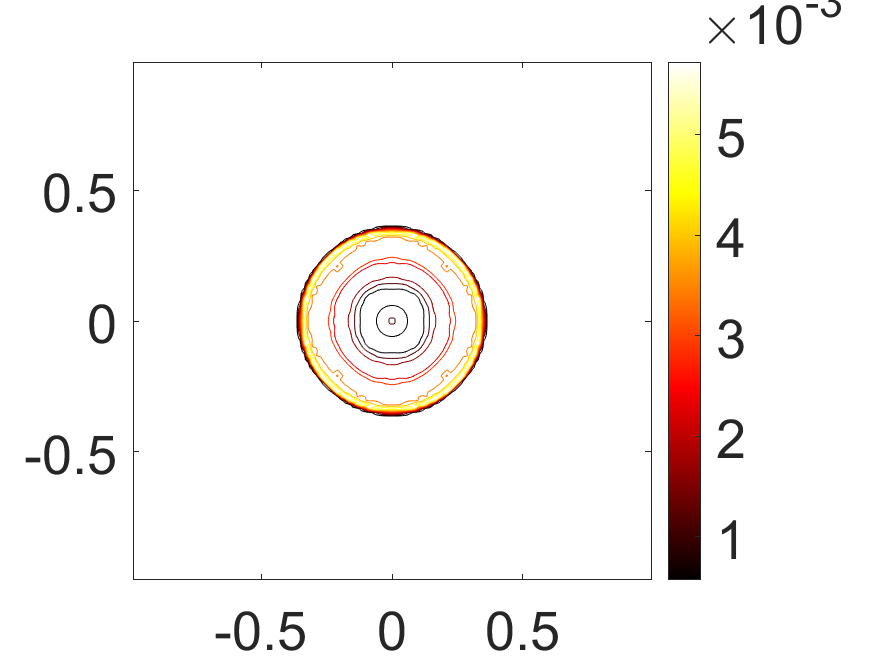} &
    \includegraphics[width=0.23\linewidth]{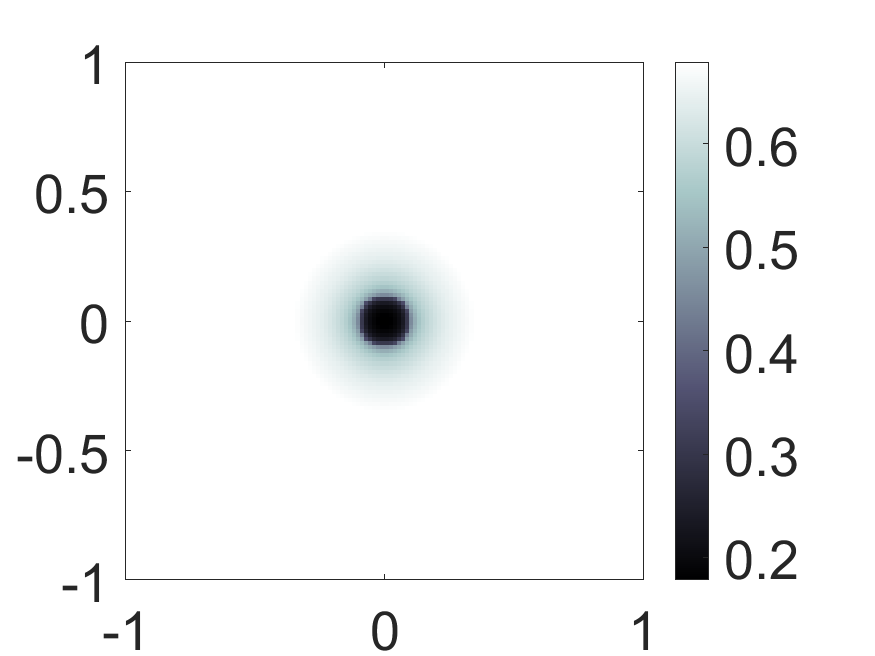} 
\end{tabular}
    \vskip-1em
    \caption{\textbf{Experiment \ref{exp:uni-ecm} --- Uniform ECM.} Top panel: Initial conditions ($t=0$) for $a=0$, $0.5$, $1$ for the density of cancer cells (1st row, columns 1-3), their isolines (2nd row, columns 1-3), and TGF-$\beta$ and ECM on column 4. Lower panel: At later time ($t=150$), EMT has given rise to higher $a$ cancer cell densities. As the ECM is degraded, they are driven by its gradient and, due to the rotationally symmetry of the system, they invade the local tissue in the form of circular invasion rings.}\label{fig:uniECM}
\end{figure}

\end{experiment}

\begin{experiment}\textbf{Generic invasion}\label{exp:generic}

The second numerical experiment is designed to illustrate the formation of isolated tumour foci in a more biologically realistic scenario in which the tissue surrounding the primary tumour is heterogeneous. We present with this experiment the full range of the dynamics of the system \eqref{mdl3}. The results are presented in Figure \ref{fig:generic} where they are organised as follows: in each one of the three major panels, that represent the time instances $t=0$, $t=150$, and $t=300$, are shown a) the densities of the cancer cells (first row, columns 1-3) along with the corresponding isolines (second row, columns 1-3) for the cancer-cell phenotypes $\alpha=0$, $0.5$, and $1$ respectively, and the densities of the TGF-$\beta$ (row 1, column 4) and the ECM (row 2, column 4). As in the previous experiment, EMT initially occurs in the centre of the spatial domain where the concentration of TGF-$\beta$ is the highest. However, the differences in the ECM concentration gradient in the surrounding tissue cause uneven migration of mesenchymal cells. Mesenchymal cells that follow the ECM gradient will find themselves ``far'' from the primary tumour at the same time find themselves in an area of lower TGF-$\beta$ concentration, which causes a change in their phenotype to more epithelial, accelerated proliferation and consequent formation of a new cluster of cancer cells. Accordingly, cancer cells of higher phenotypic value $\alpha$ emerge. These cancer cells introduce a local gradient of the local tissue by degrading the ECM faster, cf. \eqref{eta} and \eqref{mdl3}. In turn, this local gradient drives the migration of the cancer cells of higher $\alpha$ phenotypes away from their original formation location. The invasion of the cancer cells of higher phenotype $\alpha$ brings them to regions of lower TGF-$\beta$ concentration, where the MET transition is, accordingly, triggered. As a result, tumour foci in the form of cancer cell clusters appear in lower phenotypic values $\alpha$ and are disconnected from the pre-existing cancer cell concentrations.

For this experiment, the initial ECM density is set to randomly vary over the domain according to a procedure developed in \cite{hybrid_2nd}, where we refer the reader for full details. The process, in two dimensions, is illustrated in Figure~\ref{fig:ecm} and starts with a $8 \times 8$ grid/matrix with entries taken from a standard normal distribution, $\mathcal N(0,1)$. A number of bisection steps are taken until the grid size reaches the desired resolution of the domain. At every refinement stage, the new entries are obtained from interpolating the values of the previous stage matrix with the addition of a small amount of Gaussian noise. As a result, the final ECM density preserves the initial randomly chosen distribution of the $8 \times 8$ grid. The final values are scaled between the biologically rellevant minimum and maximum ECM density values.

\begin{figure}[t]
    \centering
    \includegraphics[width=0.75\textwidth]{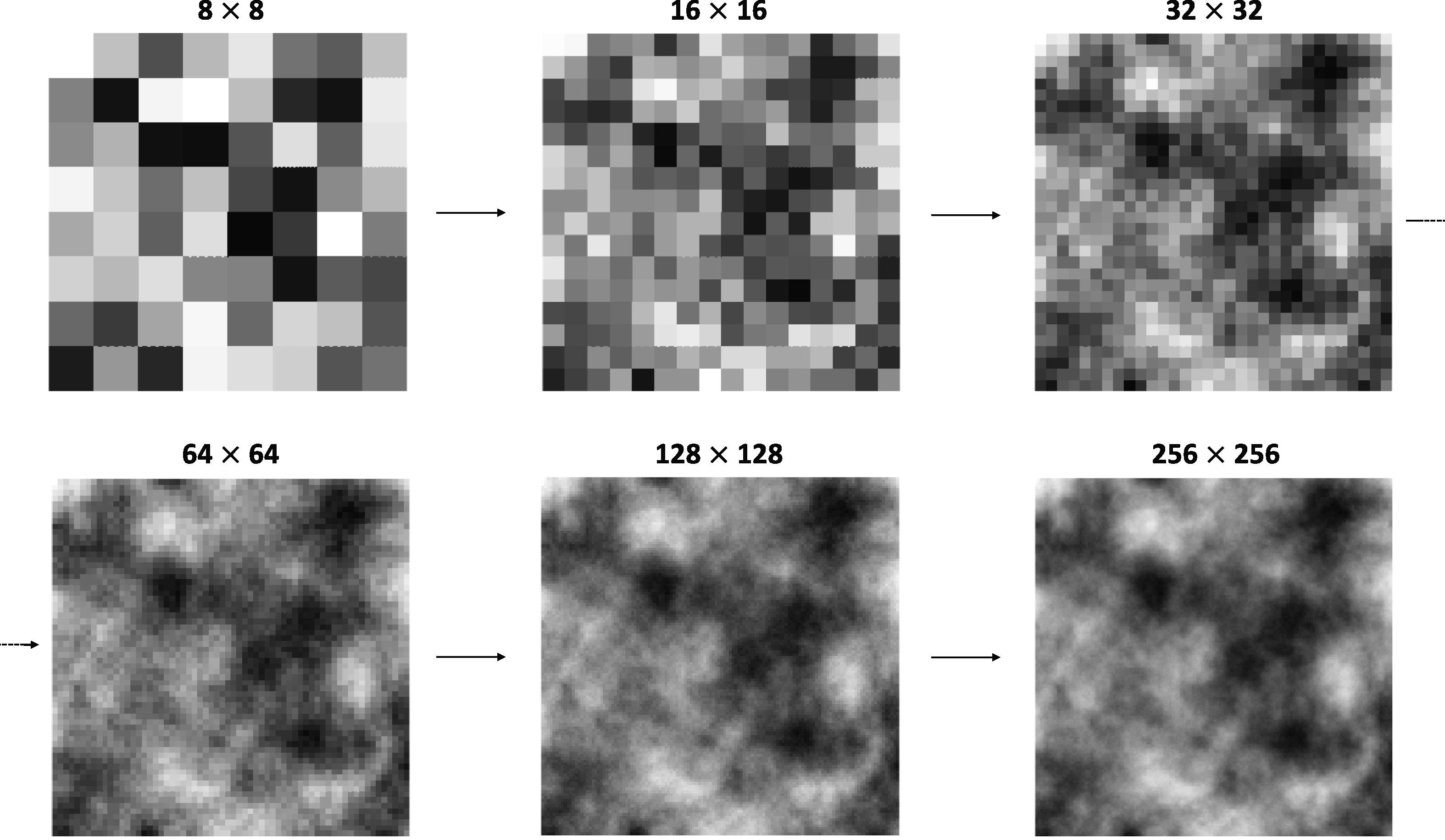}
    \caption{Construction of the initial ECM density employed in Experiments \ref{exp:generic}, \ref{exp:decoupled}, and \ref{exp:coupled}. The process starts with random values over an $8\times8$ grid. In every stage of the construction process the grid is bisected and the new values are attained by averaging the neighbouring values. The process stops when the required grid size is reached, when also scaling to the proper biological range takes place cf. \cite{hybrid_2nd}.}\label{fig:ecm}
\end{figure}

\begin{figure} 
\setlength{\tabcolsep}{1pt}
\centering
\begin{tabular}{l|ccc|c}
    \phantom{X}&\footnotesize{Cancer cells}&\footnotesize{Cancer cells}&\footnotesize{Cancer cells}&\footnotesize{TGF-$\beta$ \& ECM}\\[-0.3em] &\footnotesize{$c(t,\cdot,\alpha=0)$}&\footnotesize{$c(t,\cdot,\alpha=0.5)$}&\footnotesize{$c(t,\cdot,\alpha=1)$}&\footnotesize{$f(t, \cdot),\ v(t,\cdot)$}\\
    \hline
    \multirow{2}{*}{\rotatebox{90}{\footnotesize{$t=0$}}}&
    \includegraphics[width=0.22\linewidth,trim={1em 0 0.4em 0},clip]{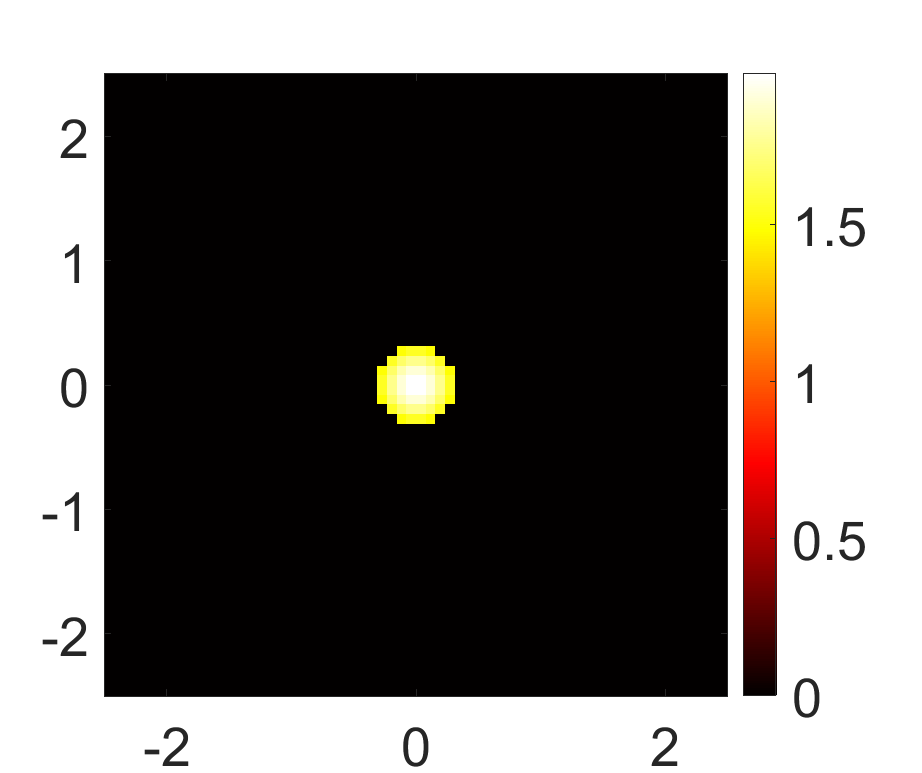} & 
    \includegraphics[width=0.22\linewidth,trim={2em 0 0.4em 0},clip]{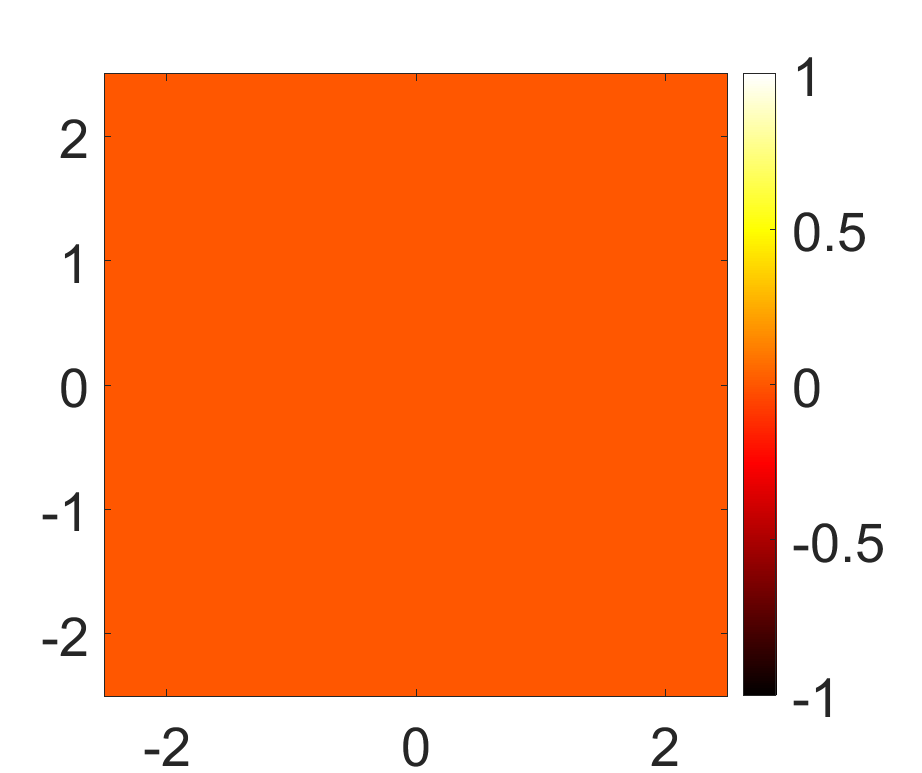} & 
    \includegraphics[width=0.22\linewidth,trim={2em 0 0 0},clip]{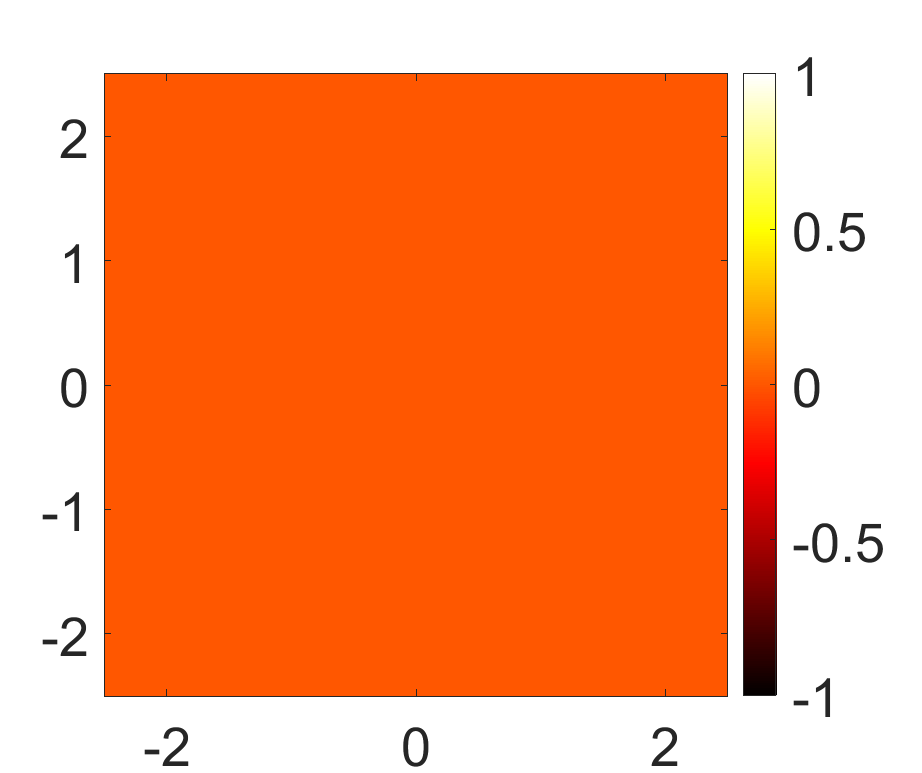} & 
    \includegraphics[width=0.22\linewidth,trim={1em 0 0.4em 0},clip]{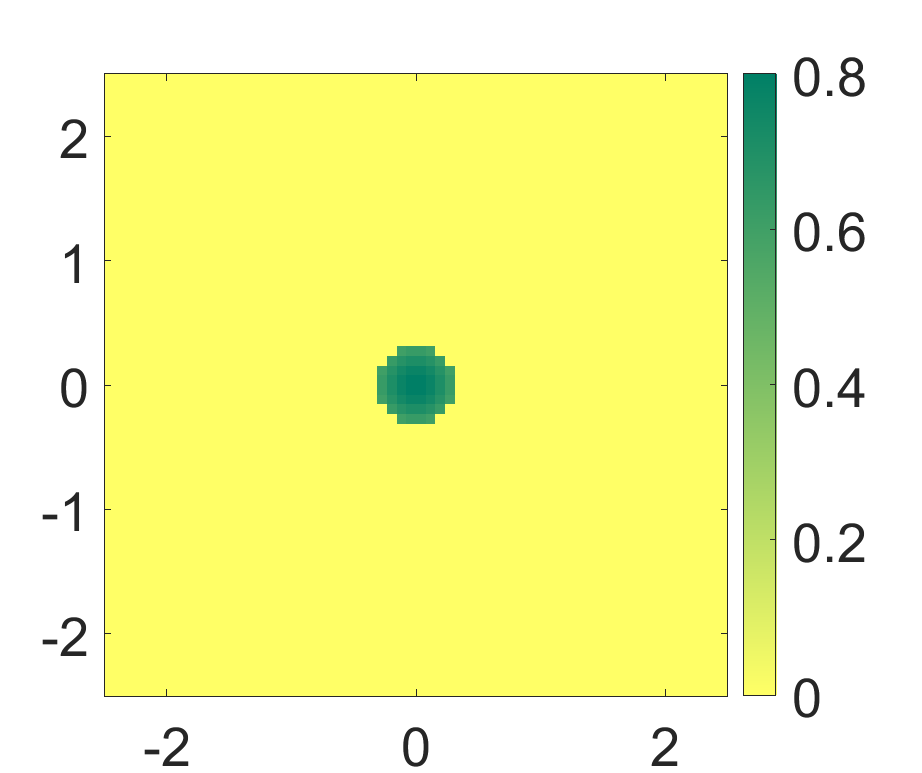}
    \\[-0.4em]
    &\includegraphics[width=0.22\linewidth,trim={1em 0 0.4em 0},clip]{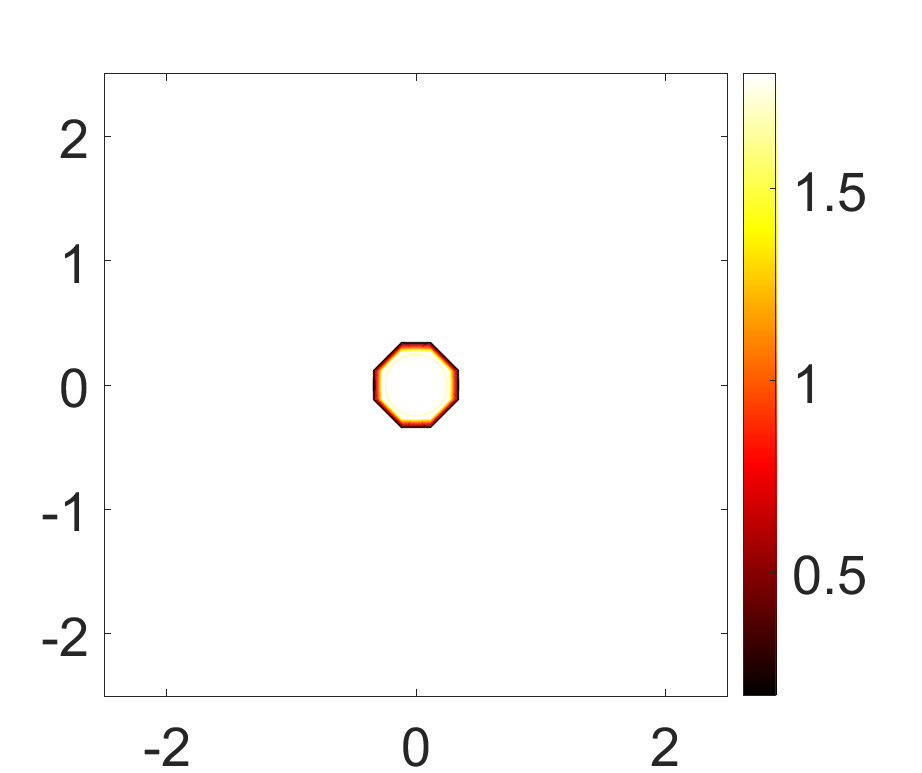} & 
    \includegraphics[width=0.22\linewidth,trim={2em 0 0.4em 0},clip]{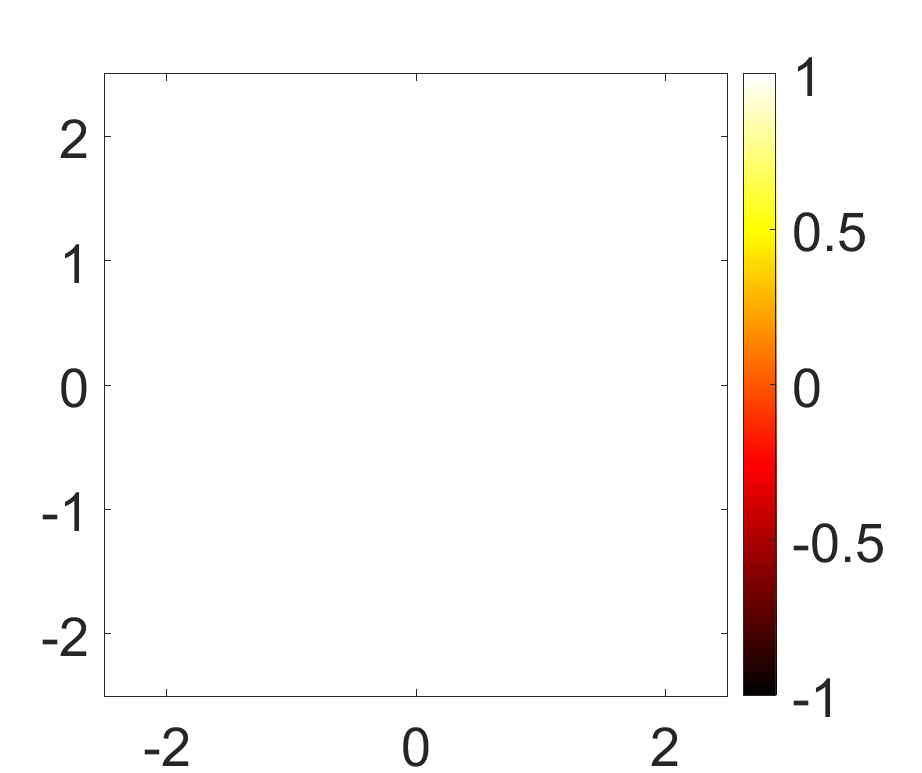} & 
    \includegraphics[width=0.22\linewidth,trim={2em 0 0 0},clip]{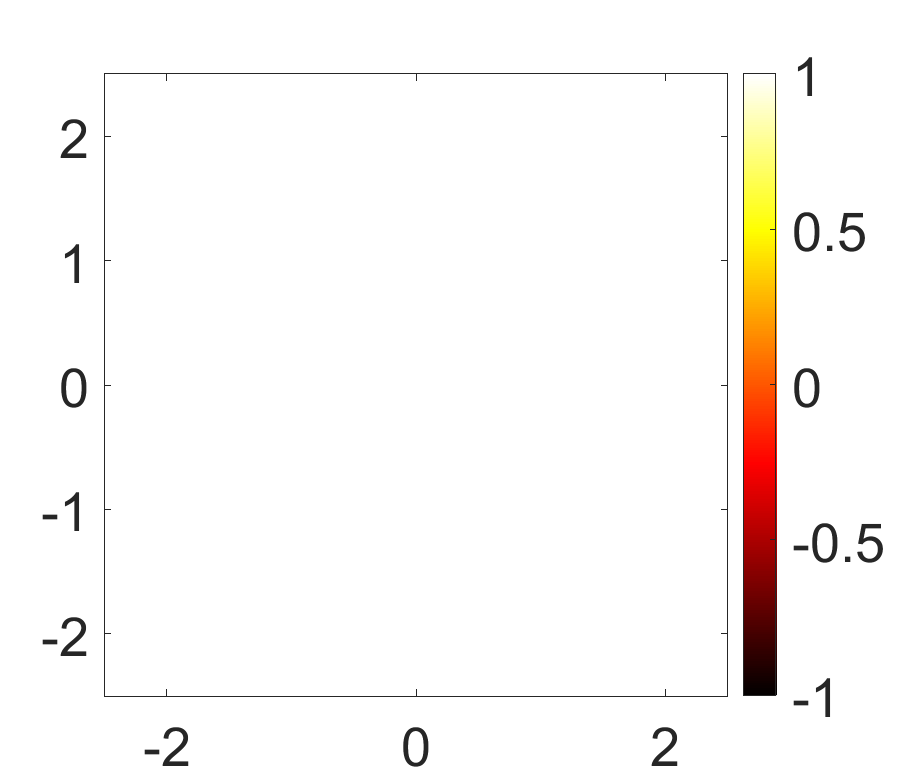} &
    \includegraphics[width=0.22\linewidth,trim={1em 0 0.4em 0},clip]{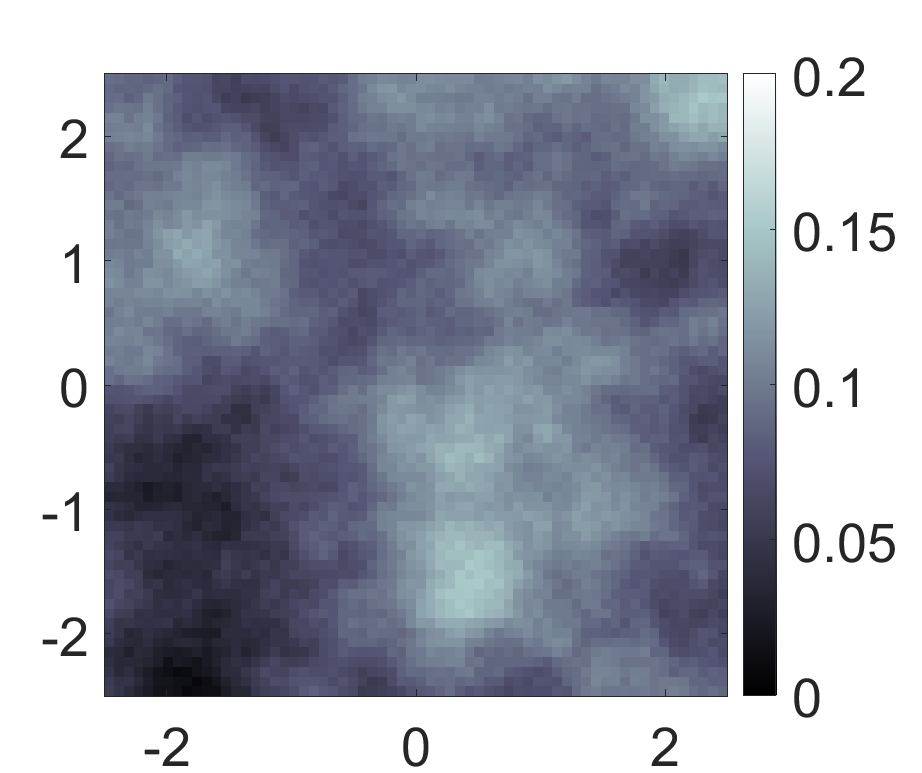} 
    \\
    \hline
    \multirow{2}{*}{\rotatebox{90}{\footnotesize{$t=150$}}}&
    \includegraphics[width=0.22\linewidth,trim={1em 0 0.4em 0},clip]{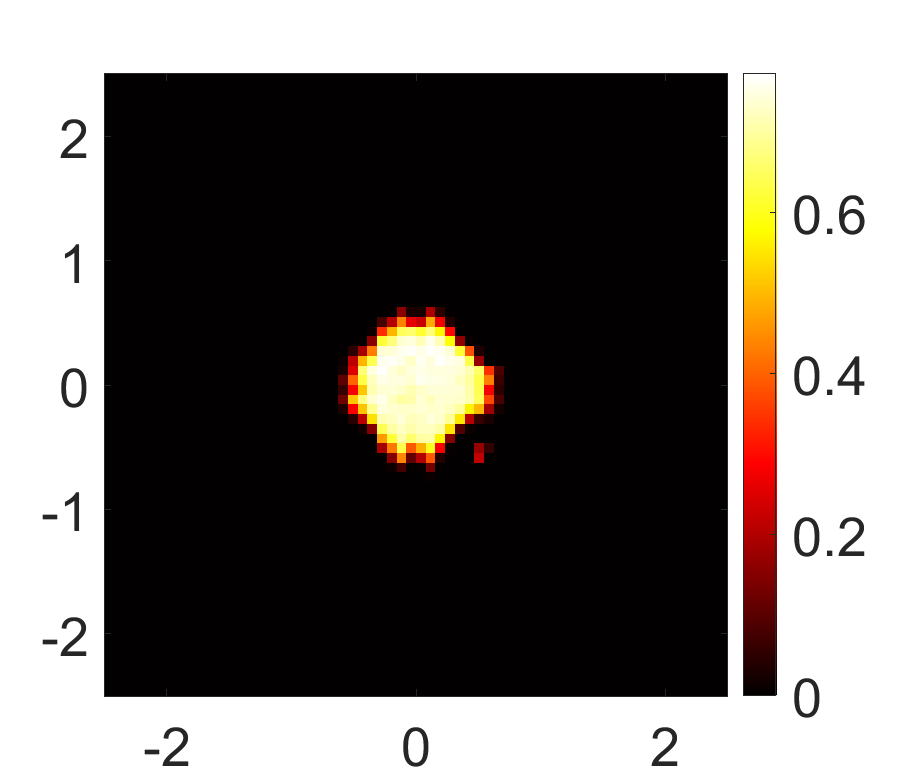} & \includegraphics[width=0.22\linewidth,trim={2em 0 0.4em 0},clip]{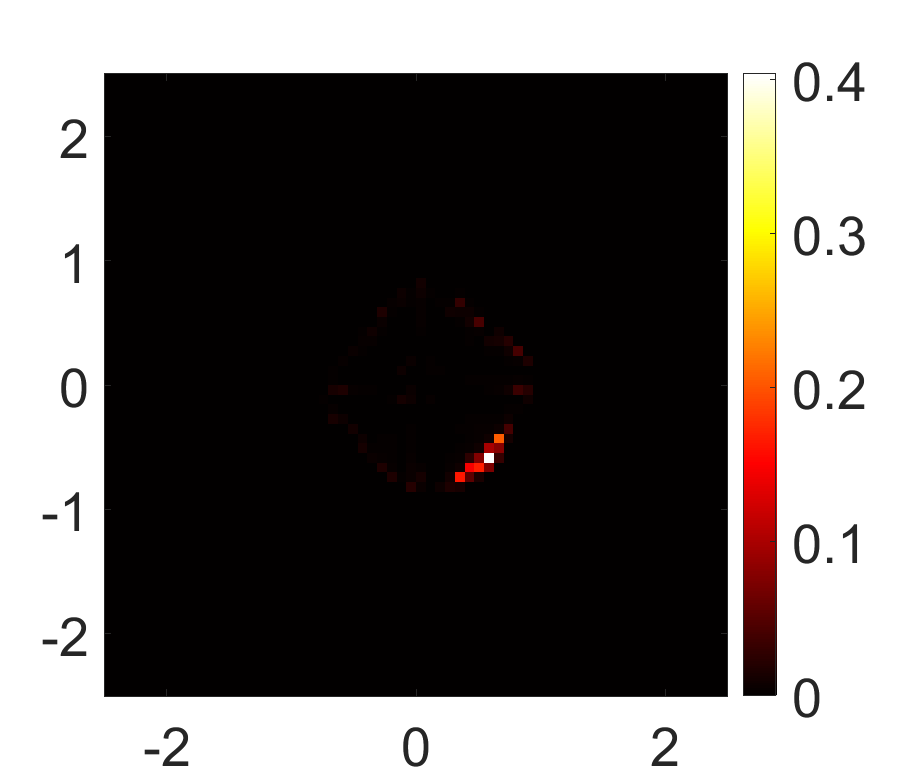} & \includegraphics[width=0.22\linewidth,trim={2em 0 0 0},clip]{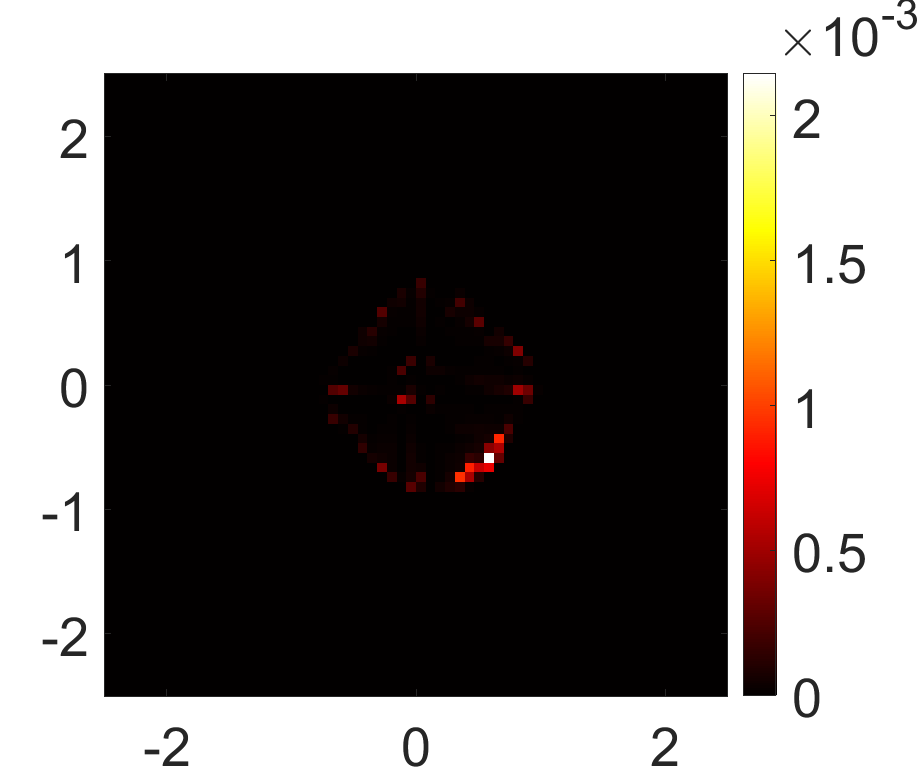} &
    \includegraphics[width=0.22\linewidth,trim={1em 0 0.4em 0},clip]{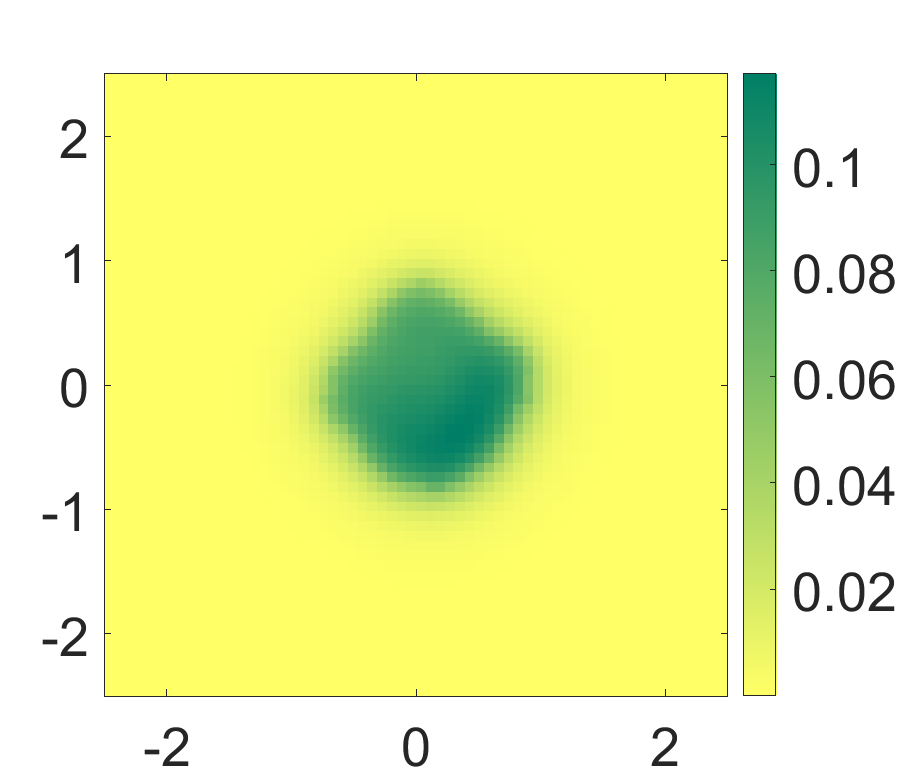}
    \\[-0.4em]
    &\includegraphics[width=0.22\linewidth,trim={1em 0 0.4em 0},clip]{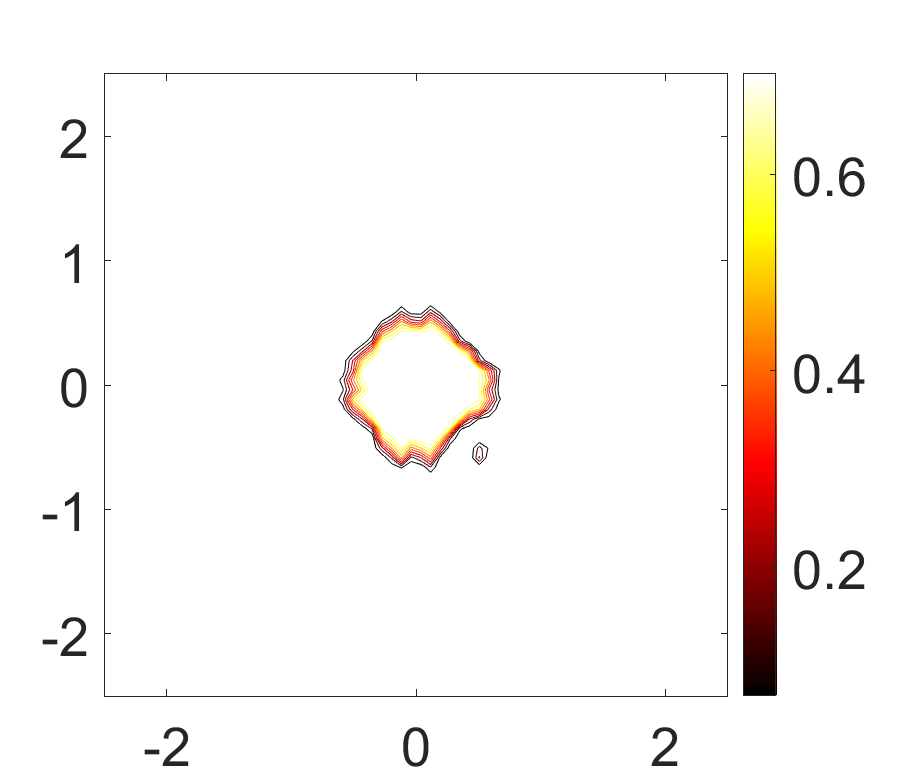} & \includegraphics[width=0.22\linewidth,trim={2em 0 0.4em 0},clip]{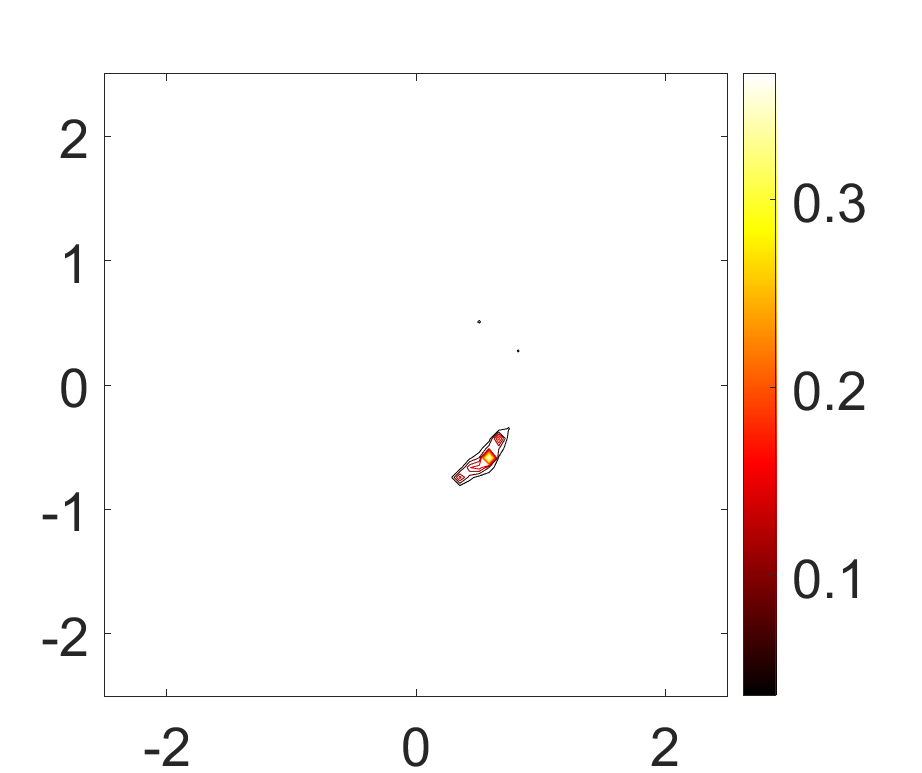} & \includegraphics[width=0.22\linewidth,trim={2em 0 0 0},clip]{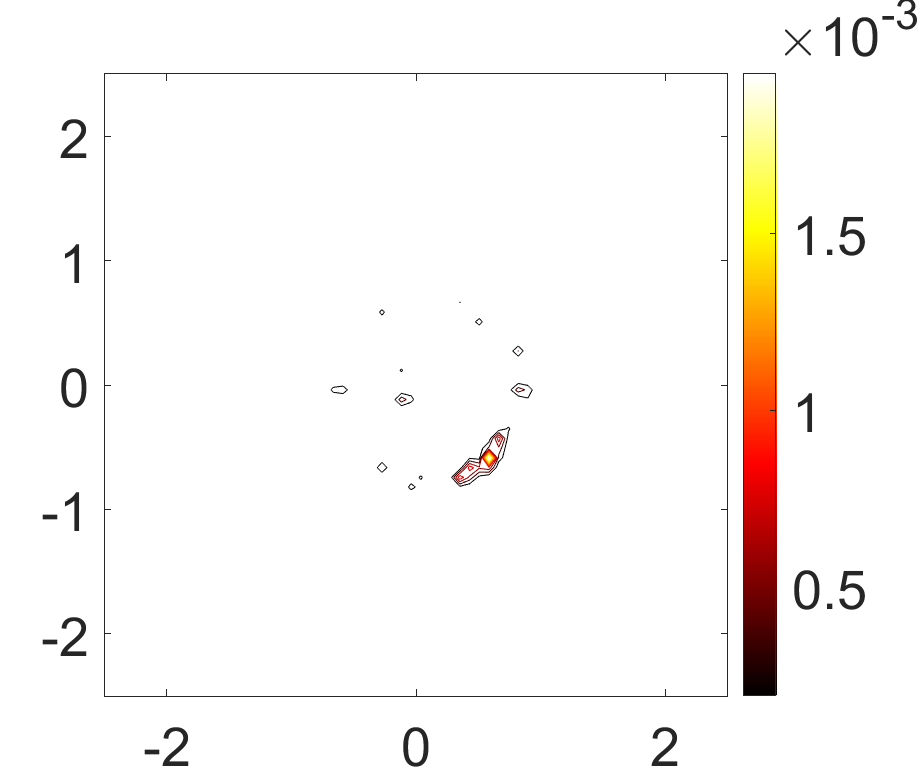} &
    \includegraphics[width=0.22\linewidth,trim={1em 0 0.4em 0},clip]{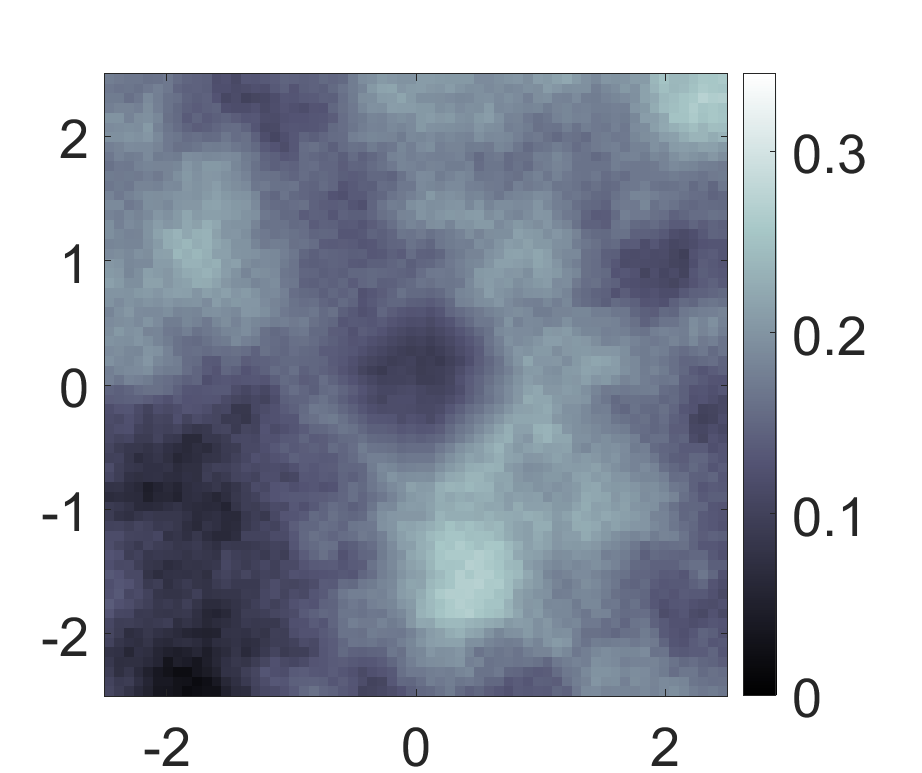} 
    \\
    \hline
    \multirow{2}{*}{\rotatebox{90}{\footnotesize{$t=300$}}}&
    \includegraphics[width=0.22\linewidth,trim={1em 0 0.4em 0},clip]{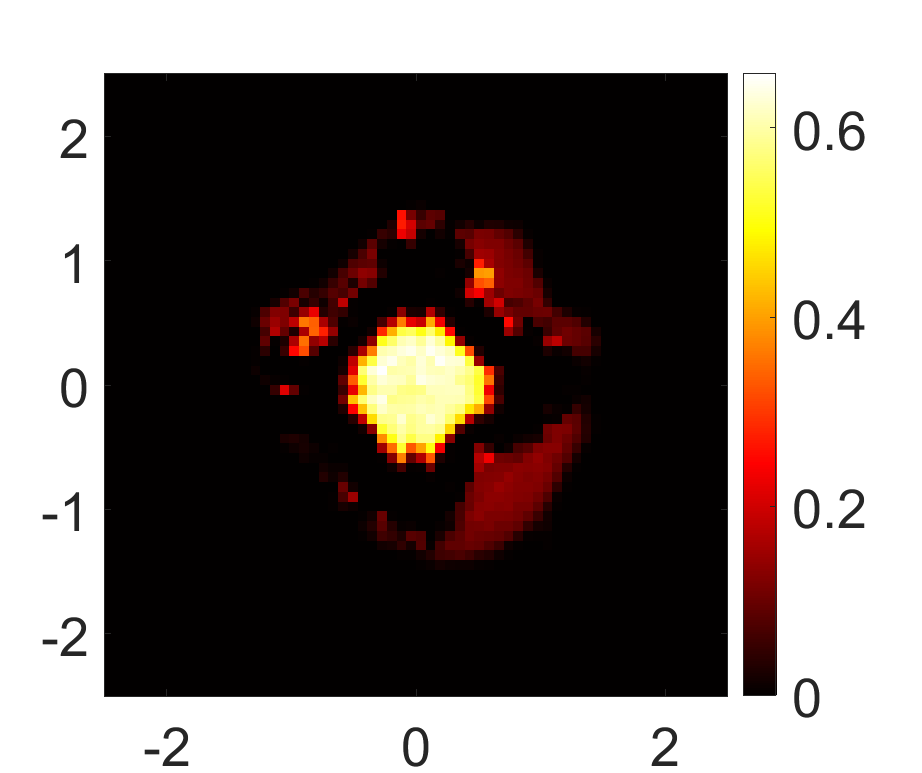} & \includegraphics[width=0.22\linewidth,trim={2em 0 0.4em 0},clip]{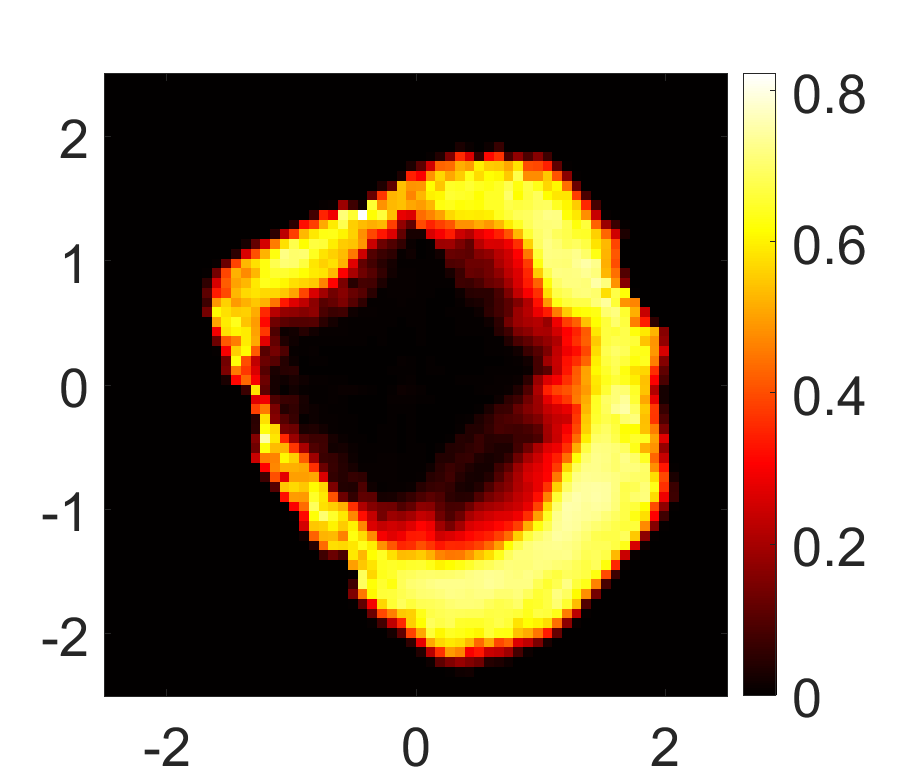} & \includegraphics[width=0.22\linewidth,trim={2em 0 0 0},clip]{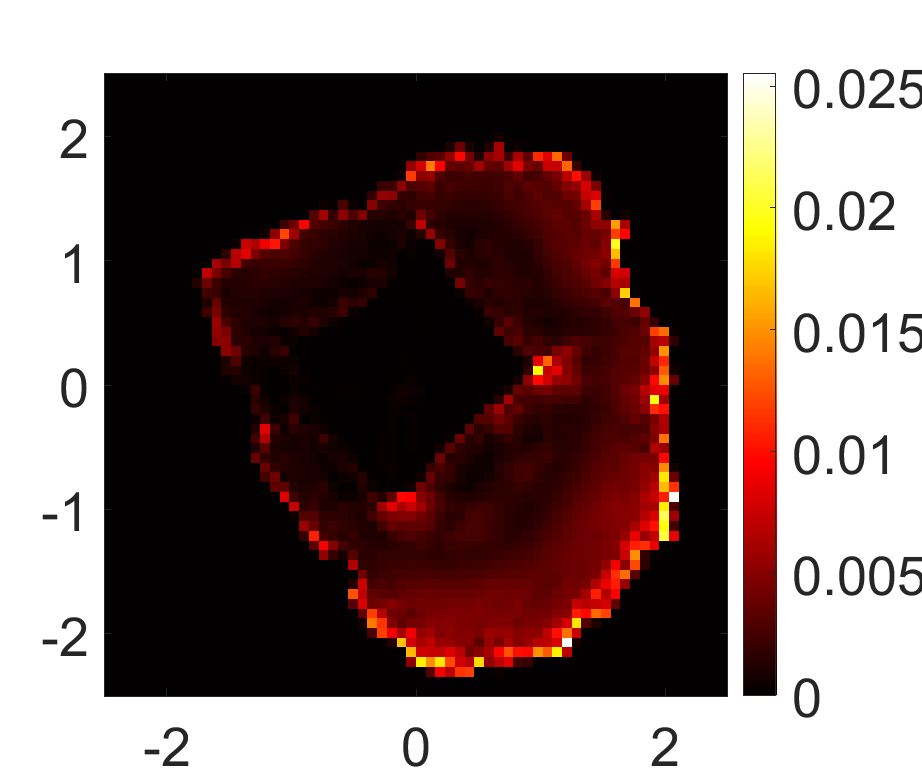} &
    \includegraphics[width=0.22\linewidth,trim={1em 0 0.4em 0},clip]{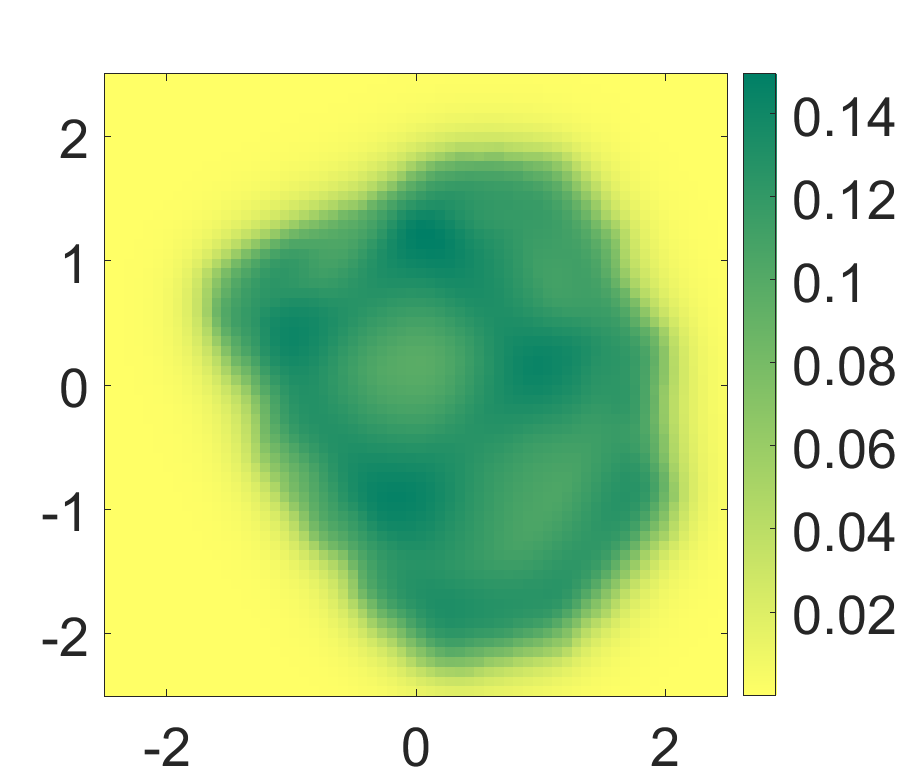}
    \\[-0.4em]
    &\includegraphics[width=0.22\linewidth,trim={1em 0 0.4em 0},clip]{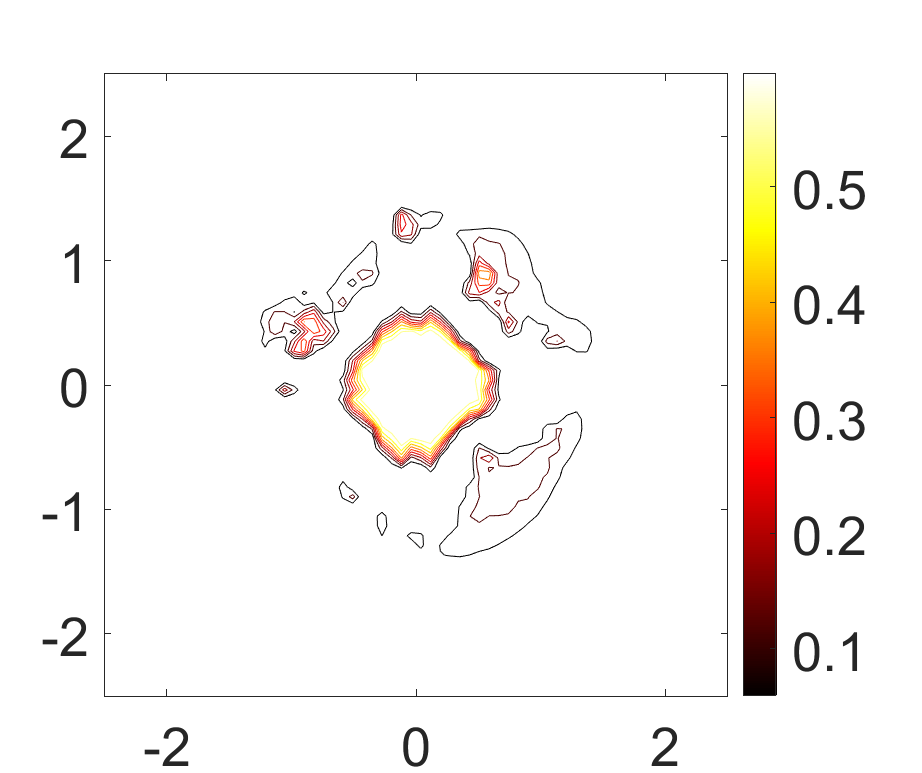} & \includegraphics[width=0.22\linewidth,trim={2em 0 0.4em 0},clip]{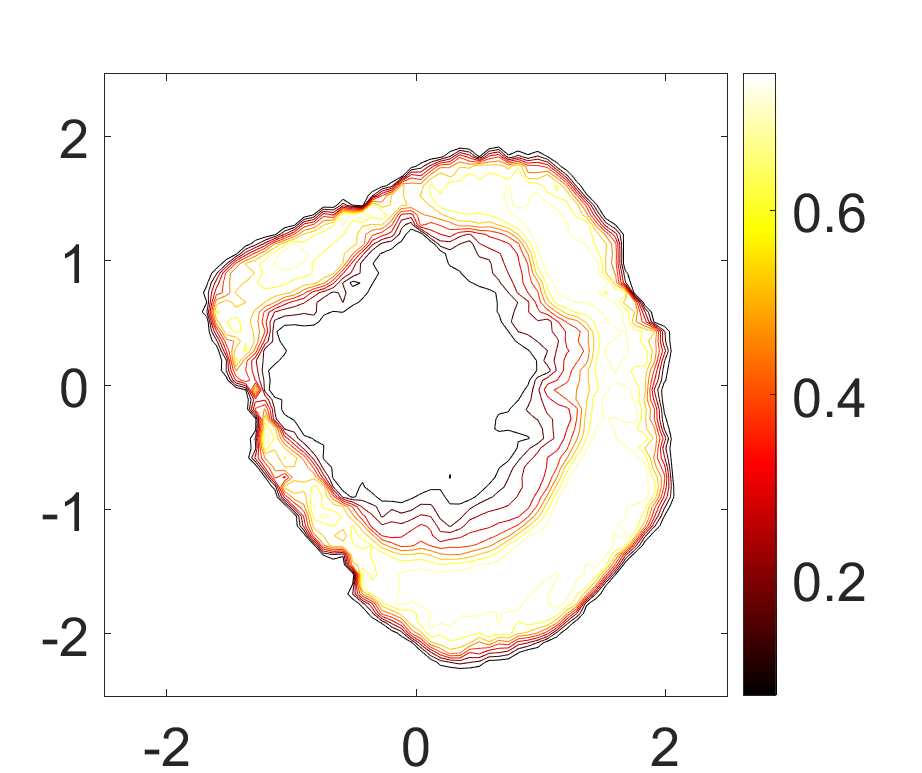} &
    \includegraphics[width=0.22\linewidth,trim={2em 0 0 0},clip]{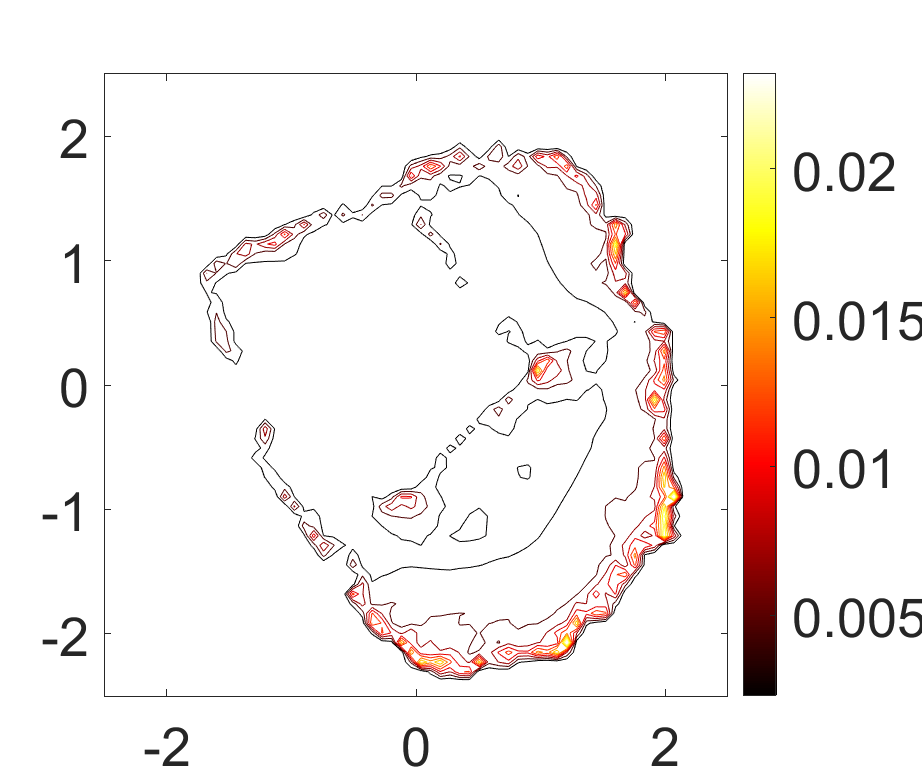} &
    \includegraphics[width=0.22\linewidth,trim={1em 0 0.4em 0},clip]{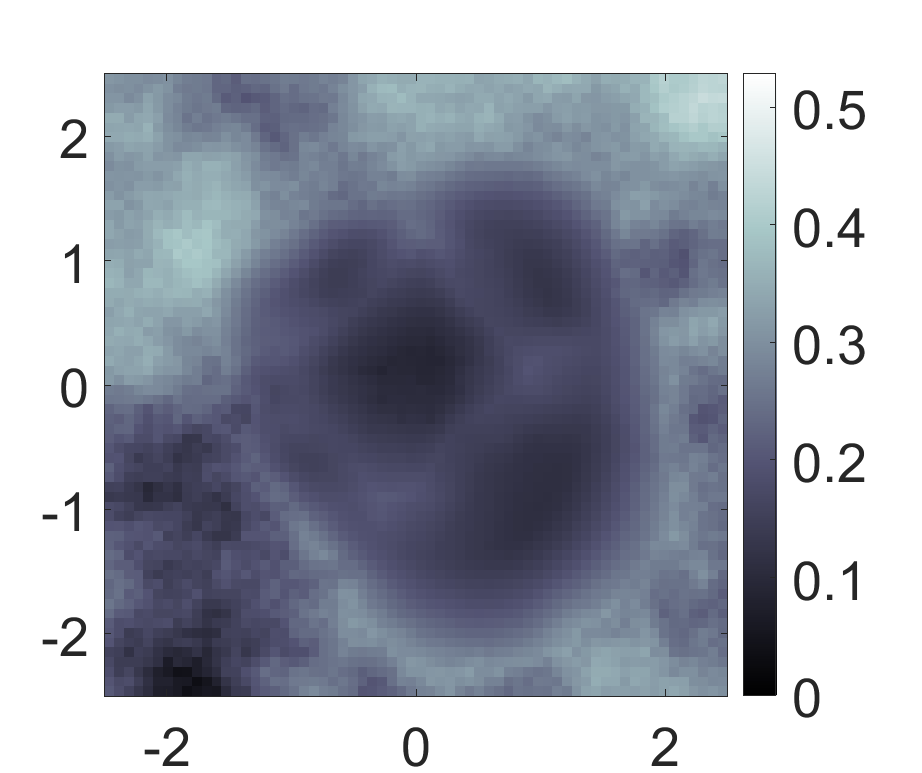}
\end{tabular}
    \vskip-1em
    \caption{\textbf{Experiment \ref{exp:generic} --- Generic invasion.} An initial concentration of epithelial ($a=0$) cancer cells (top panel) gives rise, through EMT, to mesenchymal, higher $a$, cancer cells  densities (middle and lower panels) which degrade and invade the ECM. As these move to regions of lower TGF-$\beta$ concentration (lower panel), MET is triggered and tumour foci appear at lower $a$ that are disconnected from the original location of the tumour.}\label{fig:generic}
\end{figure}

\begin{table} 
\footnotesize 
\begin{tabular}{|c|c|c|c|c|c|}
\hline
    {}&{Param.}& {Name} & {Value} & {Units} & {Reference} \\ 
\hline
\multirow{14}{*}{\rotatebox{90}{$c$-eq.}}
    &$D_c^M$&CC diffusion rate&$10^{-6}$&$\rm {mm^2}d^{-1}$& \cite{Stokes1991a} \\
    &$D_c^{sl}$&CC diffusion scale &$20$&$\rm M^2{mm^{-4}}$&(our estimate)\\
    &$D_c^{\alpha_0}$& CC diffusion midpoint&$0.5$&---&(our estimate)\\
    &$\chi_c^M$&CC haptotactic response&$0.5$&$\rm {mm^2}M^{-1}d^{-1}$&\cite{Anderson1998,Chaplain1996,Orme1996}\\
    &$\chi_c^{sl}$&CC haptotaxis scale&$20$&---&(our estimate)\\
    &$\chi_c^{\alpha_0}$&CC haptotaxis midpoint&$0.5$&---&(our estimate)\\
    &$L^\text{emt}$&EMT rate&$0.001$&$\rm {M}{mm^{-2}}d^{-1}$&(our estimate)\\
    &$\lambda^\text{emt}$&EMT scale&$30$&$\rm M^{-1}{mm^{-2}}$&(our estimate)\\
    &$f_0^\text{emt}$&midpoint EMT&$0.01$&$\rm M{mm^2}$&(our estimate)\\
    &$L^\text{met}$&MET rate&$10^{-4}$&$\rm {M}{mm^{-2}}d^{-1}$&(our estimate)\\
    &$\lambda^\text{met}$&MET scale&$30$&$\rm M^{-1}{mm^{-2}}$&(our estimate)\\
    &$f_0^\text{met}$&midpoint MET&$0.001$&$\rm M{mm^2}$&(our estimate)\\
    &$\bar{\gamma_c}$&CC proliferation rate&$1$&$\rm d^{-1}$&\cite{Stokes1991a}\\
    &$\gamma_c^\ast$&CC proliferation exponent&$7$&---&(our estimate)\\
\hline
    \multirow{3}{*}{\rotatebox{90}{$f$-eq.}}
    &$D_f$&GF diffusion rate&$0.01$ &$\rm {mm^2}d^{-1}$&\cite{Stokes1991a}\\
    &$\gamma_f$&GF production rate&$0.3$&$\rm {mm^2}M^{-1}d^{-1}$&(our estimate)\\
    &$d_f$&GF decay&$3$&$\rm d^{-1}$&cf. \cite{TGFb}\\
\hline
    \multirow{3}{*}{\rotatebox{90}{$v$-eq.}}
        &$\eta_v$&ECM degradation rate&$0.01$&$\rm {mm^2}M^{-1}d^{-1}$&cf. \cite{hybrid_2nd, Chaplain2005}\\
        &$K_\eta$&CM degradation midpoint&$10$&---&(our estimate)\\
        &$\lambda_v$&ECM reconstruction rate& $0.005$&$\rm {mm^2}M^{-1}d^{-1}$&cf. \cite{hybrid_2nd}\\
\hline
\end{tabular}
\caption{Parameters used in \textbf{Experiment \ref{exp:generic} --- Generic invasion}.} \label{tbl:generic}
\end{table}

\end{experiment}

\begin{experiment}\textbf{Decoupled EMT-MET switch functions}\label{exp:decoupled}

Being aware of the wide range of invasive strategies that can be observed in vivo we wonder whether it is the phenotype sensitivity on the TGF-$\beta$ concentration, which is in mathematical terms reflected in the shapes of functions $S_{EMT}$ and $S_{MET}$ that trigger EMT and MET, respectively, influences the macroscopic picture of the infiltrating tumour. 

As the simulations in Experiment \ref{exp:generic} exhibit---shown in Figure \ref{fig:generic}---the EMT and MET processes are driven by the $S_{EMT}$ and $S_{MET}$ switch functions \eqref{EMT-switch}, \eqref{MET-switch} respectively. In particular, their relative dependence, on the concentrations of TGF-$\beta$, has a significant impact in the final balance between of two processes. This is an important component of our proposed model and has significant biological implications. We will hence investigate it in this and the follow up experiments. 

In this experiment, in particular, we consider a parameter setting where the EMT and MET switch functions \eqref{EMT-switch} and \eqref{MET-switch} are decoupled, i.e. the MET switch is effectively activated in lower concentrations of TGF-$\beta$. Confer also the parameter values in Table \ref{tbl:decoupled} and the graphs of $S_{EMT}$ and $S_{MET}$ in Figure \ref{fig:coupled-decoupled}. 

In more detail, as can be seen in Figure \ref{fig:decoupled}, EMT gives rise to cancer cells of higher phenotype value $a$ which in turn degrade the ECM, cf. \eqref{eta}, \eqref{mdl3}, and migrate along the ECM gradient. As in the previous experiments, when the more migratory cancer cells reach regions of the domain of low TGF-$\beta$ concentration $f$, they undergo MET and give rise to cancer cell foci of lower phenotypic value $a$. These cancer cell foci resemble cancer cell islands and are disconnected from the initial body of the tumour. To make this more clear, we have plotted in Figure \ref{fig:decoupled} the isolines down to the double precision accuracy of $10^{-15}$. 

It should be noted that the parameters in this numerical experiment are such that the simulation results are pronounced enough so that, though the comparison with the ensuing experiment, the qualitative conclusions of our modelling become clear.

\begin{table}  
\footnotesize 
\begin{tabular}{|c|c|c|c|c|c|}
\hline
    {}&{Param.}& {Name} & {Value} & {Units} & {Reference} \\ 
\hline
\multirow{14}{*}{\rotatebox{90}{$c$-eq.}}
    &$D_c^M$&CC diffusion rate&$10^{-6}$&$\rm {mm^2}d^{-1}$& \cite{Stokes1991a} \\
    &$D_c^{sl}$&CC diffusion scale &$20$&$\rm M^2{mm^{-4}}$&(our estimate)\\
    &$D_c^{\alpha_0}$& CC diffusion midpoint&$0.5$&---&(our estimate)\\
    &$\chi_c^M$&CC haptotactic response&$0.4$&$\rm {mm^2}M^{-1}d^{-1}$&\cite{Anderson1998,Chaplain1996,Orme1996}\\
    &$\chi_c^{sl}$&CC haptotaxis scale&$20$&---&(our estimate)\\
    &$\chi_c^{\alpha_0}$&CC haptotaxis midpoint&$0.5$&---&(our estimate)\\
    &$L^\text{emt}$&EMT rate&$0.1$&$\rm {M}{mm^{-2}}d^{-1}$&(our estimate)\\
    &$\lambda^\text{emt}$&EMT scale&$30$&$\rm M^{-1}{mm^{-2}}$&(our estimate)\\
    &$f_0^\text{emt}$&midpoint EMT&$0.01$&$\rm M{mm^2}$&(our estimate)\\
    &$L^\text{met}$&MET rate&$20$&$\rm {M}{mm^{-2}}d^{-1}$&(our estimate)\\
    &$\lambda^\text{met}$&MET scale&$300$&$\rm M^{-1}{mm^{-2}}$&(our estimate)\\
    &$f_0^\text{met}$&midpoint MET&$0.003$&$\rm M{mm^2}$&(our estimate)\\
    &$\bar{\gamma_c}$&CC proliferation rate&$1$&$\rm d^{-1}$&\cite{Stokes1991a}\\
    &$\gamma_c^\ast$&CC proliferation exponent&$7$&---&(our estimate)\\
\hline
    \multirow{3}{*}{\rotatebox{90}{$f$-eq.}}
    &$D_f$&GF diffusion rate&$0.01$ &$\rm {mm^2}d^{-1}$&\cite{Stokes1991a}\\
    &$\gamma_f$&GF production rate&$0.3$&$\rm {mm^2}M^{-1}d^{-1}$&(our estimate)\\
    &$d_f$&GF decay&$3$&$\rm d^{-1}$&cf. \cite{TGFb}\\
\hline
    \multirow{3}{*}{\rotatebox{90}{$v$-eq.}}
        &$\eta_v$&ECM degradation rate&$0.01$&$\rm {mm^2}M^{-1}d^{-1}$&cf. \cite{hybrid_2nd, Chaplain2005}\\
        &$K_\eta$&CM degradation midpoint&$10$&---&(our estimate)\\
        &$\lambda_v$&ECM reconstruction rate& $10^{-4}$&$\rm {mm^2}M^{-1}d^{-1}$&cf. \cite{hybrid_2nd}\\
\hline
\end{tabular}
\caption{Parameters used in \textbf{Experiment \ref{exp:decoupled} --- Decoupled EMT-MET switch functions}. } \label{tbl:decoupled}
\end{table}

\begin{figure} 
\setlength{\tabcolsep}{1pt}
\centering
\begin{tabular}{l|ccc|c}
    \phantom{X}&\footnotesize{Cancer cells}&\footnotesize{Cancer cells}&\footnotesize{Cancer cells}&\footnotesize{TGF-$\beta$ \& ECM}\\[-0.3em]    &\footnotesize{$c(t,\cdot,\alpha=0)$}&\footnotesize{$c(t,\cdot,\alpha=0.5)$}&\footnotesize{$c(t,\cdot,\alpha=1)$}&\footnotesize{$f(t, \cdot),\ v(t,\cdot)$}\\
    \hline
    \multirow{2}{*}{\rotatebox{90}{\footnotesize{$t=0$}}}&
    \includegraphics[width=0.23\linewidth,trim={2em 0 0.5em 0},clip]{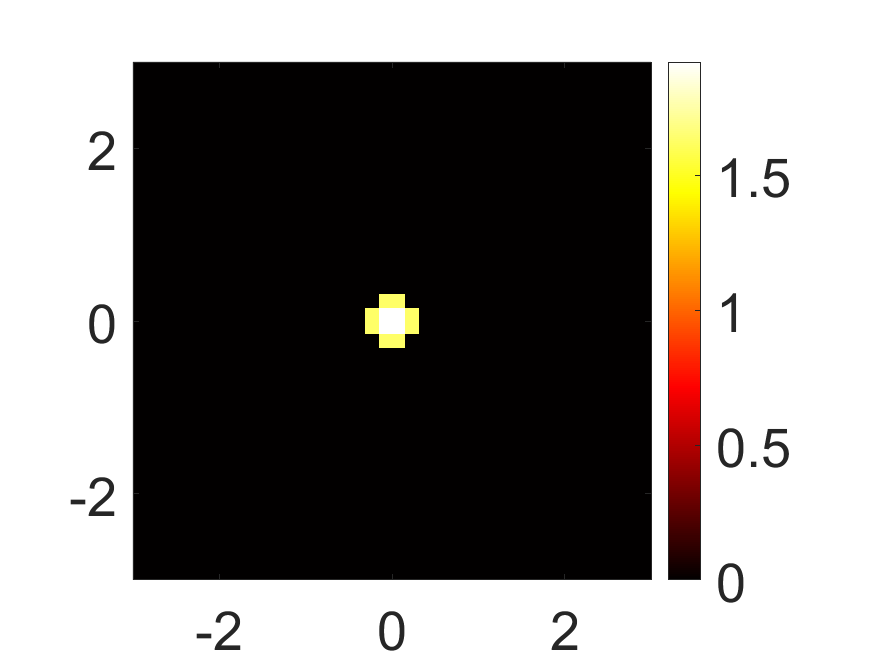} & \includegraphics[width=0.23\linewidth,trim={2em 0 0.5em 0},clip]{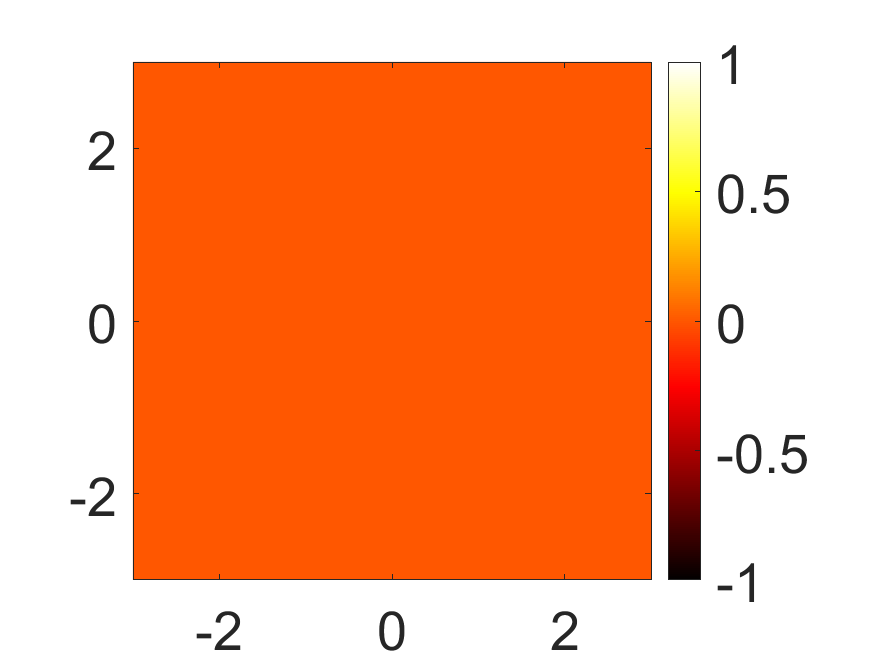} & \includegraphics[width=0.23\linewidth,trim={2em 0 0.5em 0},clip]{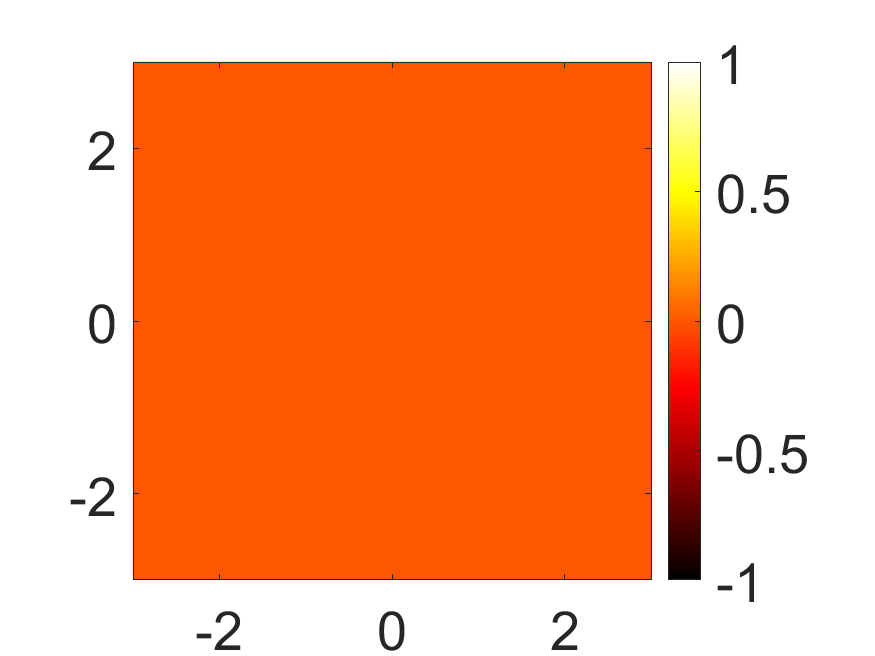} & 
    \includegraphics[width=0.23\linewidth,trim={2em 0 0.5em 0},clip]{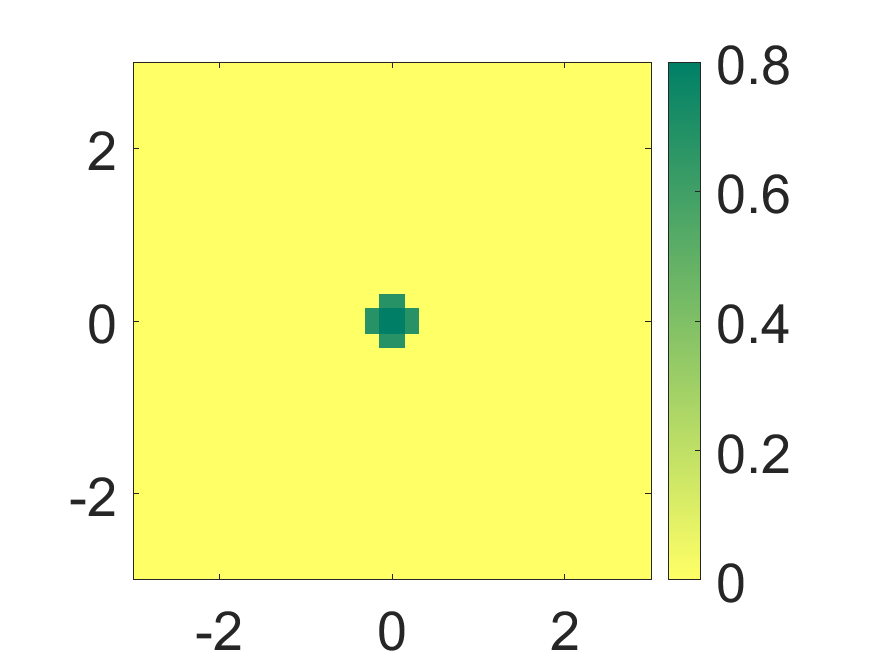}
    \\[-0.3em]
    &\includegraphics[width=0.23\linewidth,trim={2em 0 0.5em 0},clip]{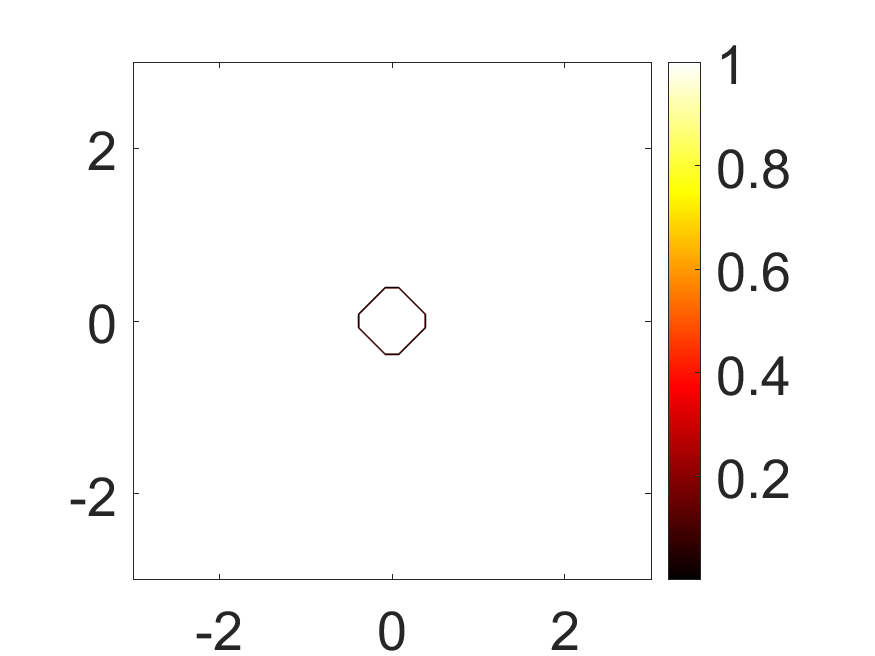} & \includegraphics[width=0.23\linewidth,trim={2em 0 0.5em 0},clip]{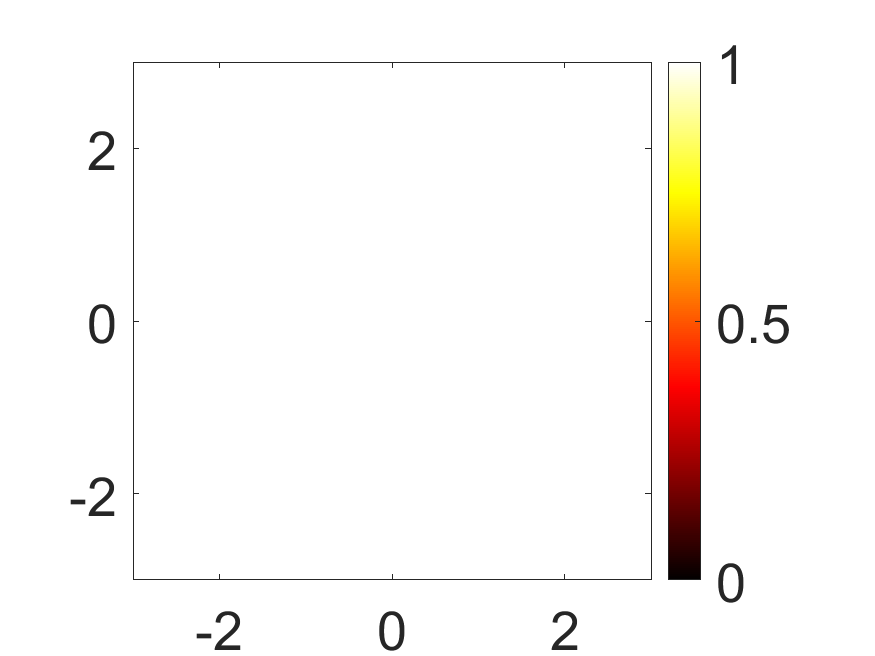} & \includegraphics[width=0.23\linewidth,trim={2em 0 0.5em 0},clip]{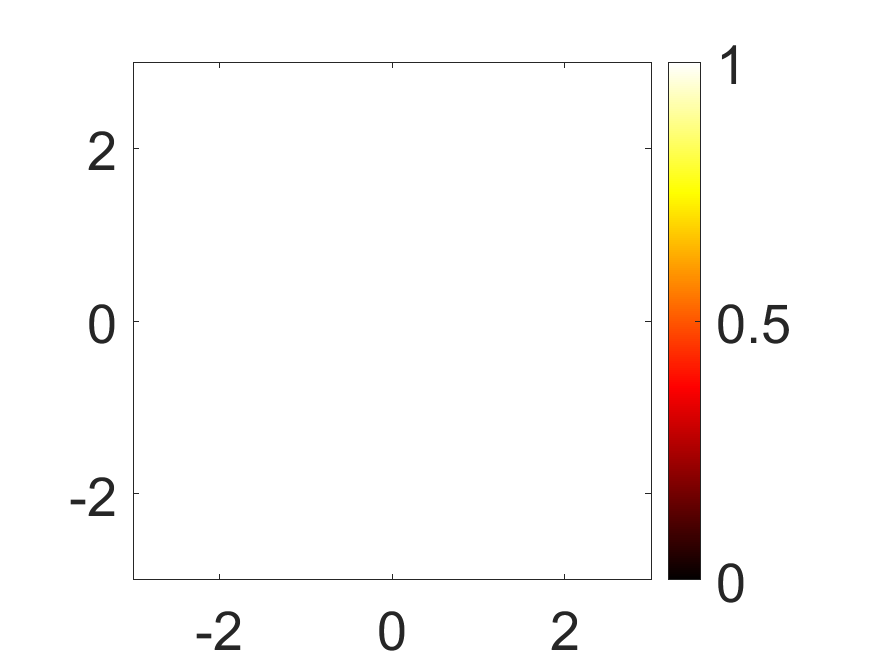} &
    \includegraphics[width=0.23\linewidth,trim={2em 0 0.5em 0},clip]{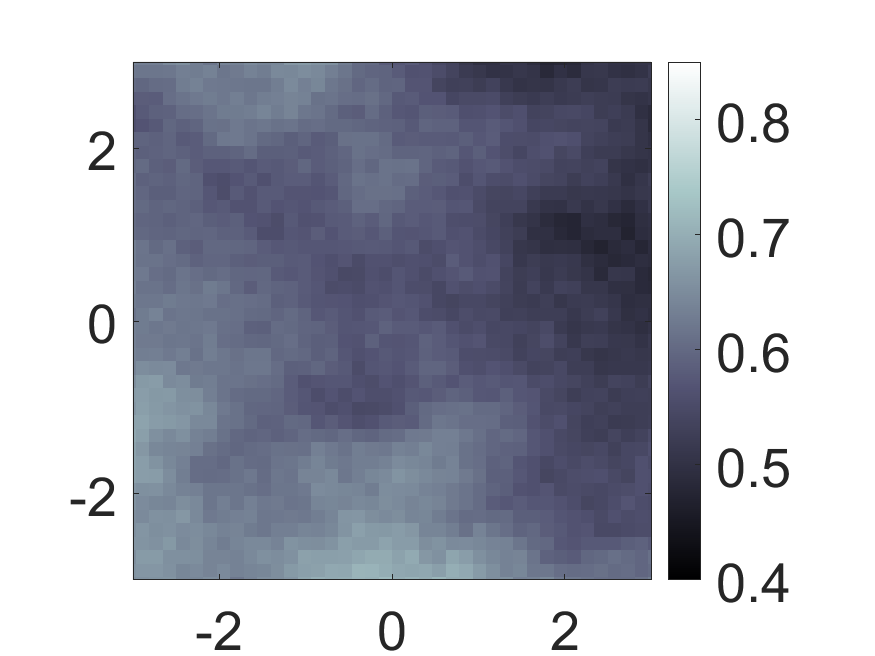} 
    \\
    \hline
    \multirow{2}{*}{\rotatebox{90}{\footnotesize{$t=150$}}}&
    \includegraphics[width=0.23\linewidth,trim={2em 0 0.5em 0},clip]{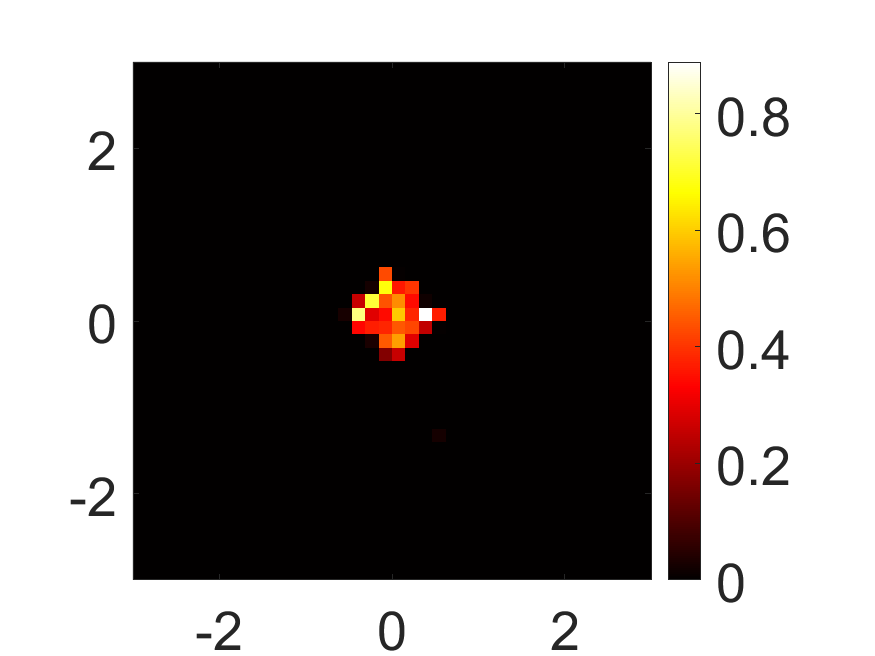} & \includegraphics[width=0.23\linewidth,trim={2em 0 0.5em 0},clip]{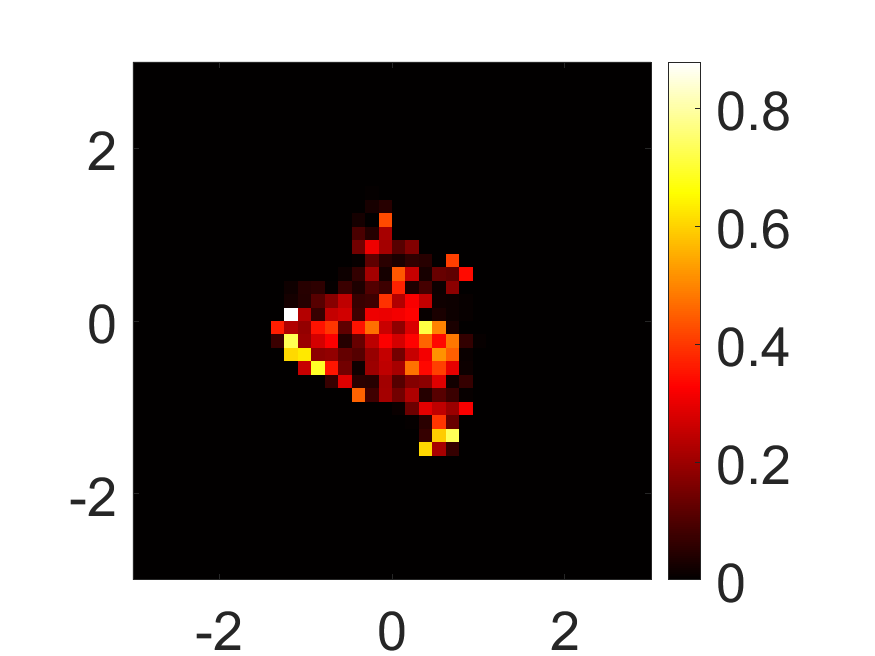} & \includegraphics[width=0.23\linewidth,trim={2em 0 0.5em 0},clip]{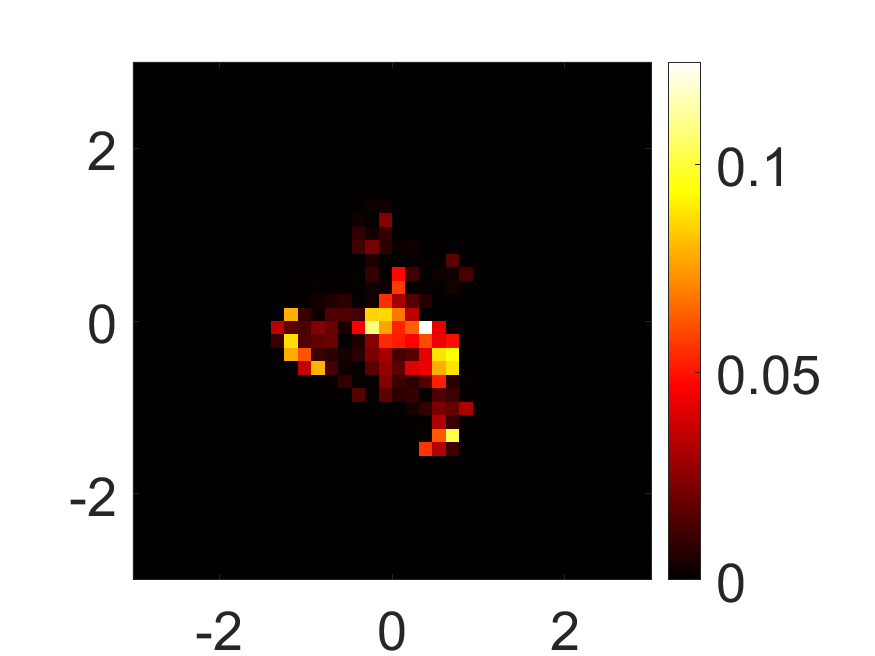} &
    \includegraphics[width=0.23\linewidth,trim={2em 0 0.5em 0},clip]{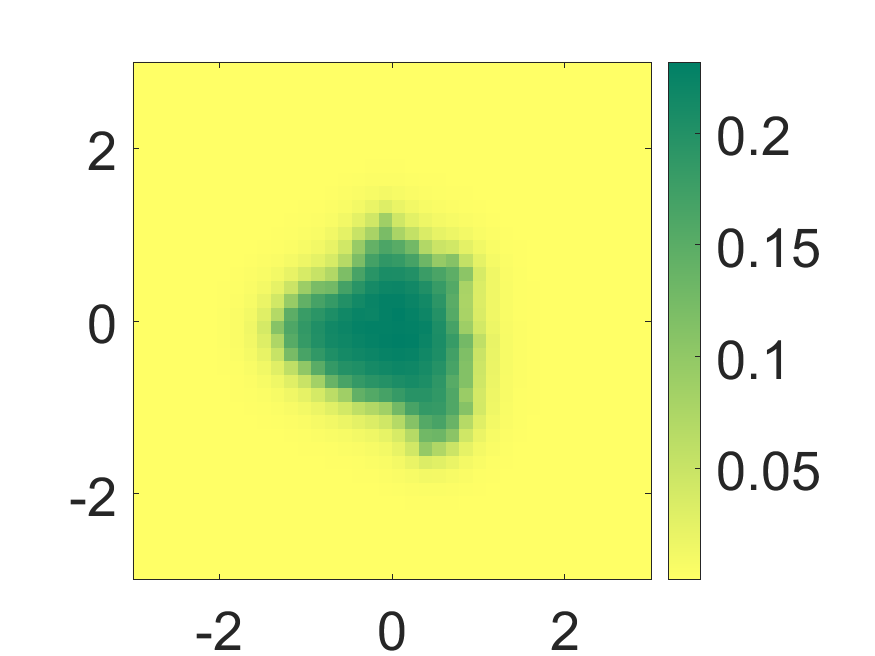}
    \\[-0.3em]
    &\includegraphics[width=0.23\linewidth,trim={2em 0 0.5em 0},clip]{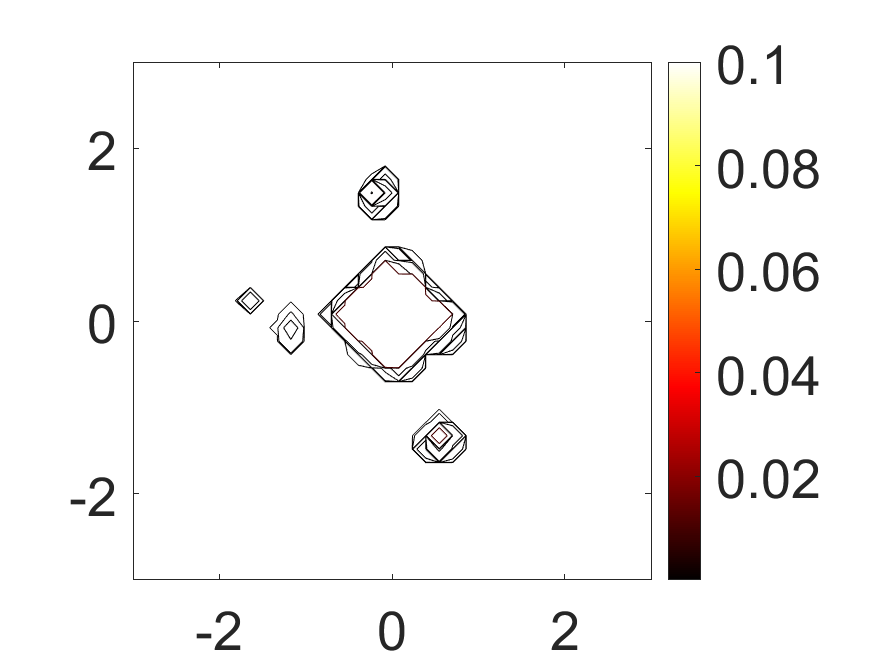} & \includegraphics[width=0.23\linewidth,trim={2em 0 0.5em 0},clip]{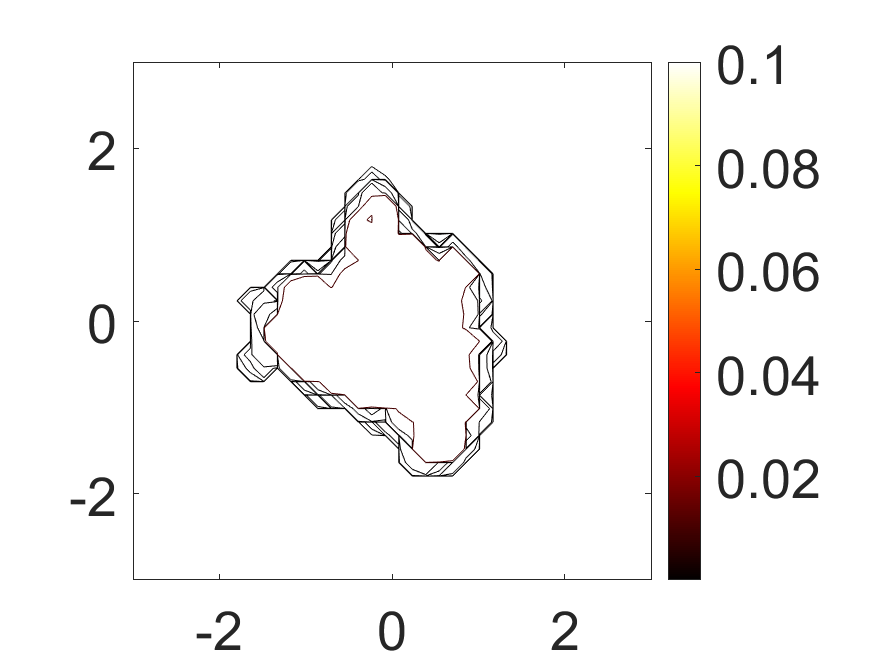} & \includegraphics[width=0.23\linewidth,trim={2em 0 0.5em 0},clip]{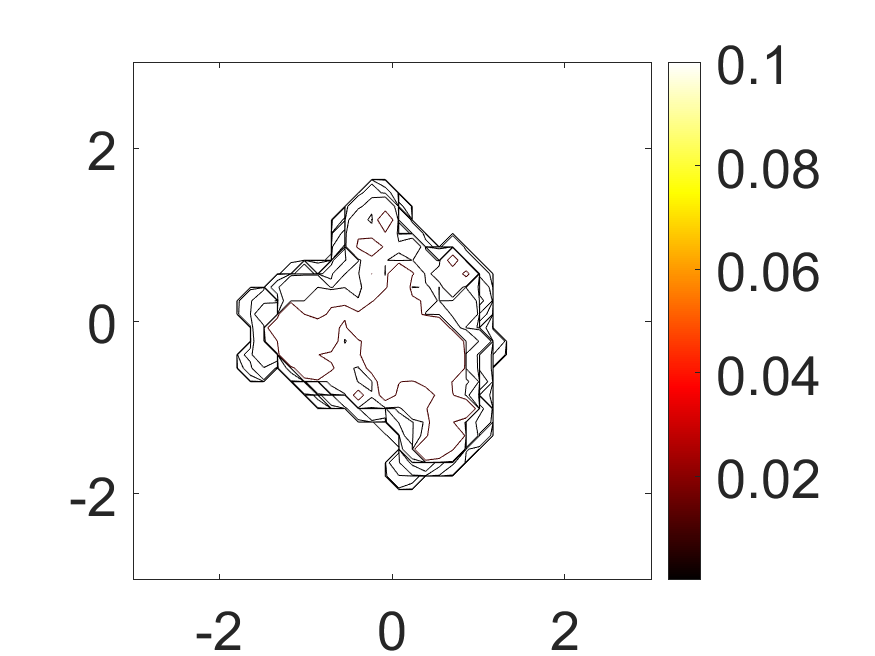} &
    \includegraphics[width=0.23\linewidth,trim={2em 0 0.5em 0},clip]{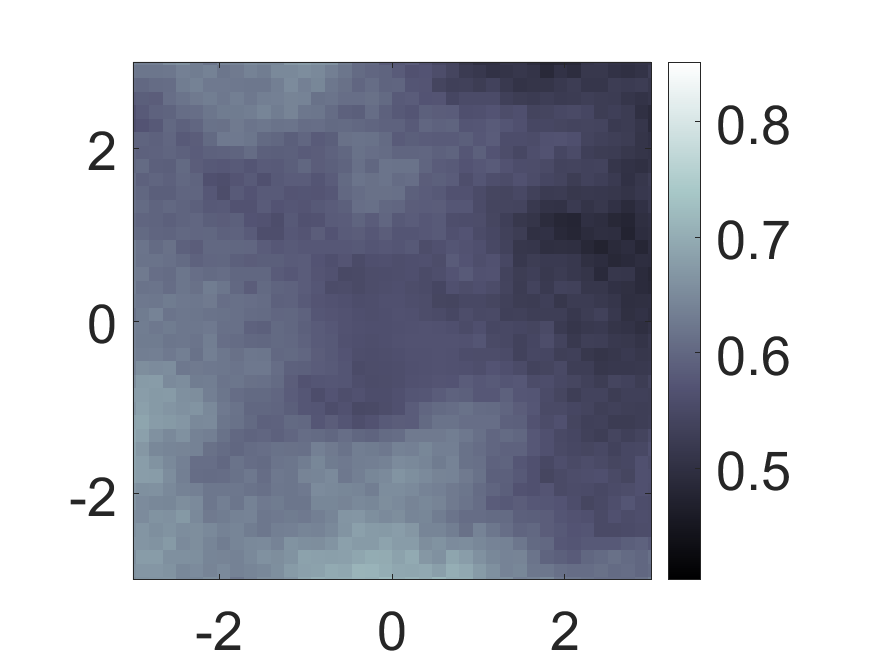} 
    \\
    \hline
    \multirow{2}{*}{\rotatebox{90}{\footnotesize{$t=300$}}}&
    \includegraphics[width=0.23\linewidth,trim={2em 0 0.5em 0},clip]{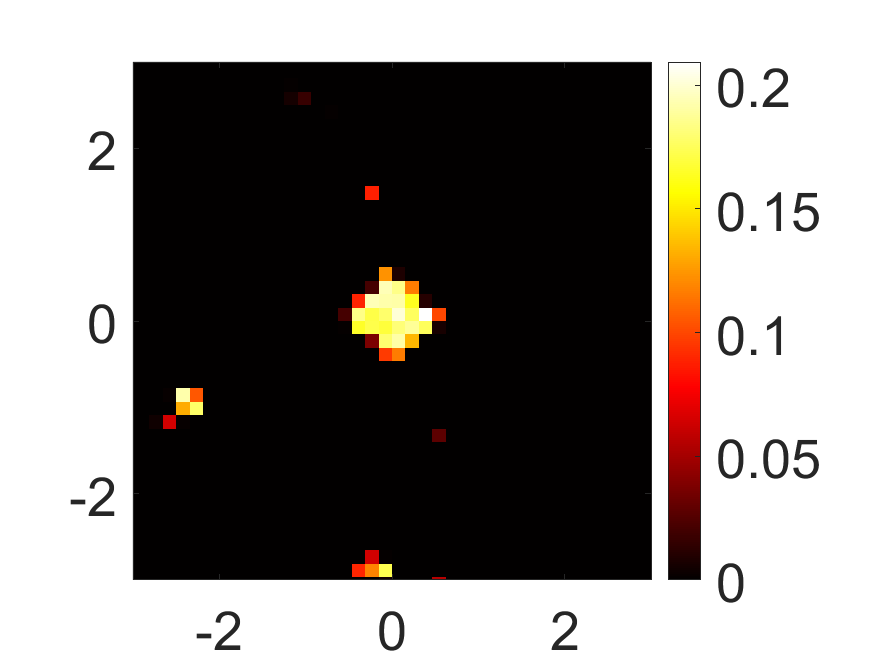} & \includegraphics[width=0.23\linewidth,trim={2em 0 0.5em 0},clip]{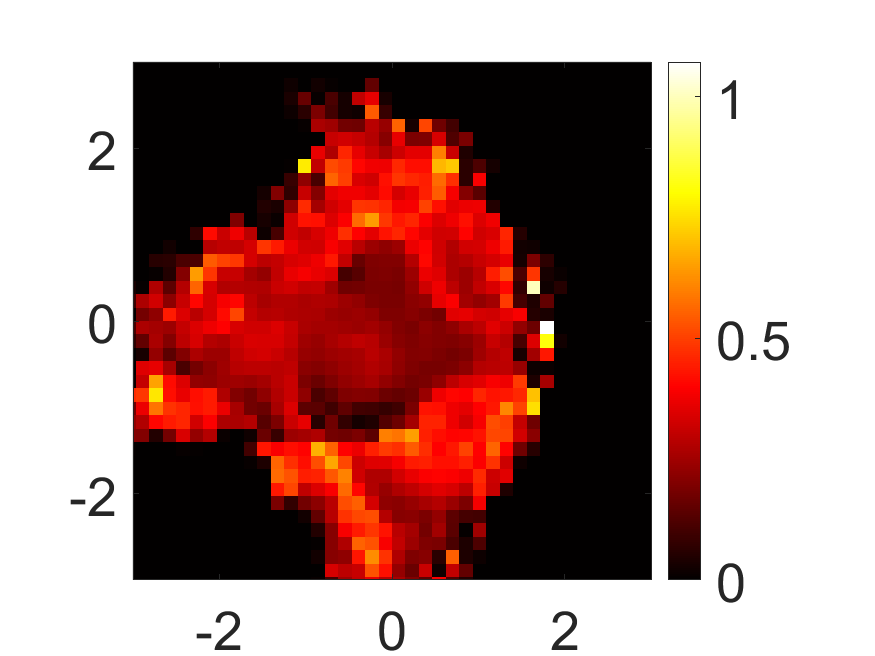} & \includegraphics[width=0.23\linewidth,trim={2em 0 0.5em 0},clip]{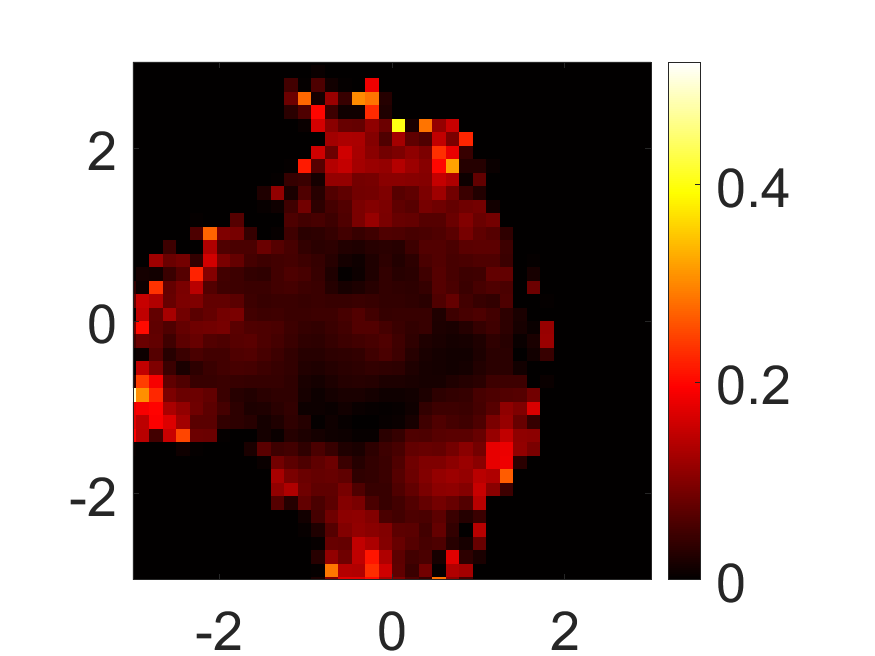} &
    \includegraphics[width=0.23\linewidth,trim={2em 0 0.5em 0},clip]{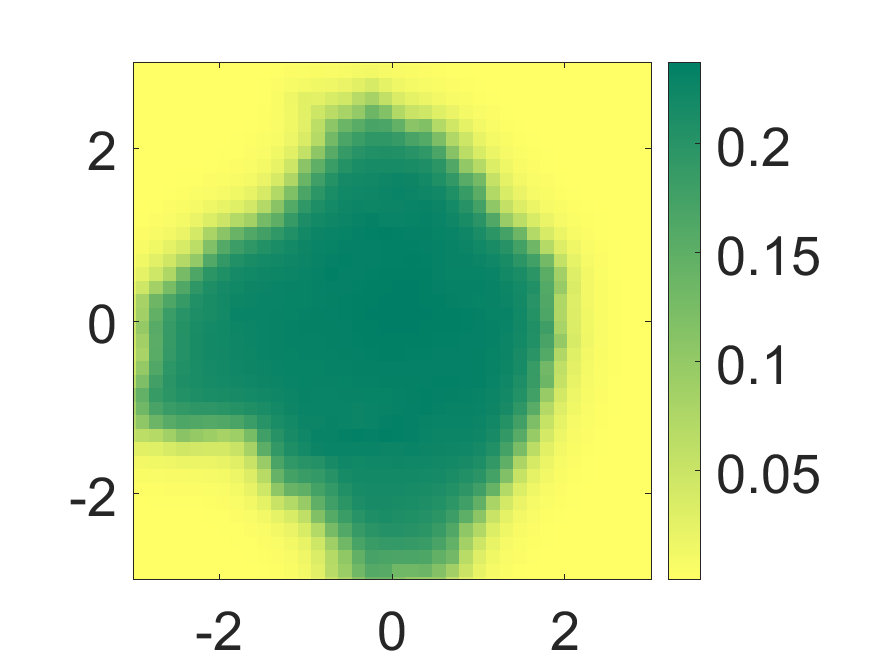}
    \\[-0.3em]
    &\includegraphics[width=0.23\linewidth,trim={2em 0 0.5em 0},clip]{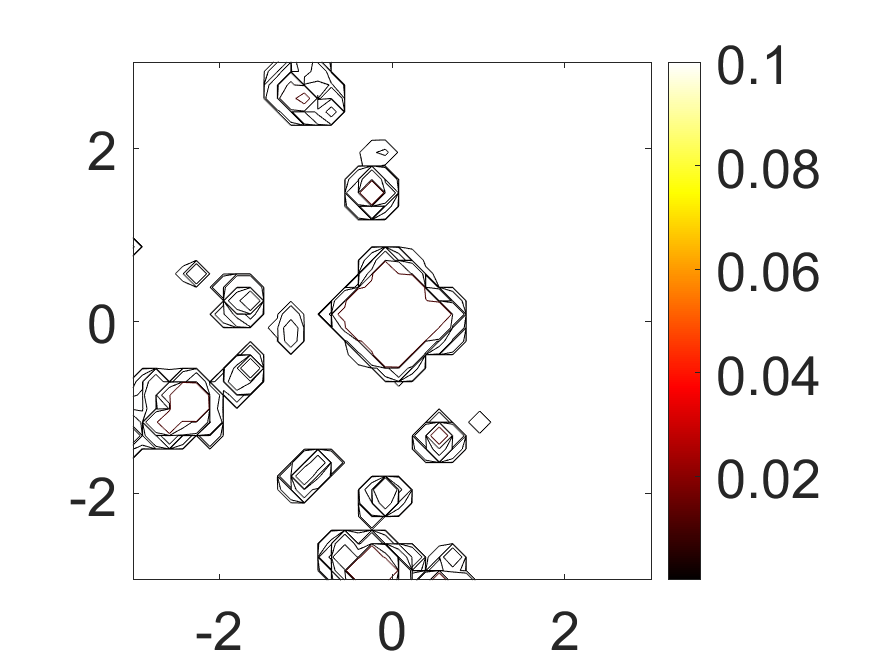} & \includegraphics[width=0.23\linewidth,trim={2em 0 0.5em 0},clip]{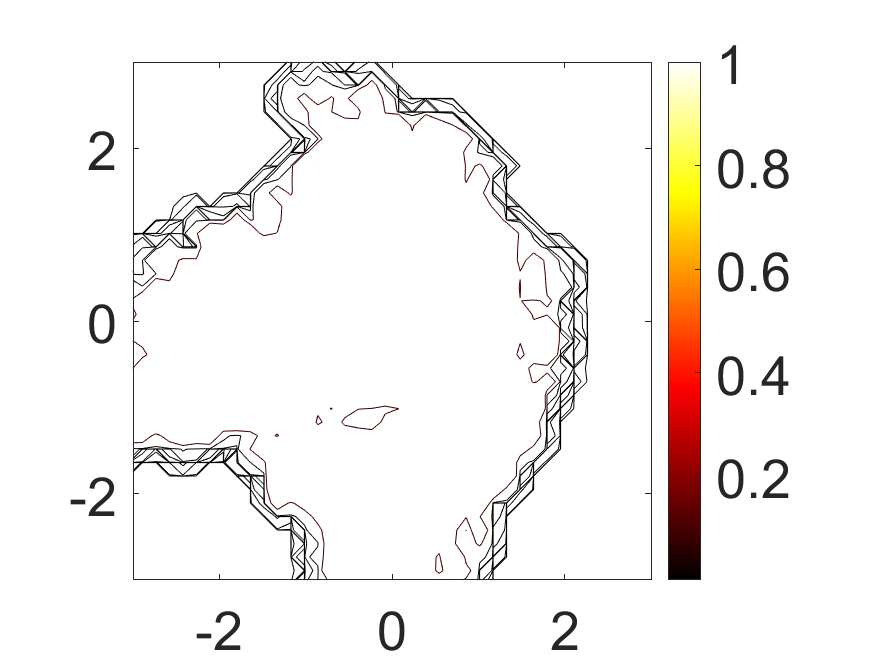} & \includegraphics[width=0.23\linewidth,trim={2em 0 0.5em 0},clip]{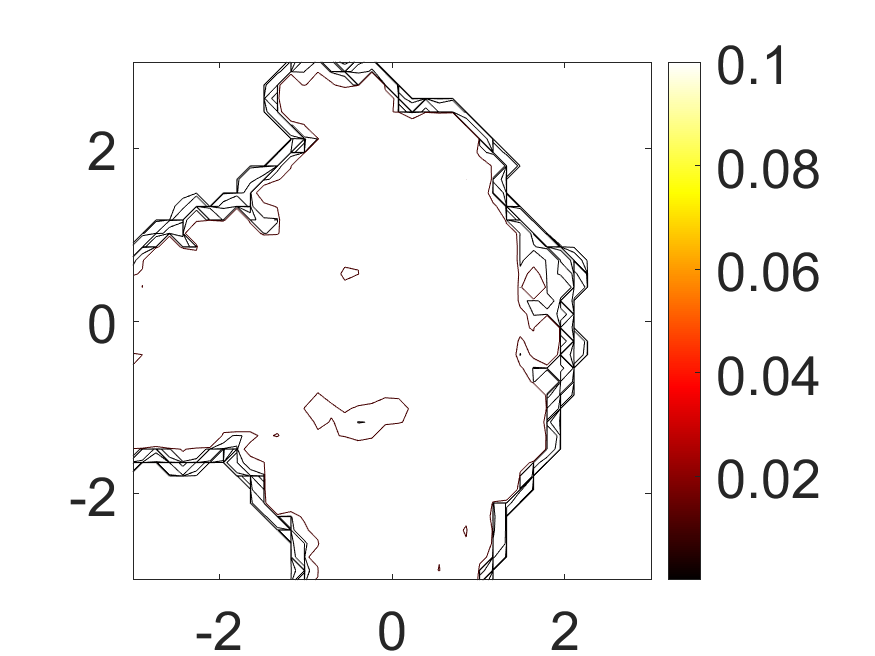} &
    \includegraphics[width=0.23\linewidth,trim={2em 0 0.5em 0},clip]{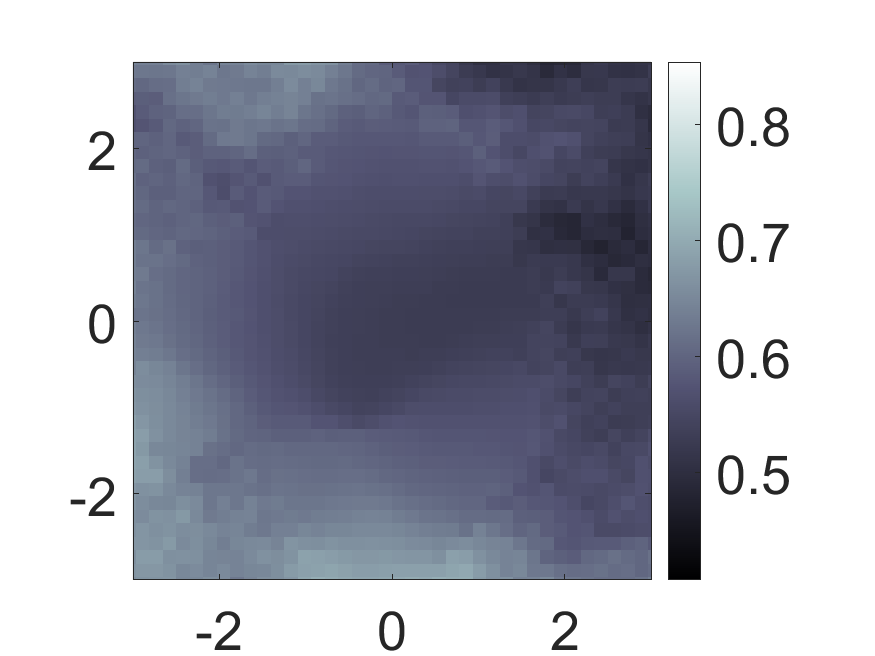}
\end{tabular}
    \vskip-1em
    \caption{\textbf{Experiment \ref{exp:decoupled} --- Decoupled EMT-MET switch functions}: Shown here are the transient dynamics of the cancer growth when the EMT and MET triggers are de-coupled as shown in Figure \ref{fig:coupled-decoupled}. The isolines are plotted down to the double precision accuracy $10^{-15}$. Characteristic of this experiment  is that the cancer foci that are formed, in lower $a$, after the EMT-MET cascade are disjoint from the main body of the tumour. }\label{fig:decoupled}
\end{figure}
 
\end{experiment} 

\begin{experiment}\textbf{Coupled EMT-MET switch functions}\label{exp:coupled}

In contrast to the previous Experiment \ref{exp:decoupled} where the EMT and MET switch functions, $S_{EMT}$ \eqref{EMT-switch} and $S_{MET}$ \eqref{MET-switch}, were decoupled, we consider here a parameter setting where $S_{EMT}$ and $S_{MET}$ are, in effect, coupled.

The parameters for this experiment are shown in Table \ref{tbl:coupled} and differ from the ones in Experiment \ref{exp:decoupled}, i.e. in Table \ref{tbl:decoupled}, only in the values of the midpoint MET parameter $f_0^\text{met}$. For more clarity we have produced the graphs of the corresponding switch functions $S_{EMT}$ and $S_{MET}$ in Figure \ref{fig:coupled-decoupled}.  

Namely, as can be seen in Figure \ref{fig:decoupled}, EMT gives rise to cancer cells of higher phenotype value $a$ that degrade the ECM and migrate along the ECM gradient. As in the previous experiments, these more migratory cancer cells reach regions of low TGF-$\beta$ concentration $f$ where they undergo MET and give rise to cancer cell foci of lower phenotypic value $a$. 

In this experiment though, the tumour foci formed by this process in the phenotype $a=0$ are connected to the initial tumour. This is not apparent when inspecting the cancer cell densities, it is rather revealed through the isolines of the cancer cell densities. This becomes apparent in Figure \ref{fig:coupled} where we have plotted the isolines down to the double precision accuracy of $10^{-15}$.

\begin{table} 
\footnotesize 
\begin{tabular}{|c|c|c|c|c|c|}
\hline
    {}&{Param.}& {Name} & {Value} & {Units} & {Reference} \\ 
\hline
\multirow{14}{*}{\rotatebox{90}{$c$-eq.}}
    &$D_c^M$&CC diffusion rate&$10^{-6}$&$\rm {mm^2}d^{-1}$& \cite{Stokes1991a} \\
    &$D_c^{sl}$&CC diffusion scale &$20$&$\rm M^2{mm^{-4}}$&(our estimate)\\
    &$D_c^{\alpha_0}$& CC diffusion midpoint&$0.5$&---&(our estimate)\\
    &$\chi_c^M$&CC haptotactic response&$0.4$&$\rm {mm^2}M^{-1}d^{-1}$&\cite{Anderson1998,Chaplain1996,Orme1996}\\
    &$\chi_c^{sl}$&CC haptotaxis scale&$20$&---&(our estimate)\\
    &$\chi_c^{\alpha_0}$&CC haptotaxis midpoint&$0.5$&---&(our estimate)\\
    &$L^\text{emt}$&EMT rate&$0.1$&$\rm {M}{mm^{-2}}d^{-1}$&(our estimate)\\
    &$\lambda^\text{emt}$&EMT scale&$30$&$\rm M^{-1}{mm^{-2}}$&(our estimate)\\
    &$f_0^\text{emt}$&midpoint EMT&$0.01$&$\rm M{mm^2}$&(our estimate)\\
    &$L^\text{met}$&MET rate&$20$&$\rm {M}{mm^{-2}}d^{-1}$&(our estimate)\\
    &$\lambda^\text{met}$&MET scale&$300$&$\rm M^{-1}{mm^{-2}}$&(our estimate)\\
    &$f_0^\text{met}$&midpoint MET&$0.01$&$\rm M{mm^2}$&(our estimate)\\
    &$\bar{\gamma_c}$&CC proliferation rate&$1$&$\rm d^{-1}$&\cite{Stokes1991a}\\
    &$\gamma_c^\ast$&CC proliferation exponent&$7$&---&(our estimate)\\
\hline
    \multirow{3}{*}{\rotatebox{90}{$f$-eq.}}
    &$D_f$&GF diffusion rate&$0.01$ &$\rm {mm^2}d^{-1}$&\cite{Stokes1991a}\\
    &$\gamma_f$&GF production rate&$0.3$&$\rm {mm^2}M^{-1}d^{-1}$&(our estimate)\\
    &$d_f$&GF decay&$3$&$\rm d^{-1}$&cf. \cite{TGFb}\\
\hline
    \multirow{3}{*}{\rotatebox{90}{$v$-eq.}}
        &$\eta_v$&ECM degradation rate&$0.01$&$\rm {mm^2}M^{-1}d^{-1}$&cf. \cite{hybrid_2nd, Chaplain2005}\\
        &$K_\eta$&CM degradation midpoint&$10$&---&(our estimate)\\
        &$\lambda_v$&ECM reconstruction rate& $10^{-4}$&$\rm {mm^2}M^{-1}d^{-1}$&cf. \cite{hybrid_2nd}\\
\hline
\end{tabular}
\caption{Parameters used in \textbf{Experiment \ref{exp:coupled} --- Coupled EMT-MET switch functions}.}\label{tbl:coupled}
\end{table}

\begin{figure} 
\setlength{\tabcolsep}{1pt}
\centering
\begin{tabular}{l|ccc|c}
    \phantom{X}&\footnotesize{Cancer cells}&\footnotesize{Cancer cells}&\footnotesize{Cancer cells}&\footnotesize{TGF-$\beta$s \& ECM}\\[-0.3em]    &\footnotesize{$c(t,\cdot,\alpha=0)$}&\footnotesize{$c(t,\cdot,\alpha=0.5)$}&\footnotesize{$c(t,\cdot,\alpha=1)$}&\footnotesize{$f(t, \cdot),\ v(t,\cdot)$}\\
    \hline
    \multirow{2}{*}{\rotatebox{90}{\footnotesize{$t=0$}}}&
    \includegraphics[width=0.23\linewidth,trim={2em 0 0.5em 0},clip]{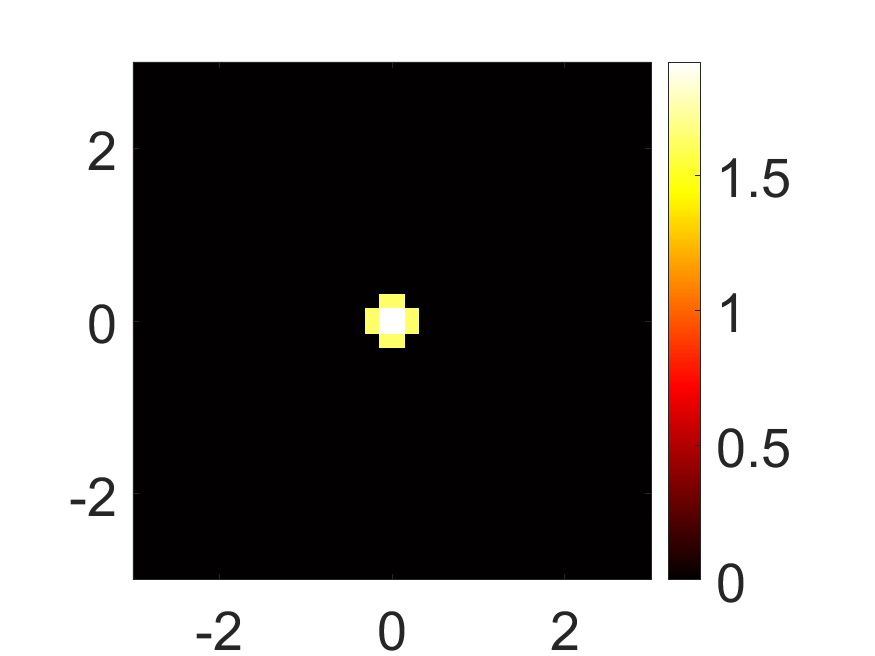} & \includegraphics[width=0.23\linewidth,trim={2em 0 0.5em 0},clip]{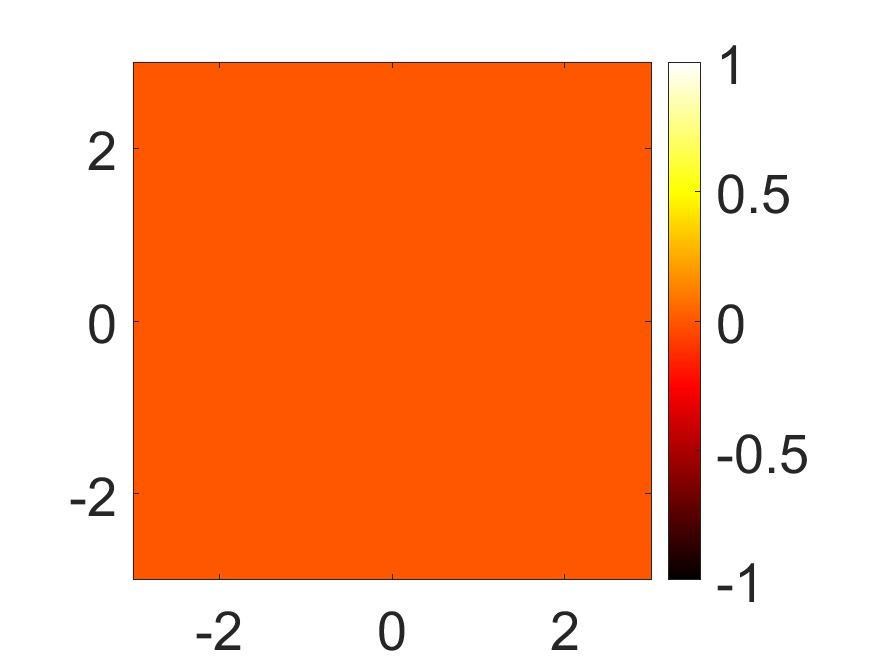} & \includegraphics[width=0.23\linewidth,trim={2em 0 0.5em 0},clip]{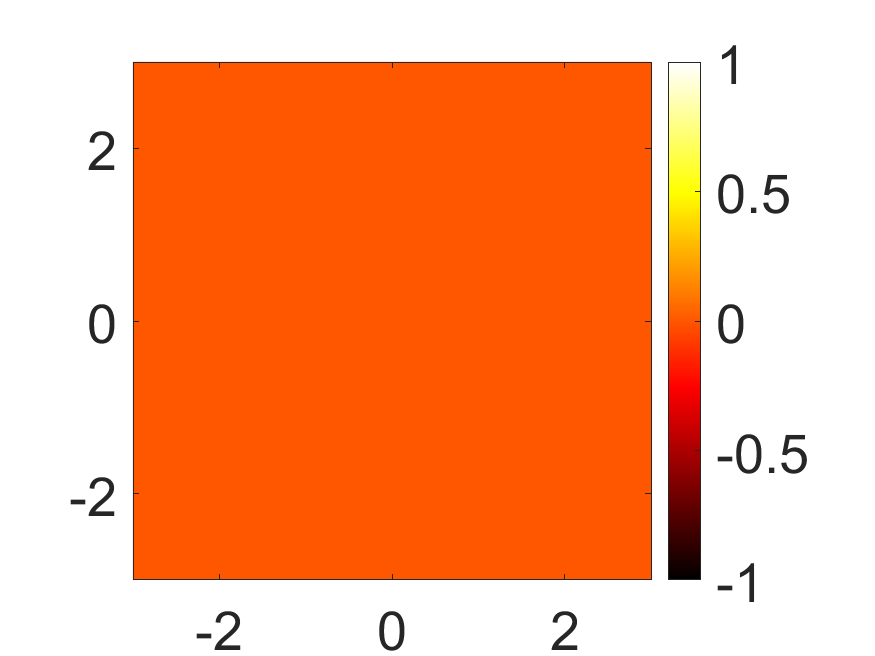} & 
    \includegraphics[width=0.23\linewidth,trim={2em 0 0.5em 0},clip]{exp5_decoupled/exp5_gf_t_1.png}
    \\[-0.4em]
    &\includegraphics[width=0.23\linewidth,trim={2em 0 0.5em 0},clip]{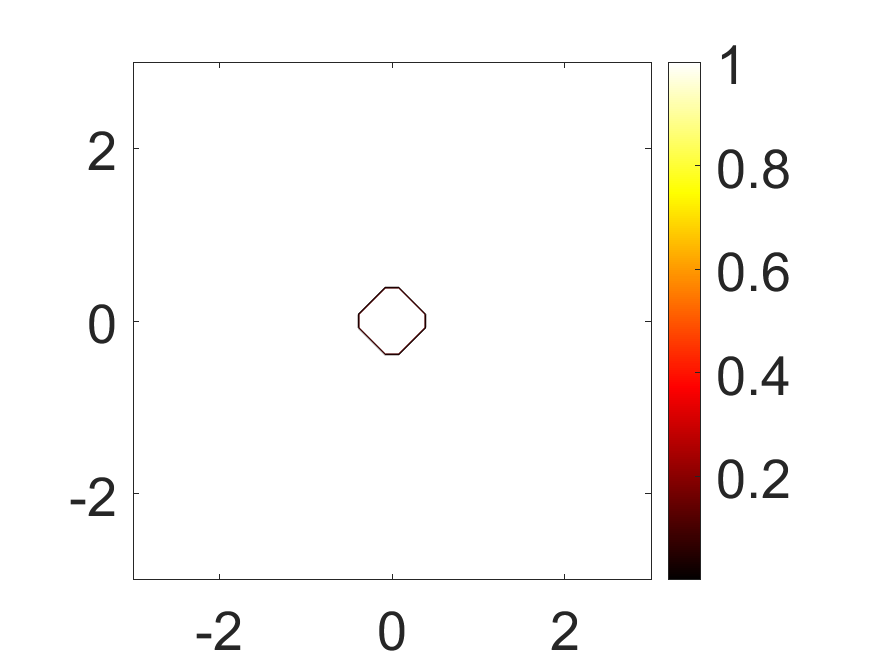} & \includegraphics[width=0.23\linewidth,trim={2em 0 0.5em 0},clip]{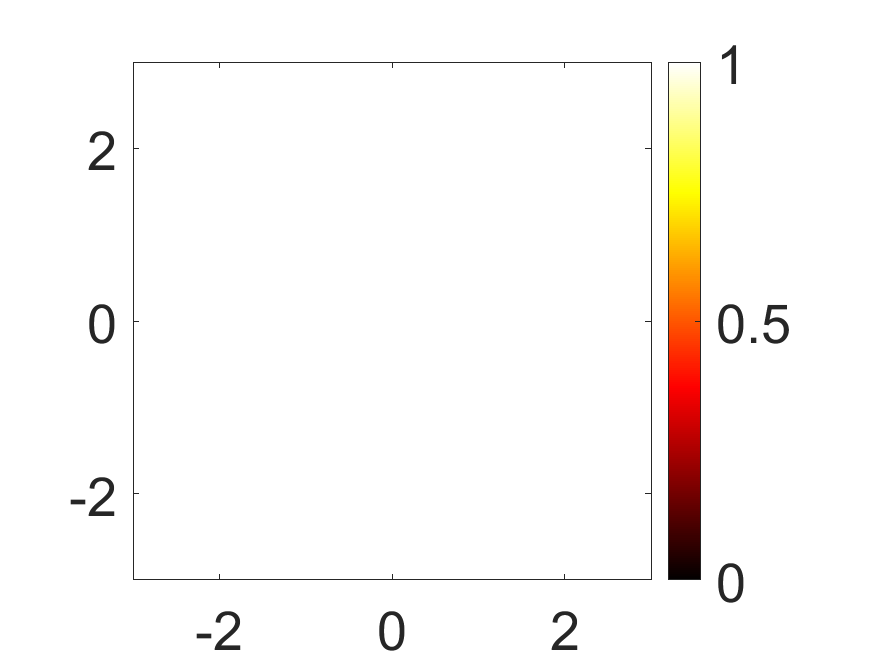} & \includegraphics[width=0.23\linewidth,trim={2em 0 0.5em 0},clip]{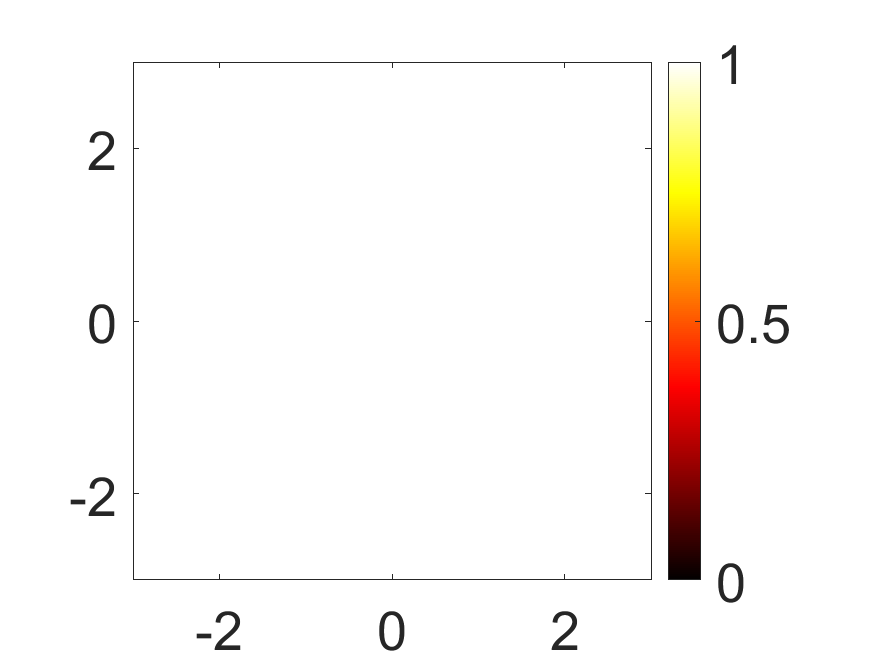} &
    \includegraphics[width=0.23\linewidth,trim={2em 0 0.5em 0},clip]{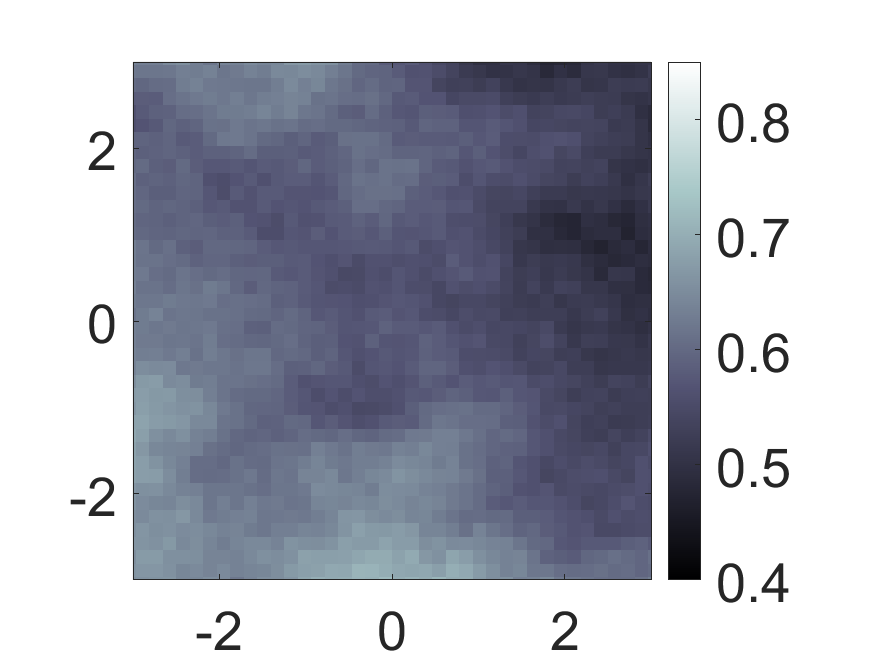} 
    \\
    \hline
    \multirow{2}{*}{\rotatebox{90}{\footnotesize{$t=150$}}}&
    \includegraphics[width=0.23\linewidth,trim={2em 0 0.5em 0},clip]{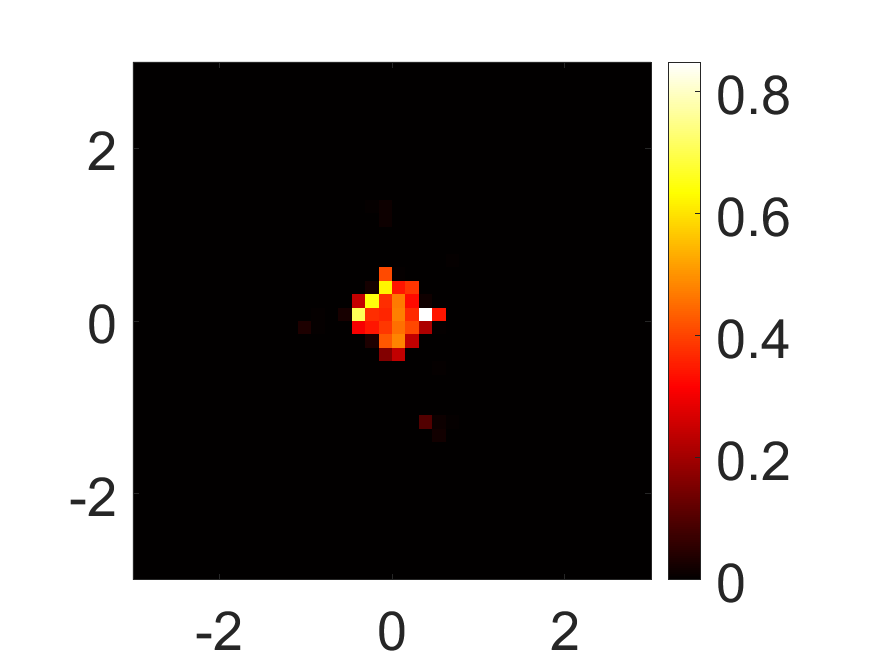} & \includegraphics[width=0.23\linewidth,trim={2em 0 0.5em 0},clip]{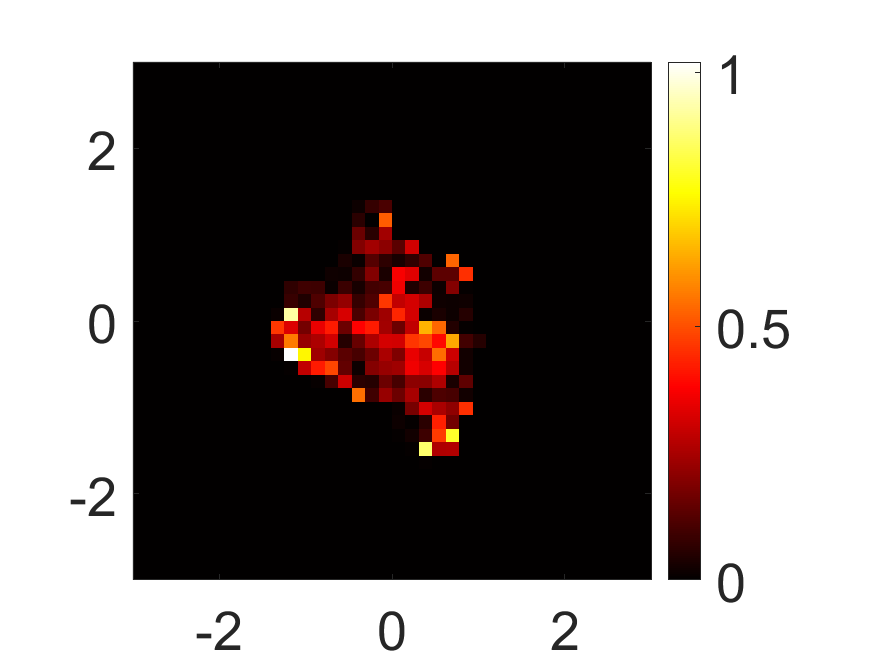} & \includegraphics[width=0.23\linewidth,trim={2em 0 0.5em 0},clip]{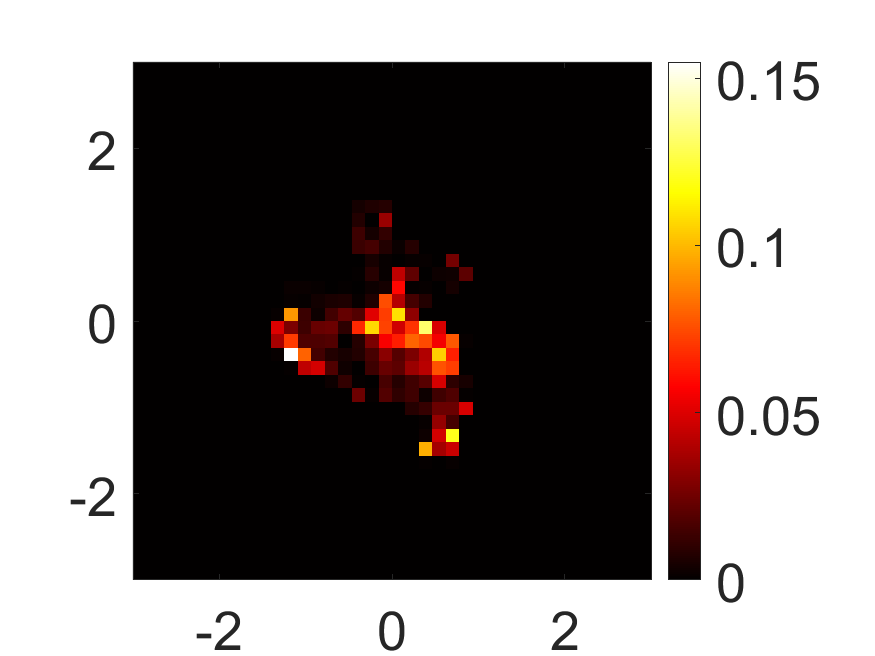} &
    \includegraphics[width=0.23\linewidth,trim={2em 0 0.5em 0},clip]{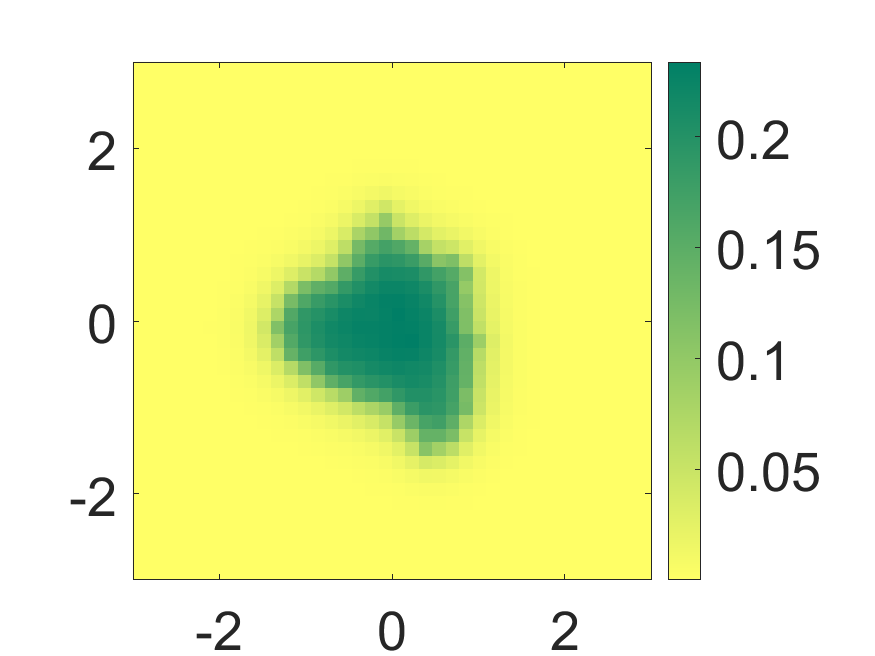}
    \\[-0.4em]
    &\includegraphics[width=0.23\linewidth,trim={2em 0 0.5em 0},clip]{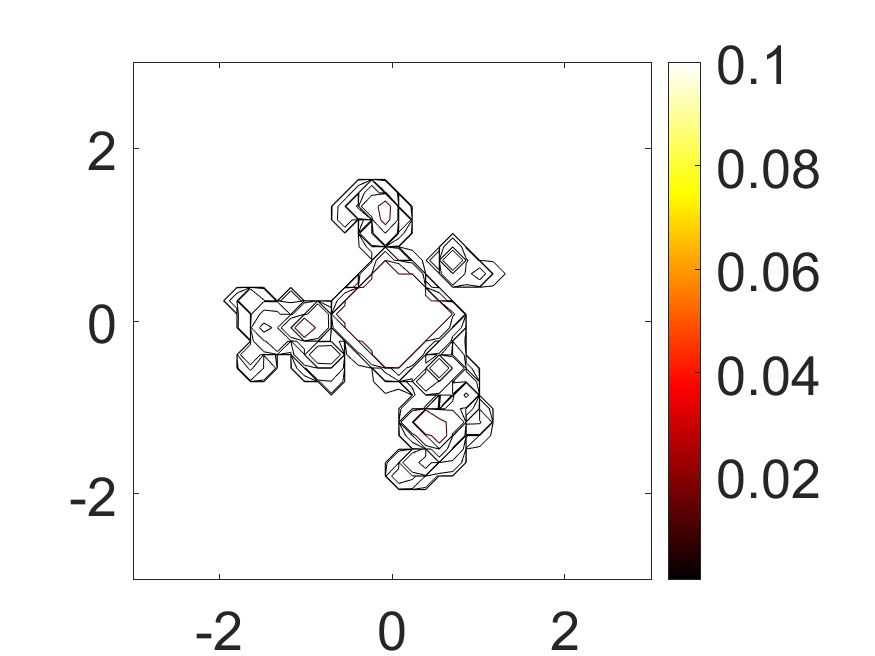} & \includegraphics[width=0.23\linewidth,trim={2em 0 0.5em 0},clip]{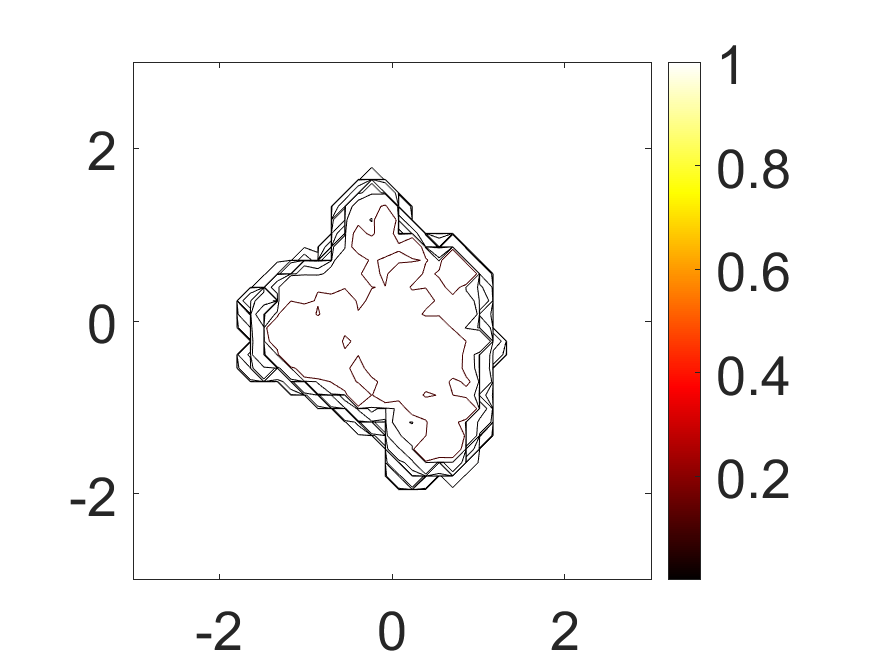} & \includegraphics[width=0.23\linewidth,trim={2em 0 0.5em 0},clip]{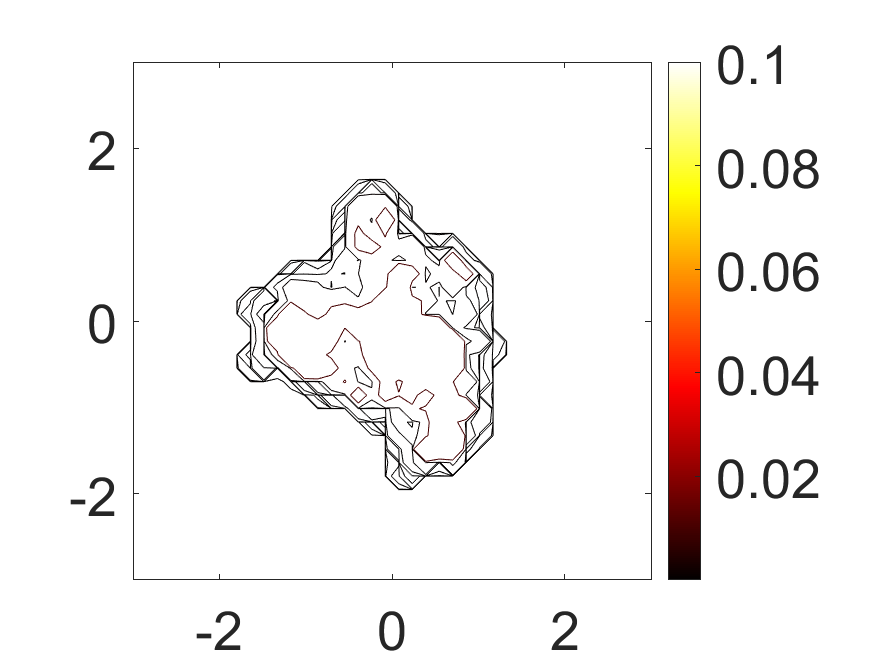} &
    \includegraphics[width=0.23\linewidth,trim={2em 0 0.5em 0},clip]{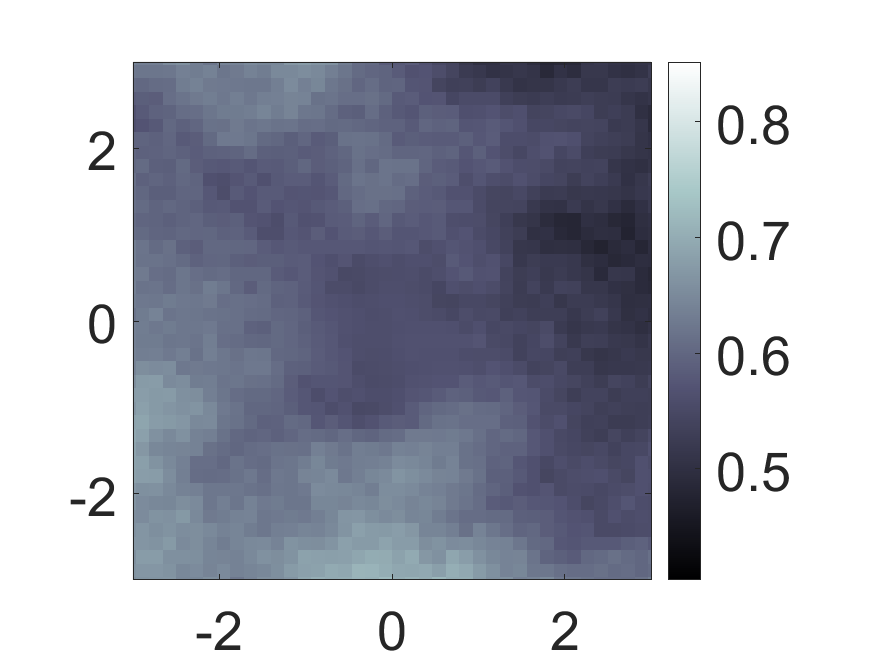} 
    \\
    \hline
    \multirow{2}{*}{\rotatebox{90}{\footnotesize{$t=300$}}}&
    \includegraphics[width=0.23\linewidth,trim={2em 0 0.5em 0},clip]{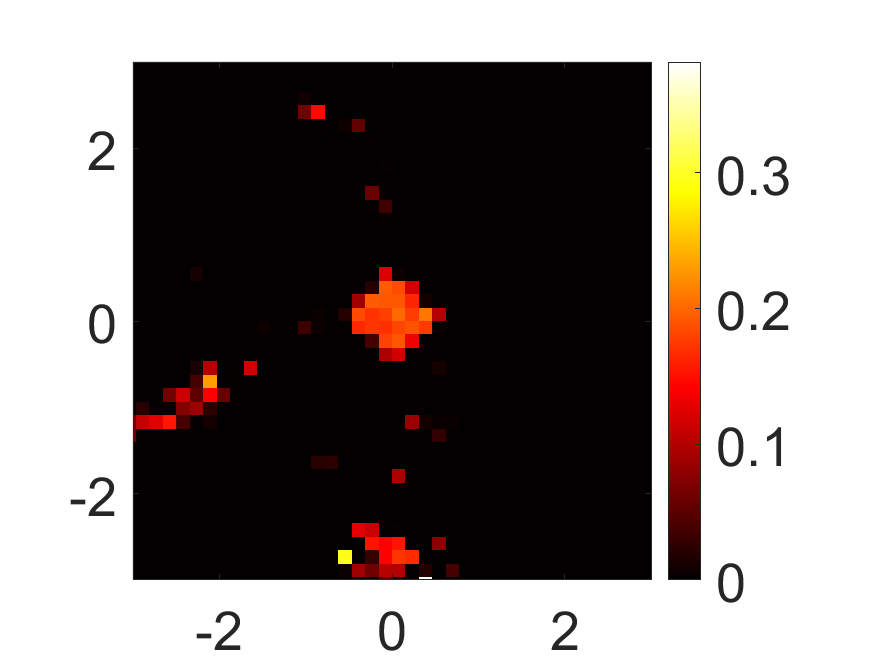} & \includegraphics[width=0.23\linewidth,trim={2em 0 0.5em 0},clip]{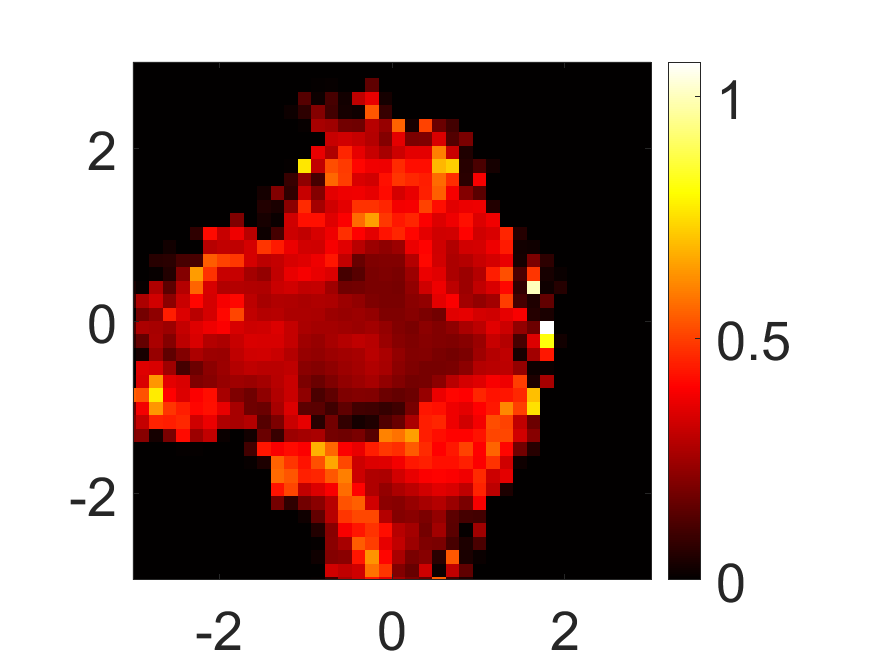} & \includegraphics[width=0.23\linewidth,trim={2em 0 0.5em 0},clip]{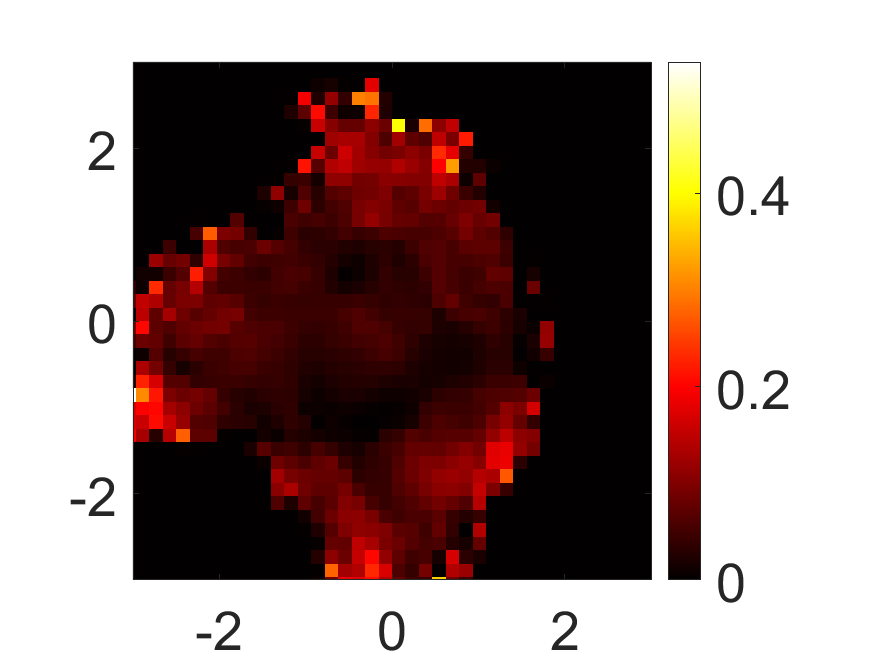} &
    \includegraphics[width=0.23\linewidth,trim={2em 0 0.5em 0},clip]{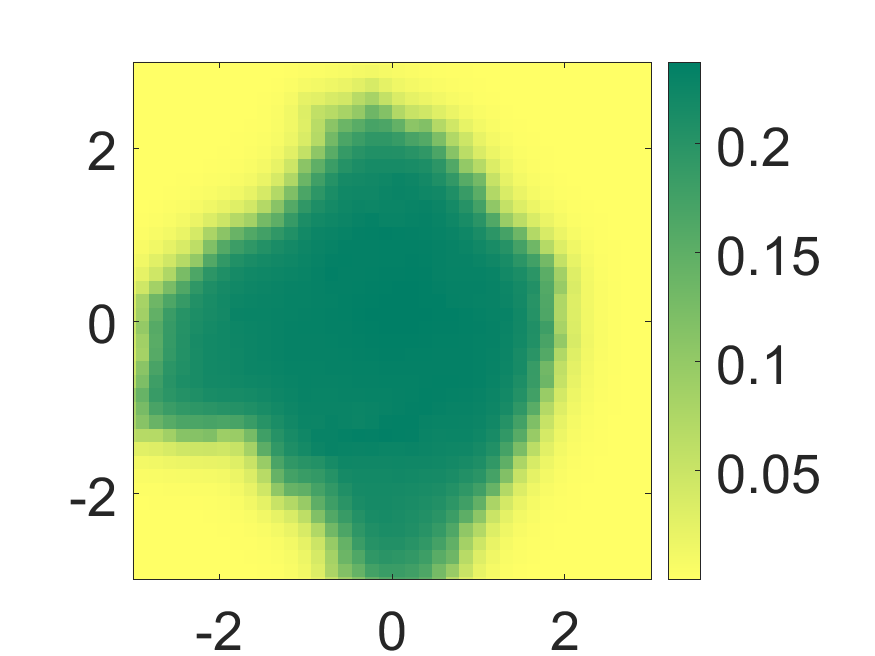}
    \\[-0.4em]
    &\includegraphics[width=0.23\linewidth,trim={2em 0 0.5em 0},clip]{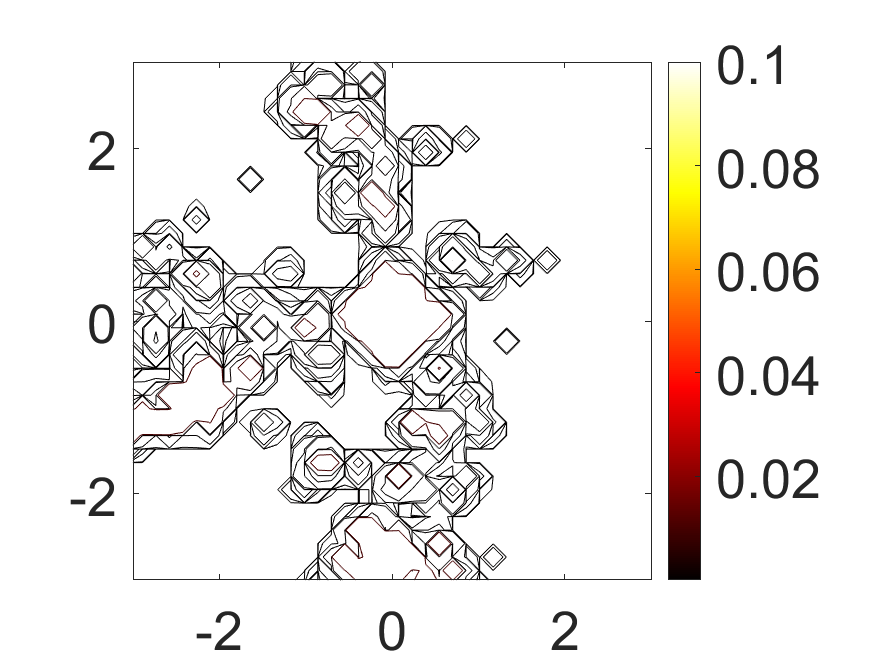} & \includegraphics[width=0.23\linewidth,trim={2em 0 0.5em 0},clip]{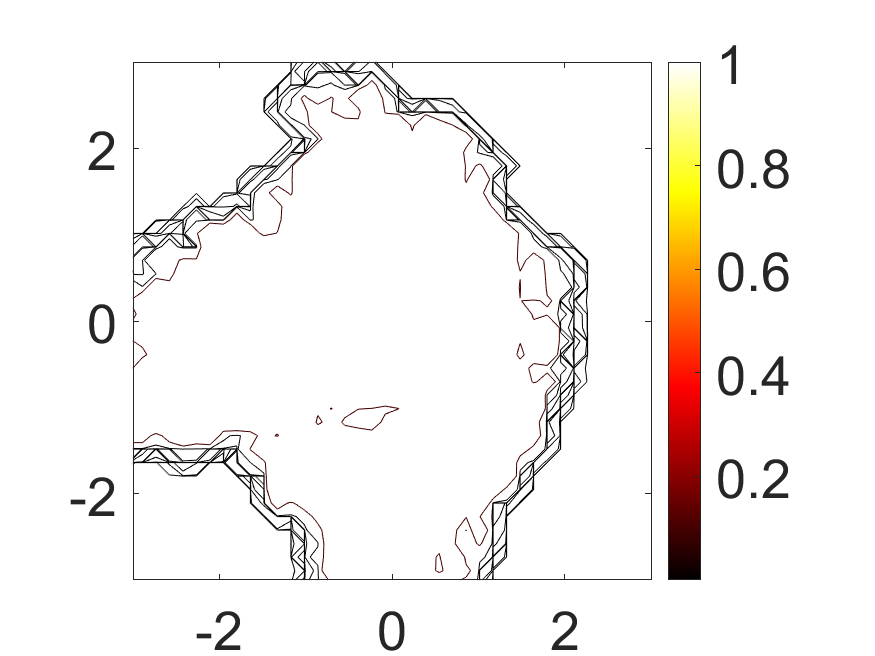} & \includegraphics[width=0.23\linewidth,trim={2em 0 0.5em 0},clip]{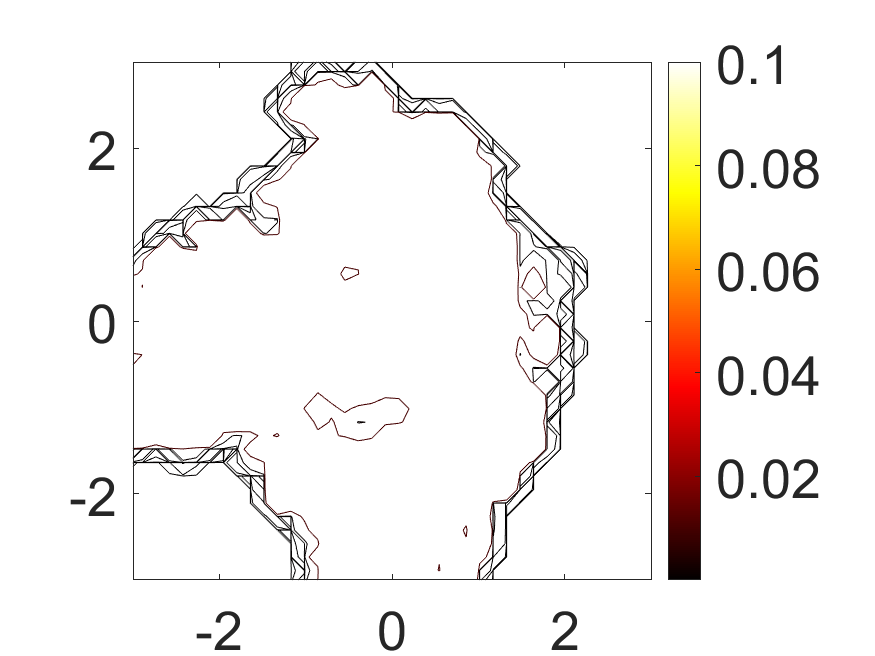} &
    \includegraphics[width=0.23\linewidth,trim={2em 0 0.5em 0},clip]{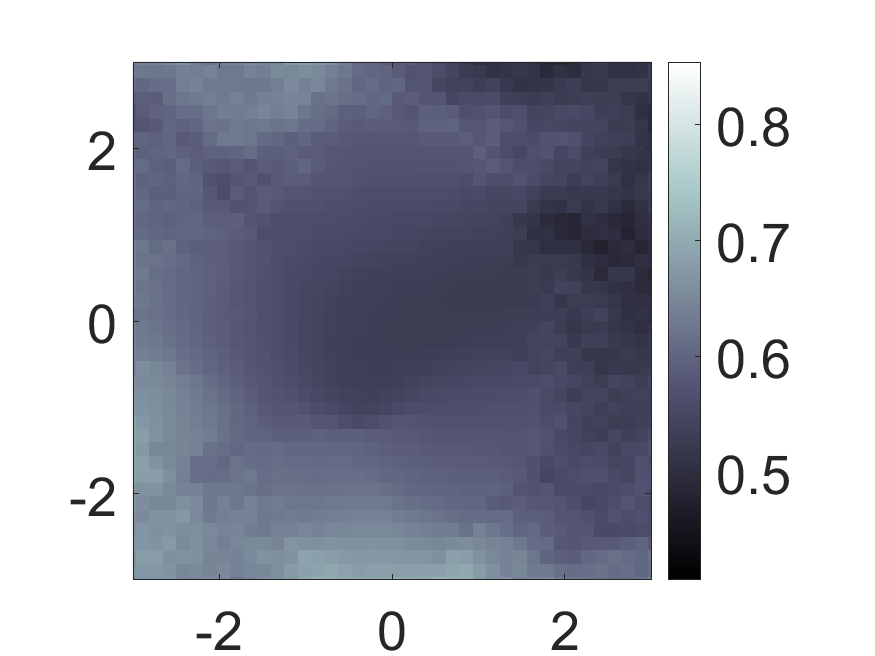}
\end{tabular}
    \vskip-1em
    \caption{\textbf{Experiment \ref{exp:coupled} --- Coupled EMT-MET switch functions:} In this experiment the EMT and MET trigger functions are coupled as detailed in Figure \ref{fig:coupled-decoupled}. As a result, the cancer cell foci that appear, after the EMT-MET cascade, in lower $a$ are connected to the main body of the tumour. Rather than in the density plots, this becomes visible with the isolines that are plotted down to the double precision accuracy $10^{-15}$.}\label{fig:coupled}
\end{figure}
 
\end{experiment} 

\begin{experiment}\textbf{Coupled versus decoupled EMT-MET switch}\label{exp:comp}

The two EMT-MET strategies described in Experiment \ref{exp:decoupled} and Experiment \ref{exp:coupled}, and presented in Figures \ref{fig:decoupled} and \ref{fig:coupled} respectively, have significant implications in the understanding of the EMT and MET processes. Their difference relies solely in the relative dependence of the switch functions $S_{EMT}$ and $S_{MET}$ on the concentration of TGF-$\beta$. See \eqref{EMT-switch}, \eqref{MET-switch}, the parameter Tables \ref{tbl:decoupled} and \ref{tbl:coupled} and Figure \ref{fig:coupled-decoupled} where we have plotted $S_{EMT}$ and $S_{MET}$. 

A direct comparison of the qualitative predictions of the two experiments is presented in Figure \ref{fig:comp} where we plot the densities and isolines of the epithelial phenotype $a=0$ cancer cells at three different time instances $t=0$, $t=50$, $t=100$.

Inspecting the simulation results using the eyeball-norm directly on the density profiles (upper panel) exhibits some difference but does not reveal any significant qualitative difference between the two. Whereas, when inspecting the corresponding isolines (lower panel); it can be seen there that the newly formed tumour foci in the decoupled EMT-MET case are indeed disjoint from the initial tumour. This is clearly not the case with the coupled EMT-MET switch scenario. It needs to be noted again, as in the previous experiments, that the isolines have been plotted down to the double precision accuracy of $10^{-15}$. 

The difference between the two scenarios can be understood as follows: in the coupled EMT-MET case, the EMT and MET processes take place (primarily) in adjacent regions of the physical domain. This is because their corresponding switch functions are ``complementary'' with respect to the concentration of the TGF-$\beta$. On the other hand, in the decoupled EMT-MET case, MET takes place (primarily) in regions of lower densities of TGF-$\beta$ than EMT does. These two regions are not adjacent as the TGF-$\beta$ diffuse freely in the environment and decay rapidly.

\begin{figure} 
\setlength{\tabcolsep}{1pt}
\centering
    \includegraphics[width=0.33\linewidth]{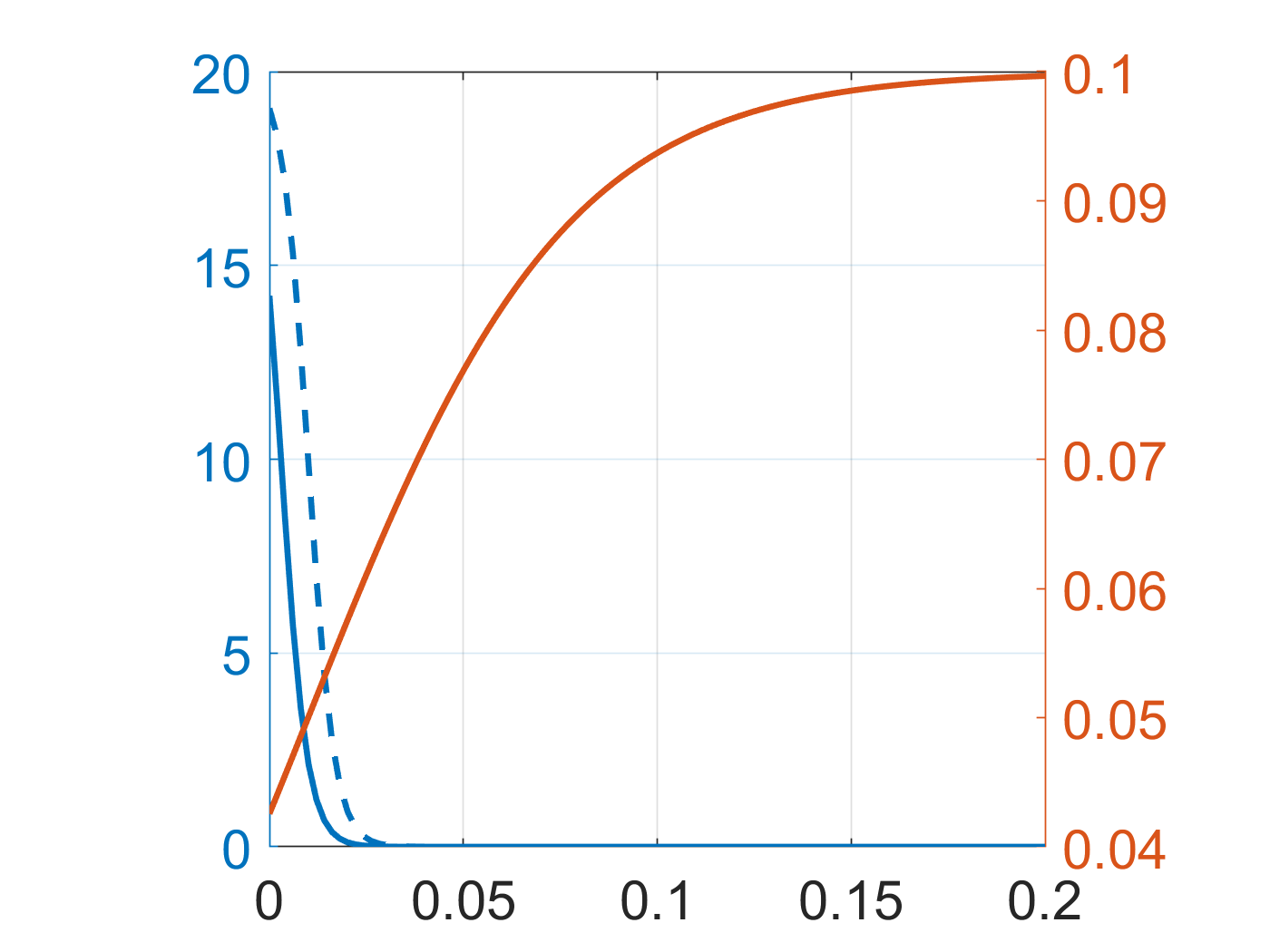}
    \vskip-1em
    \caption{Comparison of the MET switch functions \eqref{MET-switch} between the decoupled EMT-MET scenario where  $f_0^\text{met}=0.003$ (solid decreasing) and the coupled EMT-MET scenario where $f_0^\text{met}=0.01$ (dashed decreasing). In both scenarios the EMT switch function \eqref{EMT-switch} is the same with midpoint EMT rate $f_0^\text{emt}=0.01$ (increasing).}\label{fig:coupled-decoupled}
\end{figure}

\begin{figure}
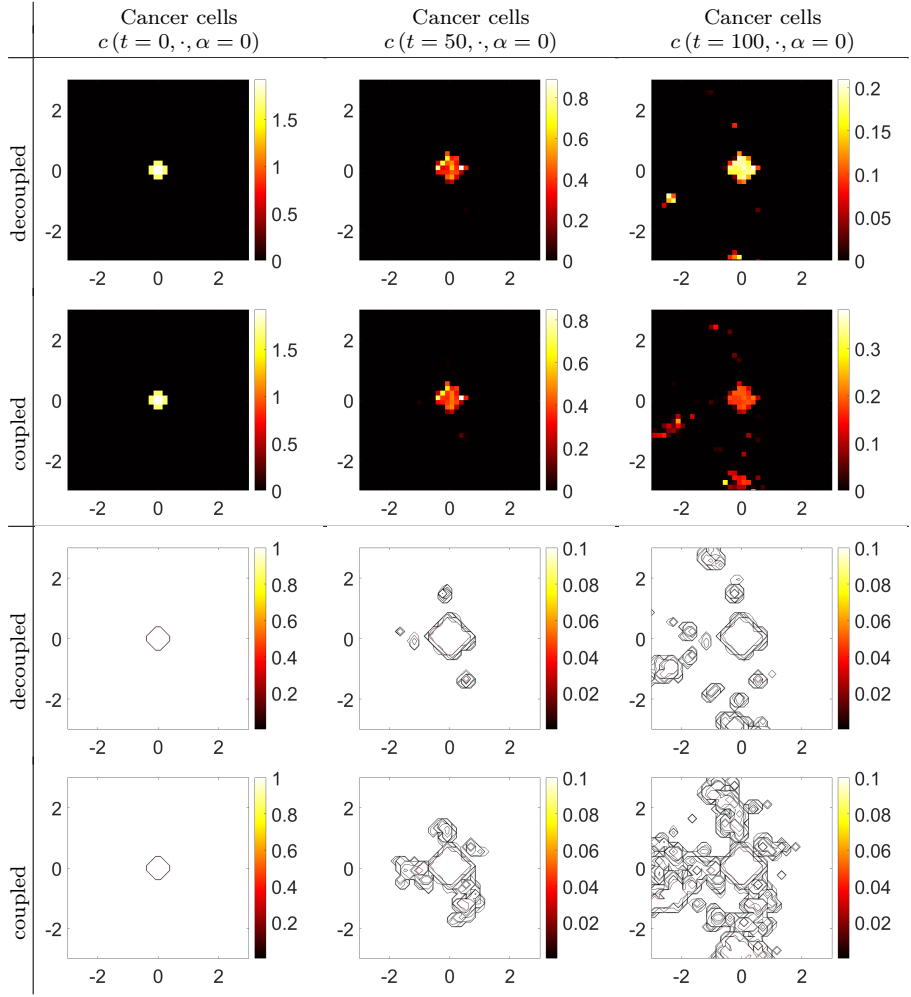
 
\setlength{\tabcolsep}{1pt}
\centering
    \begin{tabular}{l|ccc}
        \phantom{X}&\footnotesize{Cancer cells}&\footnotesize{Cancer cells}&\footnotesize{Cancer cells}
        \\[-0.3em]    &\footnotesize{$c\left(t=0,\cdot,\alpha=0\right)$}&\footnotesize{$c\left(t=50,\cdot,\alpha=0\right)$}&\footnotesize{$c\left(t=100,\cdot,\alpha=0\right)$}
        \\
        \hline
        {\rotatebox{90}{\footnotesize{\hspace{2em}decoupled}}}&
        \includegraphics[width=0.25\linewidth,trim={2em 0 0.5em 0},clip]{exp5_decoupled/exp5_ccdens_1t_1.png} & \includegraphics[width=0.25\linewidth,trim={2em 0 0.5em 0},clip]{exp5_decoupled/exp5_ccdens_1t_30.png} & \includegraphics[width=0.25\linewidth,trim={2em 0 0.5em 0},clip]{exp5_decoupled/exp5_ccdens_1t_61.png} 
        \\[-0.3em]
        {\rotatebox{90}{\footnotesize{\hspace{2em}coupled}}}&
        \includegraphics[width=0.25\linewidth,trim={2em 0 0.5em 0},clip]{exp6_coupled/exp6_ccdens_1t_1.png} & \includegraphics[width=0.25\linewidth,trim={2em 0 0.5em 0},clip]{exp6_coupled/exp6_ccdens_1t_51.png} & \includegraphics[width=0.25\linewidth,trim={2em 0 0.5em 0},clip]{exp6_coupled/exp6_ccdens_1t_101.png} 
        \\
        \hline
        {\rotatebox{90}{\footnotesize{\hspace{2em}decoupled}}}&
        \includegraphics[width=0.25\linewidth,trim={2em 0 0.5em 0},clip]{exp5_decoupled/exp5_cciso_1t_1.png} & \includegraphics[width=0.25\linewidth,trim={2em 0 0.5em 0},clip]{exp5_decoupled/exp5_cciso_1t_30.png} & \includegraphics[width=0.25\linewidth,trim={2em 0 0.5em 0},clip]{exp5_decoupled/exp5_cciso_1t_61.png} 
        \\[-0.3em]
        {\rotatebox{90}{\footnotesize{\hspace{2em}coupled}}}&
        \includegraphics[width=0.25\linewidth,trim={2em 0 0.5em 0},clip]{exp6_coupled/exp6_cciso_1t_1.png} & \includegraphics[width=0.25\linewidth,trim={2em 0 0.5em 0},clip]{exp6_coupled/exp6_cciso_1t_51.png} & \includegraphics[width=0.25\linewidth,trim={2em 0 0.5em 0},clip]{exp6_coupled/exp6_cciso_1t_101.png} 
\end{tabular}
    \vskip-1em
    \caption{\textbf{Coupled versus decoupled EMT-MET switch scenarios}: The upper two rows are the cancer cell density profiles of the two solutions for $\alpha=0$, whereas the lower two are the corresponding isolines plotted down to the double precision accuracy $10^{-15}$. These isolines exhibit that in the decoupled EMT-MET case the tumour spreads in disjoint tumour foci. This is a result of the MET switch function $S_{MET}$ that promotes MET primarily in lower concentrations of $f$, i.e. away from the location of the initial tumour.}\label{fig:comp}
\end{figure}

\end{experiment}


\section{Conclusions} \label{Sec:conclusions}


By combining biological findings and mathematical modelling, we postulate that we have reproduced, in a gross simplification, of course, the general mechanism of isolated tumour foci formation. We believe, that cell phenotype change occurring as a result of the action of TGF-$\beta$, and possibly other growth factors as well, which we assume is produced by CAF cells in the vicinity of the primary tumour, leads to the formation of neoplastic foci distant from the primary tumour. As a result of elevated TGF-$\beta$ levels at the edge of the primary tumour epithelial cells transforms into mesenchymal cells that can migrate. Mesenchymal cells migrate toward the extracellular matrix concentration gradient, which naturally means moving away from the primary tumour. This results in these cells finding themselves at some distance from the primary tumour, where the concentration of TGF-$\beta$ factor is lower, causing the mesenchymal cells to become endothelial cells again. That change in phenotype alters the behaviour of the cells, which stop migrating and become highly proliferative cells again. \\ \\
Although the mechanism we have presented relates to the formation of isolated tumour foci localized in a single tissue, we postulate that it is a mechanism that also explains the formation of metastases in distant parts of the body. This is also evidenced by new experimental findings, namely the non-plastic mesenchymal cells were observed to be unable to form metastatic outgrowths, as cancer cells need to at least partially reverse to a more epithelial phenotype for metastasis to form~\cite{Nieto2017}. Thus, EMT may be important for cancer cell dissemination but is not sufficient for metastatic colonization, which requires the reverse phenotype transition, that is MET~\cite{Nieto2017}.

What now seems to be one of the main challenges for cancer researchers is to elucidate the entire system controlling the changes in the phenotype of cancer cells that promote invasion and, consequently, cancer metastasis. It is clear that although TGF-$\beta$ appears to be the main or one of the main players responsible for regulating the plasticity of the cancer cell phenotype, it is undoubtedly not the only one. A better reconstruction of the regulatory system that governs phenotypic plasticity gives a chance to reveal the constituent susceptible to the right external forcing. A better reconstruction of the regulatory system that governs phenotypic plasticity gives a chance to reveal the constituent susceptible to the right external forcing. This also requires improving experimental techniques and designs, although progress in this area is already impressive~\cite{Poon2022}. And not to be lip service, one can cite new techniques for studying tumour invasion in vitro, in which TGF-$\beta$ was controlled release from gelatin hydrogel microspheres making it possible to study, for example, CAF activation by increased TGF-$\beta$ concentration~\cite{Nii2029}.

Multifocal breast cancer (MBC) patients are known to have a poorer prognosis compared to those with unifocal breast cancer \cite{Pekar2014}, and this has been attributed to the underestimation of tumour size in MBC cases \cite{alexander2017, moon2013}. As our model develops, it could aid in further studying MBC cases by potentially improving the accuracy of tumour size estimation and the extent of invasive spread, which may result in better prognostic outcomes for patients with MBC. In particular, our model in conjunction with magnetic resonance imaging (MRI) could prove especially valuable, given the increasing relevance of MRI in breast cancer diagnosis in recent years, \cite{johnson2011}.

\section*{Acknowledgements} Z.~Szyma\'nska acknowledge the support from the National Science Centre, Poland -- grant No. 2017/26/M/ST1/00783. 
N.~Sfakianaki's scientific visit to the University of Warsaw was partially supported by the Excellence Initiative Research University Programme at the University of Warsaw.
M.~Lachowicz is happy to acknowledge the support from the New Ideas Grant - ''R\'ownania kinetyczne w opisie zjawisk samoorganizacji'' funded by the Excellence Initiative Research University Programme at the University of Warsaw.

\bibliographystyle{elsarticle-num}
\bibliography{bibliography.bib}

\begin{thebibliography}{10}
\expandafter\ifx\csname url\endcsname\relax
  \def\url#1{\texttt{#1}}\fi
\expandafter\ifx\csname urlprefix\endcsname\relax\def\urlprefix{URL }\fi
\expandafter\ifx\csname href\endcsname\relax
  \def\href#1#2{#2} \def\path#1{#1}\fi

\bibitem{Kavya2021}
{K.S. Kavya Satheesh}, H.~Rani, M.~K. Jolly, V.~Mahadevan, {Epigenetics of
  epithelial to mesenchymal transition (EMT) in cancer}, in: P.~K. Agrawala,
  P.~Rana (Eds.), Epigenetics and Metabolomics, Vol.~28 of Translational
  Epigenetics, Academic Press, 2021, pp. 237--264.
\newblock \href {https://doi.org/10.1016/B978-0-323-85652-2.00001-4}
  {\path{doi:10.1016/B978-0-323-85652-2.00001-4}}.

\bibitem{Pinzani2011}
M.~Pinzani, {Epithelial--mesenchymal transition in chronic liver disease:
  Fibrogenesis or escape from death?}, J Hepatol 55~(2) (2011) 459--465.
\newblock \href {https://doi.org/10.1016/j.jhep.2011.02.001}
  {\path{doi:10.1016/j.jhep.2011.02.001}}.

\bibitem{Kalluri2009}
R.~Kalluri, R.~A. Weinberg, {The basics of epithelial-mesenchymal transition},
  J. Clin. Investig. 119~(6) (2009) 1420--1428.
\newblock \href {https://doi.org/10.1172/JCI39104}
  {\path{doi:10.1172/JCI39104}}.

\bibitem{Hanahan2011}
D.~Hanahan, R.~A. Weinberg, {Hallmarks of Cancer: The Next Generation}, Cell
  144~(5) (2011) 646--674.
\newblock \href {https://doi.org/10.1016/j.cell.2011.02.013}
  {\path{doi:10.1016/j.cell.2011.02.013}}.

\bibitem{Lowengrub2010}
J.~S. Lowengrub, H.~B. Frieboes, F.~Jin, Y.-L. Chuang, X.~Li, P.~Macklin, S.~M.
  Wise, V.~Cristini, {Nonlinear modelling of cancer: bridging the gap between
  cells and tumours}, Nonlinearity 23~(1) (2010) R1--R9.
\newblock \href {https://doi.org/10.1088/0951-7715/23/1/r01}
  {\path{doi:10.1088/0951-7715/23/1/r01}}.

\bibitem{Nii2029}
T.~Nii, K.~Makino, Y.~Tabata, {A cancer invasion model of cancer-associated
  fibroblasts aggregates combined with TGF-$\beta$1 release system}, Regen.
  Ther. 14 (2020) 196--204.
\newblock \href {https://doi.org/10.1016/j.reth.2020.02.003}
  {\path{doi:10.1016/j.reth.2020.02.003}}.

\bibitem{Poon2022}
S.~Poon, L.~E. Ailles, {Modeling the Role of Cancer-Associated Fibroblasts in
  Tumor Cell Invasion}, Cancers 14~(4) (2022).
\newblock \href {https://doi.org/10.3390/cancers14040962}
  {\path{doi:10.3390/cancers14040962}}.

\bibitem{Pekar2014}
G.~Pekar, M.~Gere, M.~Tarjan, D.~Hellberg, T.~Tot, Molecular phenotype of the
  foci in multifocal invasive breast carcinomas: intertumoral heterogeneity is
  related to shorter survival and may influence the choice of therapy., Cancer
  120 (2014) 26--34.
\newblock \href {https://doi.org/10.1002/cncr.28375}
  {\path{doi:10.1002/cncr.28375}}.

\bibitem{Kiesslich2013}
T.~Kiesslich, M.~Pichler, D.~Neureiter,
  \href{https://pubmed.ncbi.nlm.nih.gov/24649114}{Epigenetic control of
  epithelial-mesenchymal-transition in human cancer}, Molecular and clinical
  oncology 1~(1) (2013) 3--11.
\newblock \href {https://doi.org/10.3892/mco.2012.28}
  {\path{doi:10.3892/mco.2012.28}}.
\newline\urlprefix\url{https://pubmed.ncbi.nlm.nih.gov/24649114}

\bibitem{Orme1996}
M.~E. Orme, M.~A.~J. Chaplain, {A mathematical model of vascular tumour growth
  and invasion}, Math Comput Model 23~(10) (1996) 43--60.
\newblock \href {https://doi.org/10.1016/0895-7177(96)00053-2}
  {\path{doi:10.1016/0895-7177(96)00053-2}}.

\bibitem{perumpanani1996}
A.~J. Perumpanani, J.~A. Sherratt, J.~Norbury, H.~M. Byrne, Biological
  inferences from a mathematical model for malignant invasion, Invasion
  Metastasis 16(4-5) (1996) 209--21.

\bibitem{Andasari2011}
V.~Andasari, A.~Gerisch, G.~Lolas, A.~P. South, M.~A.~J. Chaplain,
  {Mathematical modeling of cancer cell invasion of tissue: biological insight
  from mathematical analysis and computational simulation}, J Math Biol 63~(1)
  (2011) 141--171.
\newblock \href {https://doi.org/10.1007/s00285-010-0369-1}
  {\path{doi:10.1007/s00285-010-0369-1}}.

\bibitem{Bitsouni2017}
V.~Bitsouni, M.~A.~J. Chaplain, R.~Eftimie, {Mathematical modelling of cancer
  invasion: The multiple roles of TGF-β pathway on tumour proliferation and
  cell adhesion}, Math Models Methods Appl Sci 27~(10) (2017) 1929--1962.
\newblock \href {https://doi.org/10.1142/S021820251750035X}
  {\path{doi:10.1142/S021820251750035X}}.

\bibitem{Chaplain2005}
M.~A.~J. Chaplain, G.~Lolas, {Mathematical modelling of cancer cell invasion of
  tissue: The role of the urokinase plasminogen activation system}, Math Models
  Methods Appl Sci 15~(11) (2005) 1685--1734.
\newblock \href {https://doi.org/10.1142/S0218202505000947}
  {\path{doi:10.1142/S0218202505000947}}.

\bibitem{Meral2013}
G.~Meral, C.~Surulescu, {Mathematical modelling, analysis and numerical
  simulations for the influence of heat shock proteins on tumour invasion}, J
  Math Anal Appl 408~(2) (2013) 597--614.
\newblock \href {https://doi.org/10.1016/j.jmaa.2013.06.017}
  {\path{doi:10.1016/j.jmaa.2013.06.017}}.

\bibitem{szymanska2009a}
Z.~Szyma{\'n}ska, J.~Urba{\'n}ski, A.~Marciniak-Czochra, {Mathematical
  modelling of the influence of heat shock proteins on cancer invasion of
  tissue}, J Math Biol 58~(4-5) (2009) 819--844.
\newblock \href {https://doi.org/10.1007/s00285-008-0220-0}
  {\path{doi:10.1007/s00285-008-0220-0}}.

\bibitem{Eustace2004}
B.~K. Eustace, T.~Sakurai, J.~K. Stewart, D.~Yimlamai, C.~Unger, C.~Zehetmeier,
  B.~Lain, C.~Torella, S.~W. Henning, G.~Beste, B.~T. Scroggins, L.~Neckers,
  L.~L. Ilag, D.~G. Jay, {Functional proteomic screens reveal an essential
  extracellular role for hsp90 alpha in cancer cell invasiveness}, Nat Cell
  Biol 6~(6) (2004) 507--514.
\newblock \href {https://doi.org/10.1038/ncb1131} {\path{doi:10.1038/ncb1131}}.

\bibitem{RamisConde2008}
I.~Ramis-Conde, M.~A. Chaplain, A.~R. Anderson, {Mathematical modelling of
  cancer cell invasion of tissue}, Math Comput Model 47~(5) (2008) 533--545.
\newblock \href {https://doi.org/10.1016/j.mcm.2007.02.034}
  {\path{doi:10.1016/j.mcm.2007.02.034}}.

\bibitem{Gerlee2008}
P.~Gerlee, A.~R.~A. Anderson, {A hybrid cellular automaton model of clonal
  evolution in cancer: the emergence of the glycolytic phenotype}, J Theor Biol
  250~(4) (2008) 705--722.
\newblock \href {https://doi.org/10.1016/j.jtbi.2007.10.038}
  {\path{doi:10.1016/j.jtbi.2007.10.038}}.

\bibitem{Anderson2005}
A.~R.~A. Anderson, {A hybrid mathematical model of solid tumour invasion: the
  importance of cell adhesion}, Math Med Biol 22~(2) (2005) 163--186.
\newblock \href {https://doi.org/10.1093/imammb/dqi005}
  {\path{doi:10.1093/imammb/dqi005}}.

\bibitem{Cytowski2014}
M.~Cytowski, Z.~Szyma\'nska, {Large-Scale Parallel Simulations of {3D} Cell
  Colony Dynamics}, Comput. Sci. Eng. 16~(5) (2014) 86--95.
\newblock \href {https://doi.org/10.1109/MCSE.2014.2}
  {\path{doi:10.1109/MCSE.2014.2}}.

\bibitem{Cytowski2015}
M.~Cytowski, Z.~Szyma\'nska, {Large-Scale Parallel Simulations of {3D} Cell
  Colony Dynamics: The Cellular Environment}, Comput. Sci. Eng. 17~(5) (2015)
  44--48.
\newblock \href {https://doi.org/10.1109/MCSE.2015.66}
  {\path{doi:10.1109/MCSE.2015.66}}.

\bibitem{Cytowski2017}
M.~Cytowski, Z.~Szyma{\'n}ska, P.~Umi{\'n}ski, G.~Andrejczuk, K.~Raszkowski,
  Implementation of an agent-based parallel tissue modelling framework for the
  intel mic architecture, Sci. Program. 2017 (2017) 8721612.
\newblock \href {https://doi.org/10.1155/2017/8721612}
  {\path{doi:10.1155/2017/8721612}}.

\bibitem{szymanska2018}
Z.~Szyma{\'n}ska, M.~Cytowski, E.~Mitchell, C.~K. Macnamara, M.~A.~J. Chaplain,
  Computational modelling of cancer development and growth: Modelling at
  multiple scales and multiscale modelling, Bull. Math. Biol. 80~(5) (2018)
  1366--1403.
\newblock \href {https://doi.org/10.1007/s11538-017-0292-3}
  {\path{doi:10.1007/s11538-017-0292-3}}.

\bibitem{Macklin2018}
A.~Ghaffarizadeh, R.~Heiland, S.~H. Friedman, S.~M. Mumenthaler, P.~Macklin,
  {PhysiCell: An open source physics-based cell simulator for 3-D multicellular
  systems}, PLoS Comput Biol 14~(2) (2018) e1005991.
\newblock \href {https://doi.org/10.1371/journal.pcbi.1005991}
  {\path{doi:10.1371/journal.pcbi.1005991}}.

\bibitem{Wrzosek}
M.~A.~J. Chaplain, M.~Lachowicz, Z.~Szyma\'{n}ska, D.~Wrzosek, {Mathematical
  modelling of cancer invasion: the importance of cell-cell adhesion and
  cell-matrix adhesion}, Math. Models Methods Appl. Sci. 21~(4) (2011)
  719--743.
\newblock \href {https://doi.org/10.1142/S0218202511005192}
  {\path{doi:10.1142/S0218202511005192}}.

\bibitem{gerisch2008}
A.~Gerisch, M.~A.~J. Chaplain, {Mathematical modelling of cancer cell invasion
  of tissue: Local and non-local models and the effect of adhesion}, J Theor
  Biol 250~(4) (2008) 684--704.
\newblock \href {https://doi.org/10.1016/j.jtbi.2007.10.026}
  {\path{doi:10.1016/j.jtbi.2007.10.026}}.

\bibitem{Surulescu}
M.~Eckardt, K.~J. Painter, C.~Surulescu, A.~Zhigun, {Nonlocal and local models
  for taxis in cell migration: a rigorous limit procedure}, J. Math. Biol.
  81~(6-7) (2020) 1251--1298.
\newblock \href {https://doi.org/10.1007/s00285-020-01536-4}
  {\path{doi:10.1007/s00285-020-01536-4}}.

\bibitem{CMR}
Z.~Szyma\'{n}ska, C.~M. Rodrigo, M.~Lachowicz, M.~A.~J. Chaplain, {Mathematical
  modelling of cancer invasion of tissue: the role and effect of nonlocal
  interactions}, Math. Models Methods Appl. Sci. 19~(2) (2009) 257--281.
\newblock \href {https://doi.org/10.1142/S0218202509003425}
  {\path{doi:10.1142/S0218202509003425}}.

\bibitem{trucu2014}
P.~Domschke, D.~Trucu, A.~Gerisch, M.~A.~J. Chaplain, Mathematical modelling of
  cancer invasion: implications of cell adhesion variability for tumour
  infiltrative growth patterns, J. Theor. Biol. 361 (2014) 41--60.

\bibitem{szymanska2021}
Z.~Szyma{\'n}ska, J.~Skrzeczkowski, B.~Miasojedow, P.~Gwiazda, Bayesian
  inference of a non-local proliferation model, R. Soc. Open Sci. 8~(11) (2021)
  211279.
\newblock \href {https://doi.org/10.1098/rsos.211279}
  {\path{doi:10.1098/rsos.211279}}.

\bibitem{gwiazda2023}
P.~Gwiazda, B.~Miasojedow, J.~Skrzeczkowski, Z.~Szyma{\'n}ska, {Convergence of
  the EBT method for a non-local model of cell proliferation with discontinuous
  interaction kernel}, IMA J. Numer. Anal. 43 (2023) 590--626.
\newblock \href {https://doi.org/10.1093/imanum/drab102}
  {\path{doi:10.1093/imanum/drab102}}.

\bibitem{trucu2013}
D.~Trucu, P.~Lin, M.~A.~J. Chaplain, Y.~Wang, A multiscale moving boundary
  model arising in cancer invasion, Multiscale Model. Simul. 11 (2013)
  309--335.

\bibitem{trucu2019}
R.~Shuttleworth, D.~Trucu, {Multiscale Modelling of Fibres Dynamics and Cell
  Adhesion within Moving Boundary Cancer Invasion}, Bull. Math. Biol. 81 (2019)
  2176--2219.

\bibitem{trucu2020}
R.~Shuttleworth, D.~Trucu, Multiscale dynamics of a heterotypic cancer cell
  population within a fibrous extracellular matrix, J. Theor. Biol. 486 (2020)
  1--22.

\bibitem{Chaplain2019}
M.~A.~J. Chaplain, T.~Lorenzi, A.~Lorz, C.~Venkataraman, {Mathematical
  Modelling of Phenotypic Selection Within Solid Tumours}, in: F.~A. Radu,
  K.~Kumar, I.~Berre, J.~M. Nordbotten, I.~S. Pop (Eds.), Numerical Mathematics
  and Advanced Applications ENUMATH 2017, Springer International Publishing,
  2019, pp. 237--245.
\newblock \href {https://doi.org/10.1007/978-3-319-96415-7_20}
  {\path{doi:10.1007/978-3-319-96415-7_20}}.

\bibitem{Chisholm2016}
R.~H. Chisholm, T.~Lorenzi, L.~Desvillettes, B.~D. Hughes, {Evolutionary
  dynamics of phenotype-structured populations: from individual-level
  mechanisms to population-level consequences}, Z. Angew. Math. Phys. 67~(4)
  (2016) 100.
\newblock \href {https://doi.org/10.1007/s00033-016-0690-7}
  {\path{doi:10.1007/s00033-016-0690-7}}.

\bibitem{Macfarlane2022}
F.~R. Macfarlane, X.~Ruan, T.~Lorenzi, {Individual-based and continuum models
  of phenotypically heterogeneous growing cell populations}, AIMS Bioeng. 9~(1)
  (2022) 68--92.
\newblock \href {https://doi.org/10.3934/bioeng.2022007}
  {\path{doi:10.3934/bioeng.2022007}}.

\bibitem{Almeida2019}
L.~Almeida, P.~Bagnerini, G.~Fabrini, B.~D. Hughes, T.~Lorenzi, {Evolution of
  cancer cell populations under cytotoxic therapy and treatment optimisation:
  insight from a phenotype-structured model}, ESAIM: M2AN 53~(4) (2019)
  1157--1190.
\newblock \href {https://doi.org/10.1051/m2an/2019010}
  {\path{doi:10.1051/m2an/2019010}}.

\bibitem{Lorz2013}
A.~Lorz, T.~Lorenzi, M.~Hochberg, J.~Clairambault, B.~Perthame, {Populational
  adaptive evolution, chemotherapeutic resistance and multiple anti-cancer
  therapies}, ESAIM: M2AN 47~(2) (2013) 377--399.
\newblock \href {https://doi.org/10.1051/m2an/2012031}
  {\path{doi:10.1051/m2an/2012031}}.

\bibitem{Lorenzi2018}
T.~Lorenzi, C.~Venkataraman, A.~Lorz, M.~A.~J. Chaplain, {The role of spatial
  variations of abiotic factors in mediating intratumour phenotypic
  heterogeneity}, J Theor Biol 451 (2018) 101--110.
\newblock \href {https://doi.org/10.1016/j.jtbi.2018.05.002}
  {\path{doi:10.1016/j.jtbi.2018.05.002}}.

\bibitem{Villa2021}
C.~Villa, M.~A.~J. Chaplain, T.~Lorenzi, {Modeling the Emergence of Phenotypic
  Heterogeneity in Vascularized Tumors}, SIAM J Appl Math 81~(2) (2021)
  434--453.
\newblock \href {https://doi.org/10.1137/19M1293971}
  {\path{doi:10.1137/19M1293971}}.

\bibitem{Mark_Linnea_1}
L.~C. Franssen, T.~Lorenzi, A.~E.~F. Burgess, M.~A.~J. Chaplain, {A
  mathematical framework for modelling the metastatic spread of cancer}, Bull.
  Math. Biol. 81~(6) (2019) 1965--2010.
\newblock \href {https://doi.org/10.1007/s11538-019-00597-x}
  {\path{doi:10.1007/s11538-019-00597-x}}.

\bibitem{Mark_Linnea_2}
L.~C. Franssen, M.~A.~J. Chaplain, {A mathematical multi-organ model for
  bidirectional epithelial-mesenchymal transitions in the metastatic spread of
  cancer}, IMA J. Appl. Math. 85~(5) (2020) 724--761.
\newblock \href {https://doi.org/10.1093/imamat/hxaa022}
  {\path{doi:10.1093/imamat/hxaa022}}.

\bibitem{SfakKolbe2017}
N.~Sfakianakis, N.~Kolbe, N.~Hellmann, M.~Luk\'a\v{c}ov\'a-Medvid'ov\'a, {A
  Multiscale Approach to the Migration of Cancer Stem Cells: Mathematical
  Modelling and Simulations}, Bull. Math. Biol. 79 (2017) 209--235.
\newblock \href {https://doi.org/10.1007/s11538-016-0233-6}
  {\path{doi:10.1007/s11538-016-0233-6}}.

\bibitem{Chauviere2010}
A.~Chauviere, L.~Preziosi, H.~Byrne, {A model of cell migration within the
  extracellular matrix based on a phenotypic switching mechanism}, Math Med
  Biol 27~(3) (2009) 255--281.
\newblock \href {https://doi.org/10.1093/imammb/dqp021}
  {\path{doi:10.1093/imammb/dqp021}}.

\bibitem{hybrid_1st}
N.~Sfakianakis, A.~Madzvamuse, M.~A.~J. Chaplain, {A Hybrid Multiscale Model
  for Cancer Invasion of the Extracellular Matrix}, Multiscale Model. Simul.
  18~(2) (2020) 824--850.
\newblock \href {https://doi.org/10.1137/18M1189026}
  {\path{doi:10.1137/18M1189026}}.

\bibitem{hybrid_2nd}
L.~C. Franssen, N.~Sfakianakis, M.~A.~J. Chaplain, {A novel {3D}
  atomistic-continuum cancer invasion model: in silico simulations of an in
  vitro organotypic invasion assay}, J Theor Biol 522 (2021).
\newblock \href {https://doi.org/10.1016/j.jtbi.2021.110677}
  {\path{doi:10.1016/j.jtbi.2021.110677}}.

\bibitem{deutsch2012}
H.~Hatzikirou, D.~Basanta, M.~Simon, K.~Schaller, A.~Deutsch, {`Go or Grow':
  the key to the emergence of invasion in tumour progression?}, Math. Med.
  Biol. 29 (2012) 49--65.
\newblock \href {https://doi.org/10.1093/imammb/dqq011}
  {\path{doi:10.1093/imammb/dqq011}}.

\bibitem{deutsch2012b}
K.~B\"ottger, H.~Hatzikirou, A.~Chauviere, A.~Deutsch, Investigation of the
  migration/proliferation dichotomy and its impact on avascular glioma
  invasion, Math. Model. Nat. Phenom. 7 (2012) 105--135.

\bibitem{Yang2020}
{Yang, Jing and Antin, Parker and Berx, Geert and Blanpain, C\'edric and
  Brabletz, Thomas and Bronner, Marianne and Campbell, Kyra and Cano, Amparo
  and Casanova, Jordi and Christofori, Gerhard and Dedhar, Shoukat and Derynck,
  Rik and Ford, Heide L. and Fuxe, Jonas and Garc\'ia de Herreros, Antonio and
  Goodall, Gregory J. and Hadjantonakis, Anna-Katerina and Huang, Ruby Y. J.
  and Kalcheim, Chaya and Kalluri, Raghu and Kang, Yibin and Khew-Goodall,
  Yeesim and Levine, Herbert and Liu, Jinsong and Longmore, Gregory D. and
  Mani, Sendurai A. and Massagu\'e, Joan and Mayor, Roberto and McClay, David
  and Mostov, Keith E. and Newgreen, Donald F. and Nieto, M. Angela and
  Puisieux, Alain and Runyan, Raymond and Savagner, Pierre and Stanger, Ben and
  Stemmler, Marc P. and Takahashi, Yoshiko and Takeichi, Masatoshi and
  Theveneau, Eric and Thiery, Jean Paul and Thompson, Erik W. and Weinberg,
  Robert A. and Williams, Elizabeth D. and Xing, Jianhua and Zhou, Binhua P.
  and Sheng, Guojun and On behalf of the EMT International Association
  (TEMTIA)}, {Guidelines and definitions for research on
  epithelial--mesenchymal transition}, Nat. Rev. Mol. 21~(6) (2020) 341--352.
\newblock \href {https://doi.org/10.1038/s41580-020-0237-9}
  {\path{doi:10.1038/s41580-020-0237-9}}.

\bibitem{Dongre}
A.~Dongre, R.~A. Weinberg, {New insights into the mechanisms of
  epithelial--mesenchymal transition and implications for cancer}, Nat. Rev.
  Mol. 20~(2) (2019) 69--84.
\newblock \href {https://doi.org/10.1038/s41580-018-0080-4}
  {\path{doi:10.1038/s41580-018-0080-4}}.

\bibitem{Sha_2019}
Y.~Sha, D.~Haensel, G.~Gutierrez, H.~Du, X.~Dai, Q.~Nie, {Intermediate cell
  states in epithelial-to-mesenchymal transition}, Phys. Biol. 16~(2) (2019)
  021001.
\newblock \href {https://doi.org/10.1088/1478-3975/aaf928}
  {\path{doi:10.1088/1478-3975/aaf928}}.

\bibitem{Fazilaty2019}
H.~Fazilaty, L.~Rago, K.~Kass~Youssef, O.~H. Oca{\~n}a, F.~Garcia-Asencio,
  A.~Arcas, J.~Galceran, M.~A. Nieto, {A gene regulatory network to control EMT
  programs in development and disease}, Nat. Commun. 10~(1) (2019) 5115.
\newblock \href {https://doi.org/10.1038/s41467-019-13091-8}
  {\path{doi:10.1038/s41467-019-13091-8}}.

\bibitem{Lee2022b}
J.~H. Lee, J.~Massagu{\'e}, {TGF-$\beta$ in developmental and fibrogenic EMTs},
  Semin Cancer Biol 86~(Pt 2) (2022) 136--145.
\newblock \href {https://doi.org/10.1016/j.semcancer.2022.09.004}
  {\path{doi:10.1016/j.semcancer.2022.09.004}}.

\bibitem{Friedl2011}
P.~Friedl, S.~Alexander, {Cancer invasion and the microenvironment: plasticity
  and reciprocity}, Cell 147~(5) (2011) 992--1009.
\newblock \href {https://doi.org/10.1016/j.cell.2011.11.016}
  {\path{doi:10.1016/j.cell.2011.11.016}}.

\bibitem{Ort1998}
M.~Oft, K.-H. Heider, H.~Beug, {TGF$\beta$ signaling is necessary for carcinoma
  cell invasiveness and metastasis}, Curr. Biol. 8~(23) (1998) 1243--1252.
\newblock \href {https://doi.org/10.1016/S0960-9822(07)00533-7}
  {\path{doi:10.1016/S0960-9822(07)00533-7}}.

\bibitem{Xu2009}
J.~Xu, S.~Lamouille, R.~Derynck, {TGF-$\beta$-induced epithelial to mesenchymal
  transition}, Cell Res. 19~(2) (2009) 156--172.
\newblock \href {https://doi.org/10.1038/cr.2009.5}
  {\path{doi:10.1038/cr.2009.5}}.

\bibitem{Santos2022}
M.~Santos, M.~Ferreira, P.~Oliveira, N.~Mendes, A.~Andr{\'e}, A.~F. Vieira,
  J.~B. Nunes, J.~Carvalho, S.~Rocha, M.~Azevedo, D.~Ferreira, I.~Reis,
  J.~Vinagre, J.~Paredes, A.~Heravi-Moussavi, J.~Lima, V.~M{\'a}ximo,
  A.~Burleigh, C.~Roskelley, F.~Carneiro, D.~Huntsman, C.~Oliveira,
  {Epithelial-Mesenchymal Plasticity Induced by Discontinuous Exposure to
  TGF-$\beta$1 Promotes Tumour Growth}, Biology 11~(7) (2022).
\newblock \href {https://doi.org/10.3390/biology11071046}
  {\path{doi:10.3390/biology11071046}}.

\bibitem{Du2021}
W.~Du, X.~Liu, M.~Yang, W.~Wang, J.~Sun, {The Regulatory Role of PRRX1 in
  Cancer Epithelial-Mesenchymal Transition}, Onco. Targets. Ther. 14 (2021)
  4223--4229.
\newblock \href {https://doi.org/10.2147/OTT.S316102}
  {\path{doi:10.2147/OTT.S316102}}.

\bibitem{Giampieri2009}
S.~Giampieri, C.~Manning, S.~Hooper, L.~Jones, C.~S. Hill, E.~Sahai, {Localized
  and reversible TGF$\beta$ signalling switches breast cancer cells from
  cohesive to single cell motility}, Nat. Cell Biol. 11~(11) (2009) 1287--1296.
\newblock \href {https://doi.org/10.1038/ncb1973} {\path{doi:10.1038/ncb1973}}.

\bibitem{Hardin2014}
H.~Hardin, Z.~Guo, W.~Shan, C.~Montemayor-Garcia, S.~Asioli, X.-M. Yu, A.~D.
  Harrison, H.~Chen, R.~V. Lloyd, {The roles of the epithelial-mesenchymal
  transition marker PRRX1 and miR-146b-5p in papillary thyroid carcinoma
  progression}, Am. J. Pathol. 184~(8) (2014) 2342--2354.
\newblock \href {https://doi.org/10.1016/j.ajpath.2014.04.011}
  {\path{doi:10.1016/j.ajpath.2014.04.011}}.

\bibitem{Ocana2012}
O.~H. Oca{\~n}a, R.~C{\'o}rcoles, {\'A}.~Fabra, G.~Moreno-Bueno, H.~Acloque,
  S.~Vega, A.~Barrallo-Gimeno, A.~Cano, M.~A. Nieto, {Metastatic Colonization
  Requires the Repression of the Epithelial-Mesenchymal Transition Inducer
  Prrx1}, Cancer Cell 22~(6) (2012) 709--724.
\newblock \href {https://doi.org/10.1016/j.ccr.2012.10.012}
  {\path{doi:10.1016/j.ccr.2012.10.012}}.

\bibitem{Vazquez2006}
J.~L. Vazquez, {The Porous Medium Equation: Mathematical Theory}, Oxford
  University Press, 2006.
\newblock \href {https://doi.org/10.1093/acprof:oso/9780198569039.001.0001}
  {\path{doi:10.1093/acprof:oso/9780198569039.001.0001}}.

\bibitem{Byrne1999}
H.~M. Byrne, M.~A.~J. Chaplain, G.~J. Pettet, D.~L.~S. McElwain, A mathematical
  model of trophoblast invasion, Comput. Math. Method M. 1 (1999) 275--28.
\newblock \href {https://doi.org/10.1080/10273669908833026}
  {\path{doi:10.1080/10273669908833026}}.

\bibitem{Mascheroni2016}
P.~Mascheroni, C.~Stigliano, M.~Carfagna, D.~P. Boso, L.~Preziosi, P.~Decuzzi,
  B.~A. Schrefler, Predicting the growth of glioblastoma multiforme spheroids
  using a multiphase porous media model, Biomech. Model. Mechanobiol. 15(5)
  (2016) 1215--28.
\newblock \href {https://doi.org/10.1007/s10237-015-0755-0}
  {\path{doi:10.1007/s10237-015-0755-0}}.

\bibitem{Mascheroni2017}
P.~Mascheroni, D.~Boso, L.~Preziosi, B.~A. Schrefler, Evaluating the influence
  of mechanical stress on anticancer treatments through a multiphase porous
  media model, J. Theor. Biol. (2017) 179--188\href
  {https://doi.org/10.1016/j.jtbi.2017.03.027}
  {\path{doi:10.1016/j.jtbi.2017.03.027}}.

\bibitem{Winkler2011}
Y.~Tao, M.~Winkler, A chemotaxis-haptotaxis model: The roles of nonlinear
  diffusion and logistic source, SIAM J. Math. ANal. 43~(2) (2011) 685--704.
\newblock \href {https://doi.org/10.1137/100802943}
  {\path{doi:10.1137/100802943}}.

\bibitem{Byrne2001}
H.~M. Byrne, M.~A.~J. Chaplain, G.~J. Pettet, D.~L.~S. McElwain, An analysis of
  a mathematical model of trophoblast invasion, Appl. Math. Lett. 14~(8) (2001)
  1005--1010.
\newblock \href {https://doi.org/10.1016/S0893-9659(01)00079-9}
  {\path{doi:10.1016/S0893-9659(01)00079-9}}.

\bibitem{Sahai2020}
E.~Sahai, I.~Astsaturov, E.~Cukierman, D.~G. DeNardo, M.~Egeblad, R.~M. Evans,
  D.~Fearon, F.~R. Greten, S.~R. Hingorani, T.~Hunter, R.~O. Hynes, R.~K. Jain,
  T.~Janowitz, C.~Jorgensen, A.~C. Kimmelman, M.~G. Kolonin, R.~G. Maki, R.~S.
  Powers, E.~Pur{\'e}, D.~C. Ramirez, R.~Scherz-Shouval, M.~H. Sherman,
  S.~Stewart, T.~D. Tlsty, D.~A. Tuveson, F.~M. Watt, V.~Weaver, A.~T.
  Weeraratna, Z.~Werb, {A framework for advancing our understanding of
  cancer-associated fibroblasts}, Nat. Rev. Cancer 20~(3) (2020) 174--186.
\newblock \href {https://doi.org/10.1038/s41568-019-0238-1}
  {\path{doi:10.1038/s41568-019-0238-1}}.

\bibitem{Liu2019}
T.~Liu, C.~Han, S.~Wang, P.~Fang, Z.~Ma, L.~Xu, R.~Yin, {Cancer-associated
  fibroblasts: an emerging target of anti-cancer immunotherapy}, J. Hematol.
  Oncol. 12~(1) (2019) 86.
\newblock \href {https://doi.org/10.1186/s13045-019-0770-1}
  {\path{doi:10.1186/s13045-019-0770-1}}.

\bibitem{Ping2021}
Q.~Ping, R.~Yan, X.~Cheng, W.~Wang, Y.~Zhong, Z.~Hou, Y.~Shi, C.~Wang, R.~Li,
  {Cancer-associated fibroblasts: overview, progress, challenges, and
  directions}, Cancer Gene Ther. 28~(9) (2021) 984--999.
\newblock \href {https://doi.org/10.1038/s41417-021-00318-4}
  {\path{doi:10.1038/s41417-021-00318-4}}.

\bibitem{Yu2014}
Y.~Yu, C.-H. Xiao, L.~D. Tan, Q.-S. Wang, X.-Q. Li, Y.-M. Feng,
  {Cancer-associated fibroblasts induce epithelial-mesenchymal transition of
  breast cancer cells through paracrine TG-$\beta$ signalling}, Br J Cancer
  110~(3) (2014) 724--732.
\newblock \href {https://doi.org/10.1038/bjc.2013.768}
  {\path{doi:10.1038/bjc.2013.768}}.

\bibitem{Bornes2021}
L.~Bornes, G.~Belthier, J.~van Rheenen, {Epithelial-to-Mesenchymal Transition
  in the Light of Plasticity and Hybrid E/M States}, J. Clin. Med. 10~(11)
  (2021).
\newblock \href {https://doi.org/10.3390/jcm10112403}
  {\path{doi:10.3390/jcm10112403}}.

\bibitem{Winkler2020}
J.~Winkler, A.~Abisoye-Ogunniyan, K.~J. Metcalf, Z.~Werb, {Concepts of
  extracellular matrix remodelling in tumour progression and metastasis}, Nat.
  Commun. 11~(1) (2020) 5120.
\newblock \href {https://doi.org/10.1038/s41467-020-18794-x}
  {\path{doi:10.1038/s41467-020-18794-x}}.

\bibitem{Peixoto2019}
P.~Peixoto, A.~Etcheverry, M.~Aubry, A.~Missey, C.~Lachat, J.~Perrard,
  E.~Hendrick, R.~Delage-Mourroux, J.~Mosser, C.~Borg, J.-P. Feugeas, M.~Herfs,
  M.~Boyer-Guittaut, E.~Hervouet, {EMT is associated with an epigenetic
  signature of ECM remodeling genes}, Cell Death Dis. 10~(3) (2019) 205.
\newblock \href {https://doi.org/10.1038/s41419-019-1397-4}
  {\path{doi:10.1038/s41419-019-1397-4}}.

\bibitem{Gallego-Rentero2021}
M.~Gallego-Rentero, M.~Guti{\'e}rrez-P{\'e}rez, M.~Fern{\'a}ndez-Guarino,
  M.~Mascaraque, M.~Portillo-Esnaola, Y.~Gilaberte, E.~Carrasco,
  {\'A}.~Juarranz, {TGF$\beta$1 Secreted by Cancer-Associated Fibroblasts as an
  Inductor of Resistance to Photodynamic Therapy in Squamous Cell Carcinoma
  Cells}, Cancers 13~(22) (2021).
\newblock \href {https://doi.org/10.3390/cancers13225613}
  {\path{doi:10.3390/cancers13225613}}.

\bibitem{SfakKolb2016}
N.~Kolbe, J.~Katuchova, N.~Sfakianakis, N.~Hellmann,
  M.~Luk\'a\v{c}ov\'a-Medvid'ov\'a, {A study on time discretization and
  adaptive mesh refinement methods for the simulation of cancer invasion: The
  urokinase model}, Appl. Math. Comput. 273 (2022) 353--376.
\newblock \href {https://doi.org/10.1016/j.amc.2015.08.023}
  {\path{doi:10.1016/j.amc.2015.08.023}}.

\bibitem{glioma_2022}
A.~Dietrich, N.~Kolbe, N.~Sfakianakis, C.~Surulescu, {Multiscale Modeling of
  Glioma Invasion: From Receptor Binding to Flux-Limited Macroscopic PDEs},
  Multiscale Model. Simul. 20 (2022) 685--713.
\newblock \href {https://doi.org/10.1137/21M1412104}
  {\path{doi:10.1137/21M1412104}}.

\bibitem{SfakKolb2022}
N.~Kolbe, N.~Sfakianakis, {An adaptive rectangular mesh administration and
  refinement technique with application in cancer invasion models}, J. Comput.
  Appl. Math. 416 (2022) 114442.
\newblock \href {https://doi.org/10.1016/j.cam.2022.114442}
  {\path{doi:10.1016/j.cam.2022.114442}}.

\bibitem{Kolbe_2021}
N.~Kolbe, N.~Sfakianakis, C.~Stinner, C.~Surulescu, J.~Lenz, Modeling multiple
  taxis: Tumor invasion with phenotypic heterogeneity, haptotaxis, and
  unilateral interspecies repellence, Discrete and Continuous Dynamical Systems
  - B 26~(1) (2021) 443--481.
\newblock \href {https://doi.org/10.3934/dcdsb.2020284}
  {\path{doi:10.3934/dcdsb.2020284}}.

\bibitem{van1977towards}
B.~Van~Leer, Towards the ultimate conservative difference scheme. {IV}. {A} new
  approach to numerical convection, J. Comput. Phys. 23~(3) (1977) 276--299.
\newblock \href {https://doi.org/10.1016/0021-9991(77)90095-X}
  {\path{doi:10.1016/0021-9991(77)90095-X}}.

\bibitem{christopher2001additive}
C.~A. Kennedy, M.~H. Carpenter, Additive {R}unge-{K}utta schemes for
  convection-diffusion-reaction equations, Appl. Numer. Math. 1~(44) (2003)
  139--181.
\newblock \href {https://doi.org/10.1016/S0168-9274(02)00138-1}
  {\path{doi:10.1016/S0168-9274(02)00138-1}}.

\bibitem{Krylov.1931}
A.~N. Krylov, On the numerical solution of the equation by which in technical
  questions frequencies of small oscillations of material systems are
  determined, {Izvestiya Akademii Nauk SSSR, Otdel. mat. i estest. nauk.}
  VII~(4) (1931) 491--539.

\bibitem{vdVorst.1992}
H.~A. van~der Vorst, {Bi-CGSTAB}: A fast and smoothly converging variant of
  {Bi-CG} for the solution of nonsymmetric linear systems, SIAM J. Sci. Comput.
  13~(2) (1992) 631--644.
\newblock \href {https://doi.org/10.1137/0913035} {\path{doi:10.1137/0913035}}.

\bibitem{TGFb}
I.~Fuentes-Calvo, C.~Mart\'inez-Salgado, {TGFB1 (transforming growth factor,
  beta 1)}, \url{http://atlasgeneticsoncology.org/gene/42534/}, {Atlas Genet
  Cytogenet Oncol Haematol} (2013).

\bibitem{Stokes1991a}
C.~L. Stokes, D.~A. Lauffenburger, S.~K. Williams, {Migration of individual
  microvessel endothelial cells: stochastic model and parameter measurement}, J
  Cell Sci. 99~(2) (1991) 419--30.
\newblock \href {https://doi.org/10.1242/jcs.99.2.419}
  {\path{doi:10.1242/jcs.99.2.419}}.

\bibitem{Stokes1990}
C.~L. Stokes, S.~K. Williams, D.~A. Lauffenburger, {Chemotaxis of human
  microvessel endothelial cells in response to acidic fibroblast growth
  factor}, Lab. Invest. 63 (1990) 657--668.

\bibitem{Terranova1985}
V.~P. Terranova, R.~DiFlorio, R.~M. Lyall, S.~Hic, R.~Friesel, T.~Maciag,
  {Human endothelial cells are chemotactic to endothelial cell growth factor
  and heparin}, J Cell Biol. 101~(6) (1985) 2330--2334.
\newblock \href {https://doi.org/10.1083/jcb.101.6.2330}
  {\path{doi:10.1083/jcb.101.6.2330}}.

\bibitem{Anderson1998}
A.~R.~A. Anderson, M.~A.~J. Chaplain, {Continuous and discrete mathematical
  models of tumor-induced angiogenesis}, Bull. Math. Biol. 60 (1998) 857--899.
\newblock \href {https://doi.org/10.1006/bulm.1998.0042}
  {\path{doi:10.1006/bulm.1998.0042}}.

\bibitem{Chaplain1996}
M.~A.~J. Chaplain, {Avascular growth, angiogenesis and vascular growth in solid
  tumours: The mathematical modelling of the stages of tumour development},
  Math Comput Model 23~(6) (1996) 47--87.
\newblock \href {https://doi.org/10.1016/0895-7177(96)00019-2}
  {\path{doi:10.1016/0895-7177(96)00019-2}}.

\bibitem{Nieto2017}
M.~A. Nieto, Context-specific roles of emt programmes in cancer cell
  dissemination, Nat. Cell Biol. 19~(5) (2017) 416--418.
\newblock \href {https://doi.org/10.1038/ncb3520} {\path{doi:10.1038/ncb3520}}.

\bibitem{alexander2017}
M.~Alexander, G.~Acosta~Gonzalez, S.~Malerba, T.~Hochman, J.~D. Goldberg, D.~F,
  Multifocal invasive ductal cancer: Distinguishing independent tumor foci from
  multiple satellites., Int. J. Surg. Pathol. 25(4) (2017) 298--303.
\newblock \href {https://doi.org/10.1177/1066896916676586}
  {\path{doi:10.1177/1066896916676586}}.

\bibitem{moon2013}
H.~G. Moon, W.~Han, J.~Y. Kim, S.~J. Kim, J.~H. Yoon, S.~J. Oh, J.~H. Yu, N.~D.
  Y, Effect of multiple invasive foci on breast cancer outcomes according to
  the molecular subtypes: a report from the korean breast cancer society., Ann.
  Oncol. 24 (2013) 2298--2304.
\newblock \href {https://doi.org/10.1093/annonc/mdt187}
  {\path{doi:10.1093/annonc/mdt187}}.

\bibitem{johnson2011}
L.~Johnson, S.~Pinder, M.~Douek, Multiple foci of invasive breast cancer: can
  breast mri influence surgical management?, Breast Cancer Res. Treat. 128(1)
  (2011) 1--5.
\newblock \href {https://doi.org/10.1007/s10549-011-1491-5}
  {\path{doi:10.1007/s10549-011-1491-5}}.

\end{thebibliography}

\end{document}